\documentclass[preprint]{elsarticle}

\usepackage[utf8]{inputenc}

\journal{\href{https://doi.org/10.1016/j.jlamp.2018.12.006}{Journal of Logical and Algebraic Methods in Programming}\ }

\usepackage{multirow}
\usepackage[table]{xcolor}
\usepackage{hhline}
\usepackage{tikz} 
 \usetikzlibrary{matrix,chains,graphs,backgrounds,positioning,fit,decorations.pathmorphing,shapes,arrows,calc,shadows}
\usepackage{graphicx}
\usepackage{xspace}
\usepackage{amstext}
\usepackage{amsmath}
\usepackage{amssymb}
  \numberwithin{equation}{section} 
\usepackage{stmaryrd}
\usepackage{verbatim}
\usepackage[ruled,vlined,linesnumbered]{algorithm2e}
\usepackage{bm}
\usepackage[all,cmtip]{xy}
\usepackage[hidelinks]{hyperref} 

\newdefinition{defn}{Definition}[section]
\newdefinition{example}[defn]{Example}
\newdefinition{assumption}[defn]{Assumption}
\newtheorem{theorem}[defn]{Theorem}
\newtheorem{corollary}[defn]{Corollary}

\newtheorem{lemma}[defn]{Lemma}
\newtheorem{fact}[defn]{Fact}

\newproof{proof}{Proof}

\newcommand{\figref}[1]{\hyperref[#1]{Fig.~\ref*{#1}}}
\newcommand{\tabref}[1]{\hyperref[#1]{Table~\ref*{#1}}}
\newcommand{\sectref}[1]{\hyperref[#1]{Sect.~\ref*{#1}}}
\newcommand{\sectionref}[1]{\hyperref[#1]{Section~\ref*{#1}}}
\newcommand{\stepref}[1]{\hyperref[#1]{Step~\ref*{#1}}}
\newcommand{\defref}[1]{\hyperref[#1]{Def.~\ref*{#1}}}
\newcommand{\thmref}[1]{\hyperref[#1]{Thm.~\ref*{#1}}}
\newcommand{\propref}[1]{\hyperref[#1]{Prop.~\ref*{#1}}}
\newcommand{\lemmaref}[1]{\hyperref[#1]{Lemma~\ref*{#1}}}
\newcommand{\exref}[1]{\hyperref[#1]{Example~\ref*{#1}}}
\newcommand{\obsref}[1]{\hyperref[#1]{Observation~\ref*{#1}}}
\newcommand{\caseref}[1]{\hyperref[#1]{Case~\ref*{#1}}}
\newcommand{\corelref}[1]{\hyperref[#1]{Cor.~\ref*{#1}}}
\newcommand{\conjref}[1]{\hyperref[#1]{Conjecture~\ref*{#1}}}
\newcommand{\factref}[1]{\hyperref[#1]{Fact~\ref*{#1}}}
\newcommand{\algref}[1]{\hyperref[#1]{Alg.~\ref*{#1}}}
\newcommand{\footnoteref}[1]{\hyperref[#1]{footnote~\ref*{#1}}}
\newcommand{\alineref}[1]{\hyperref[#1]{line~\ref*{#1}}}
\newcommand{\funref}[2][]{%
  \def\test{#1}%
  {\normalfont\hyperref[#2]{\ref*{#2}\ifx\test\empty\else$(#1)$\fi}}%
}
\newcommand{\assref}[1]{\hyperref[#1]{Assumption~\ref*{#1}}}

\def\LL(#1){\ensuremath{\mathrm{LL}(#1)}}
\def\LR(#1){\ensuremath{\mathrm{LR}(#1)}}
\def\LALR(#1){\ensuremath{\mathrm{LALR}(#1)}}
\def\SLR(#1){\ensuremath{\mathrm{SLR}(#1)}}
\def\SLL(#1){\ensuremath{\mathrm{SLL}(#1)}}
\def\SLRo(#1){\ensuremath{\mathrm{SLR}^\bullet(#1)}}
\def\SLLo(#1){\ensuremath{\mathrm{SLL}^\bullet(#1)}}

\def\PSR{\ensuremath{\mathrm{PSR}}\xspace}
\def\PTD{\ensuremath{\mathrm{PTD}}\xspace}
\def\Fi{\mathit{First}}
\def\Fo{\mathit{Follow}}
\def\FoA{\mathit{Follow}^*}
\def\FiFo#1(#2)(#3){\Fi_{#1}(#3) \cdot_{#1} \Fo_{#1}(#2)}

\def\inputbinding{input binding}
\def\Inputbinding{Input binding}
\def\dCFAstate{dCFA state}
\def\concretestate{state}
\def\pow#1{2^#1}
\def\pfun(#1,#2){(#1 \pto #2)}
\newcommand{\emptyseq}{\varepsilon}        
\newcommand{\Lab}{\Sigma}  
\newcommand{\TLab}{\mathcal{T}}  
\newcommand{\NLab}{\mathcal{N}}  
\newcommand{\arity}{\mathit{arity}}   
\newcommand{\lab}{\ell}    
\newcommand{\Nat}{\mathbb{N}}

\newcommand{\D}{{\mathcal{D}}}

\renewcommand{\L}{{\mathcal{L}}}

\newcommand{\R}{\mathcal{R}}

\def\pto{\rightharpoonup}
\def\dom{\mathit{dom}}
\newcommand{\To}{\drm}
\newcommand{\too}{\displaystyle\operatornamewithlimits{\to}}
\newcommand{\Too}{\displaystyle\operatornamewithlimits{\drm}}
\newcommand{\lrQ}[1]{\begin{array}{|@{\,}l@{\,\to\,}l@{\,}|} \hline%
   #1 \hline \end{array}}
\newcommand{\LRo}[3]{#1 & #2 \,{\centerdot}\, #3 \\}
\newcommand{\lro}[3]{#1 \to #2 \,{\centerdot}\, #3}

\newcommand{\set}[1]{\{ #1 \}}
\newcommand{\CDOT}{{\,\centerdot\,}}
\newcommand{\lpar}{{[}}
\newcommand{\rpar}{{]}}

\newcommand{\xyrightarrow}[3]{\mathop{\!\!\!\xymatrix@C=#1{{}\ar@{#2}[r]_{#3}&{}}\!\!\!}\nolimits}
\newcommand{\xyRightarrow}[3]{\mathop{\!\!\!\xymatrix@C=#1{{}\ar@<1pt>@{#2}[r]\ar@<-1pt>@{#2}[r]_{#3}&{}}\!\!\!}\nolimits}
\newcommand{\xyarrow}[2]{%
  \sbox{0}{$\scriptstyle#2$}%
  \mathop{\!\!\!\xymatrix@C\dimexpr\wd0+4pt\relax{{}\ar@{#1}[r]_{#2}&{}}\!\!\!}%
}
\newcommand{\xyArrow}[2]{%
  \sbox{0}{$\scriptstyle#2$}%
  \mathop{\!\!\!\xymatrix@C\dimexpr\wd0+4pt\relax{{}\ar@<1pt>@{#1}[r]\ar@<-1pt>@{#1}[r]_{#2}&{}}\!\!\!}%
}
\def\vdasharrow#1#2{\xyrightarrow{#1}{|-}{#2}}
\def\vDasharrow#1#2{\xyRightarrow{#1}{|-}{#2}}
\def\vtildearrow#1#2{\xyrightarrow{#1}{|~}{#2}}
\def\vTildearrow#1#2{\xyRightarrow{#1}{|~}{#2}}
\def\buc(#1,#2){#1 \CDOT #2}
\def\lstep#1{\xyarrow{|-}{#1}}
\def\cstep{\vdasharrow{3mm}{}}
\def\shift{\lstep{\mathsf{sh}}}
\def\literal#1{\bm{#1}}
\let\der\Rightarrow
\let\drm\Rightarrow
\def\epsilon{\varepsilon}
\def\emptyset{\varnothing}
\def\theta{\vartheta}
\def\rho{\varrho}
\def\phi{\varphi}
\def\Lit#1{\mathit{Lit}_{#1}}
\def\autoconf(#1,#2,#3){#3 {\scriptstyle\lozenge} [#1]^{#2}}
\def\Autoconf(#1,#2,#3){#3 & {\scriptstyle\lozenge} & [#1]^{#2}}
\def\autostep{\vtildearrow{4mm}{}}
\def\consumestep{\vtildearrow{4mm}{\textsf{\upshape go}}}
\def\expandstep{\vtildearrow{4mm}{\textsf{\upshape cl}}}
\def\atrans#1{\mathbin{\xrightarrow{#1}}}
\def\qrule#1{(#1)}
\def\unknown{\text{--}}
\def\leave(#1){\mathit{leave}(#1)}
\def\startstate{Q_0}
\def\acceptstate{Q_{\mathsf{A}}}
\def\startLit{\literal Z}
\def\pseudoSym{\mathit{Start}}
\def\pseudoLit{\literal{Start}}
\def\graph(#1,#2){\langle {#1}, {#2}\rangle}
\def\Gr#1{\mathcal{G}_{#1}}
\def\lit#1{\mathit{lit}(#1)}
\def\pstate(#1,#2){\langle #1, #2 \rangle}
\def\similar{\approx}
\def\cfaconf(#1,#2){#1 {\scriptstyle\blacklozenge} #2}
\def\cfastep{\vTildearrow{4mm}{}}
\def\psrstack#1{\mathcal{#1}}
\def\psrgraph#1{\mathit{graph}(#1)}
\def\mappedStates{\mathcal{Q}_{\mathrm C}}
\def\perm{\bowtie}
\def\id{\mathrm{id}}
\def\Tgt{I}
\def\clps{\vartriangleright} 
\def\eoi{\$}
\def\nCFA{\mathfrak A}
\def\dCFA{\mathfrak C}
\def\startSym{Z}
\def\startedgeL{\mathit{root}}
\def\edgeL{e}

\def\startEdge#1{\startedgeL^{#1}}
\def\nontT#1{T^{#1}}
\def\nontM{M}
\def\nontA{T}
\def\nontB{B}
\def\edge#1#2{\edgeL^{#1 #2}}

\def\startLi{\startSym()}
\def\pseudoLi{\pseudoSym()}
\def\startedgeLi#1{\startedgeL(#1)}
\def\TnontLi#1{T(#1)}
\def\edgeLi#1#2{\edgeL(#1,#2)}

\newcommand{\persuade}{\ensuremath{\mathit{per}}}
\newcommand{\try}{\ensuremath{\mathit{try}}}
\newcommand{\believe}{\ensuremath{\mathit{bel}}}
\newcommand{\items}{\mathcal I}
\def\persuadeL#1#2#3#4{\persuade(#1,#2,#3,#4)}
\def\tryL#1#2#3{\try(#1,#2,#3)}
\def\believeL#1#2#3{\believe(#1,#2,#3)}
\def\Mlit#1#2{M(#1,#2)}
\def\params{\mathit{params}}
\def\parama{a}
\def\paramb{b}
\def\paramc{c}
\newcommand{\fits}[2]{\mathrel{\mathrm{fits}^#1_#2}}
\def\startstate{Q_0}
\def\psrtop#1{\mathit{top}(#1)}

\def\step#1{\,\displaystyle\mathop{\vdash}\limits_{#1}\,}
\newcommand{\pstep}[1][]{%
  \def\test{#1}%
  \ifx\test\empty\models\else\displaystyle\mathop{\models}\limits_{#1}\fi%
}
\newcommand{\psrstep}[1][]{%
  \def\test{#1}%
  \ifx\test\empty\Vdash\else\displaystyle\mathop{\Vdash}\limits_{#1}\fi%
}
\def\trans#1{\stackrel{#1}{\to}} %

\DontPrintSemicolon
\SetKwProg{Proc}{procedure}{}{}
\SetKwInOut{Input}{Input}
\SetKwInOut{Output}{Output}
\SetKwFor{Loop}{loop}{}{}
\SetAlgoHangIndent{1em}
\SetNlSty{textsf}{}{}

\def\lpstep#1{\mathrel{\vDasharrow{3mm}{#1}}}
\def\acceptstack{{\psrstack S}_{\mathrm{A}}}
\def\Rest{\mathit{Success}}
\def\Any{\mathit{Any}}
\def\Known{\bullet}
\def\psrtr{\mathit{tr}}
\def\psrit{\mathit{it}}
\def\psrbuc(#1,#2,#3){#1 \CDOT #2 | #3}
\def\lcfastep#1{\vTildearrow{4mm}{#1}}

\def\substi#1#2{{#1}/{#2}}

\pgfdeclarelayer{background}
\pgfdeclarelayer{foreground}
\pgfsetlayers{background,main,foreground}
\tikzset{%
  x=8mm,y=8mm,>=latex, 
  arm/.style={-,very thick,draw=gray!50},  
  link/.style={->,decorate,
         decoration={snake,amplitude=0.5pt,segment length=3pt,post length=2pt}},
  read/.style={<-,decorate,
         decoration={snake,amplitude=0.5pt,segment length=3pt,pre length=2pt}},
  rcv/.style={-,double,draw=black},
  o/.style={circle,draw,inner sep=0.0pt,minimum size=1mm},
  O/.style={o,minimum size=2mm},
  b/.style={o,minimum size=1mm,fill},
  st/.style={o,minimum size=0.8mm,fill=white},
  adaptive/.style={st,draw,double},
  bst/.style={st,double},
  glass/.style={draw opacity=0,fill opacity=0},     
  satin/.style={draw opacity=0.3,fill opacity=0.3},
  opaque/.style={draw opacity=1,fill opacity=1},
  elab/.style={text=black,font=\scriptsize},
  schema/.style={rounded rectangle,minimum size=4mm,
                 draw,fill=white,
                 inner sep=2pt,font=\sffamily\scriptsize},
  blob/.style={rectangle,rounded corners=6mm,fill opacity=0.4,draw,very thin},
  term/.style={rectangle,minimum size=1.5mm,draw,fill=white,
               inner sep=1pt,font=\scriptsize},
  rule/.style={shape=diamond,aspect=1.4,minimum height=5mm,minimum width=7mm
              ,draw,fill=gray!50,inner sep=0.5pt,font=\sffamily\scriptsize},
  ctxt/.style={term,fill=green!25},
  root/.style={term,fill=red!25},
  nsort/.style={rectangle,minimum width=4mm,minimum height=4mm,fill=white,
                inner sep=2pt,font=\scriptsize\it},
  op/.style={nsort,
             draw
           },
  graphscheme/.style={shape=ellipse, draw,fill=white,
                      inner sep=2pt,minimum size=3mm},
  hy/.style={op},
  nont/.style={op,fill=gray!30},
  pred/.style={nont,fill=green!25},
  var/.style={op,draw=blue!80!black,fill=blue!20,
              font=\scriptsize},
  idf/.style={font=\sffamily\tiny},
  subgraph/.style={draw,fill=gray!20,inner sep=2pt},
  mult/.style={circle,minimum size=2.5mm,
               draw,fill=gray!20,
               inner sep=0pt,font=\tiny},
  numb/.style={font=\tiny},
  branch/.style={diamond, aspect=1.8,
                      minimum height=3.5mm,minimum width=6mm,
                      semithick,draw=black,
                      inner sep=0pt},
  assign/.style={rectangle,aspect=1.8,minimum height=3mm,minimum width=5mm,
                 draw,semithick,
                 inner sep=1.5pt,},
  triang/.style={isosceles triangle,minimum width=2mm,minimum height=1.0mm,
                 draw=black,inner sep=0pt},
  Triang/.style={triang,fill=black},
  start/.style={triang,shape border rotate=270},
  stop/.style={triang,shape border rotate=90},
  Start/.style={triang,shape border rotate=270,fill=black},
  Stop/.style={triang,shape border rotate=90,fill=black},
  Startinner/.style={Start, fill=gray!50},
  Stopinner/.style={Startinner,densely dotted,double distance=1pt},
  State/.style={Sinner, densely dotted},
  Scompartment/.style={Sinner,densely dashed,minimum size=4mm},
  Scondition/.style={rounded rectangle,fill=gray!25},
  LRq/.style={rectangle,fill=white,inner sep=0pt,font=\small},
  LRf/.style={rectangle,draw,fill=white,inner sep=1pt,font=\small}
  }
\makeatletter

\def\DownEdge[#1]{\tikz \draw (0pt,8pt) edge[#1] (0pt,0pt);}

\def\enlargebb{\node [glass,fit= (current bounding box),inner sep=2pt] {};}

\def\Just#1%
  {\tikz \draw (0pt,0pt) edge[#1] (12pt,0pt); 
   \end{graph}}
\def\Graph{\@ifnextchar[{\@Graph}{\@GrapH}}
\def\@Graph[#1]#2{%
  \BOX{%
    \begin{tikzpicture}[x=8mm,y=8mm,label distance=-2pt,>=latex,#1]
     #2%
    \end{tikzpicture}%
  } 
}
\def\@GrapH#1{%
  \BOX{%
    \begin{tikzpicture}[x=8mm,y=8mm,label distance=-2pt,>=latex]
     #1%
    \end{tikzpicture}%
  } 
}

\def\proGraph{\@ifnextchar[{\@proGraph}{\@pro@Graph}}
\def\@proGraph[#1]#2{\BOX{%
  \begin{tikzpicture}[x=8mm,y=-7mm,>=latex,every label/.style={elab},#1]
     #2
    \enlargebb
   \end{tikzpicture}}}
\def\@pro@Graph#1{\BOX{%
  \begin{tikzpicture}[x=8mm,y=-7mm,>=latex,every label/.style={elab}]
    #1
    \enlargebb
  \end{tikzpicture}}}

\def\uvar(#1)#2{\@ifnextchar[{\@uvar(#1)#2}{%
  \node (#1-node) at ($(#1)+(0,1)$) {#2};
  \path[arm] (#1) edge (#1-node);
}}
\def\@uvar(#1)#2[#3]{%
  \node (#1-node) at ($(#1)+(#3,1)$) {#2};
  \path[arm] (#1) edge (#1-node);
}
\def\ustar#1#2{\@ifnextchar[{\@ustar{#1}{#2}}{\Graph{%
      \node (r) [term] at (1,1) {#1};
      \node (h) [nont] at (1,2) {#2};
      \path (r) edge[arm] (h);
}}}
\def\@ustar#1#2[#3]{\Graph[#3]{%
      \node (r) [term] at (1,1) {#1};
      \node (h) [nont] at (1,2) {#2};
      \path (r) edge[arm] (h);
}}

\def\custar#1#2#3{\@ifnextchar[{\@custar(#1)#2}{\proGraph{%
      \node (r) [term] at (1,1) {#1};
      \node (h) [nont] at (1,2) {#2};
      \path (r) edge[arm] (h);
      \node (c) [term] at (1,3) {#3};
}}}
\def\@custar#1#2#3[#4]{\proGraph[#4]{%
      \node (r) [term] at (1,1) {#1};
      \node (h) [nont] at (1,2) {#2};
      \path (r) edge[arm] (h);
      \node (c) [term] at (1,3) {#3};
}}
\makeatother

\newcommand{\BOX}[1]{\begin{array}{@{}c@{}}#1\end{array}}

\begin{document}

\begin{frontmatter}

\title{\begin{picture}(0,0)
  \put(-35,50){\parbox{\textwidth}{\sf\small\copyright\ 2018. This manuscript version is made available under the CC-BY-NC-ND 4.0 license \href{http://creativecommons.org/licenses/by-nc-nd/4.0}{http://creativecommons.org/licenses/by-nc-nd/4.0}}}
\end{picture}Formalization and Correctness of Predictive Shift-Reduce Parsers for Graph Grammars based on Hyperedge Replacement\tnoteref{short}}
\tnotetext[short]{This paper formalizes the concepts described in~\cite{Drewes-Hoffmann-Minas:17} and provides
detailed correctness proofs for them.}


\author[FD]{Frank Drewes}
\ead{drewes@cs.umu.se}
\address[FD]{Institutionen för datavetenskap, 
             Umeå universitet, SE-901 87 Umeå, Sweden}

\author[BH]{Berthold Hoffmann}
\ead{hof@uni-bremen.de}
\address[BH]{Fachbereich 3---Informatik, Universität Bremen, 
  D-28334 Bremen, Germany}

\author[MN]{Mark Minas}
\ead{mark.minas@unibw.de}
\address[MN]{Institut für Softwaretechnologie, 
Fakultät für Informatik, \\
Universität der Bundeswehr München 
D-85577 Neubiberg, Germany}

\begin{abstract}
  Hyperedge replacement (HR) grammars can generate NP-complete graph
  languages, which makes parsing hard even for fixed HR languages.
  Therefore, we study predictive shift-reduce (\PSR) parsing that
  yields efficient parsers for a subclass of HR grammars, by
  generalizing the concepts of \SLR(1) string parsing to graphs.
  We formalize the construction of \PSR parsers and show that it is
  correct.
  \PSR parsers run in linear space and time, and are more efficient
  than the predictive top-down (\PTD) parsers recently developed by
  the authors.
\end{abstract}


\end{frontmatter}



\section{Introduction\label{sec:intro}}

Everywhere in science and beyond, diagrams occur as a means of illustration
and explanation.
In computer science and engineering, they are also used as primary
source of information: they form visual specification languages
with a precise syntax and semantics.
For instance, the diagrams of the \emph{Uniform Modeling Language}
\textsc{uml} specify software artifacts. (See \url{www.uml.org}.)
When diagram languages shall be processed by
computers, techniques of compiler construction 
have to be transferred to the domain of diagrams.
A processor of a textual language parses its syntax, which is
specified by a context-free Chomsky grammar, in order to construct an
abstract hierarchical representation that can then be further
interpreted or translated.
The syntax of a diagram language is its structure. To analyze the structure of diagrams, one thus needs grammars to specify their syntax,
and parsers for these grammars that perform the analysis. A successfully parsed diagram can eventually be processed further.
Since diagrams can be represented as graphs, their syntax
can be captured by graph grammars.

Here we consider hyperedge replacement
(HR) graph grammars.%
\footnote{Other graph grammars and parsing algorithms are
  discussed in \sectref{s:concl}.}  Hyperedges are a
generalization of edges that may connect any number of nodes, not just
two.  In a host graph $g$, the replacement of a hyperedge $\literal e$ by a
graph $\gamma$ glues the nodes connected to $\literal e$ to distinguished nodes of
$\gamma$. The context-free case, where the replacement depends just on the
label of $\literal e$ (a nonterminal symbol) is well studied \cite{habel:92}.
Unfortunately, hyperedge replacement can
generate NP-complete graph languages
\cite{Aalbersberg-Ehrenfeucht-Rozenberg:86}. In other words, even for
fixed HR languages parsing is hard. Moreover, even if restrictions are
employed that guarantee an HR language~$L$ to be in P, the degree of the polynomial
depends on $L$; see~\cite{Lautemann:90}.%
\footnote{The polynomial algorithm for a restricted class of
 (fixed) HR grammars presented in~\cite{Lautemann:90} was refined in~\cite{chiang-et-al:2013} and
  implemented in the system \emph{Bolinas} for semantic parsing in natural language
 processing.} %
Only under rather strong restrictions the problem is known to be
solvable in cubic time \cite{Vogler:91,Drewes:93c}.

Since even a cubic algorithm would not scale to diagrams occurring in
realistic applications, the authors have recently transferred results
of context-free string parsing to graphs: Simple LL-parsing
(\SLL(k) for short, \cite{Lewis-Stearns:68}), a top-down parsing
method that applies to a subclass of unambiguous context-free string
grammars (using $k$ symbols of lookahead), has been lifted to
\emph{predictive top-down parsing} of graphs (PTD parsing for
short, \cite{Drewes-Hoffmann-Minas:15}); the program generating PTD parsers approximates Parikh images
of auxiliary grammars in order to determine whether a grammar is
PTD-parsable \cite{Drewes-Hoffmann-Minas:16}, and generates parsers
that run in quadratic time, and in many cases in linear time.

In this paper, we devise---somewhat complementary---efficient bottom-up parsers
for HR grammars, called \emph{predictive shift-reduce (PSR) parsers},
which extend \SLR(1) parsers \cite{DeRemer:71}, a member of the \LR(k)
family of deterministic bottom-up parsers for context-free
string grammars \cite{Knuth:65}.
We formalize the construction and \emph{modus operandi} of PSR
parsers and show their correctness.
In \sectref{s:HRG} we recall basic notions of HR grammars.
To support intuition, we briefly recall \SLR(1) string parsing in \sectref{s:PSRintro}. In Sections~\ref{s:naive}--\ref{s:PSR},
we work out in detail how it can be lifted to PSR parsing:

\sectionref{s:naive} develops a naïve shift-reduce parser for HR grammars and shows its correctness. This parser is a stack automaton that, one by one, reads the edges of the input graph and simply ``guesses'' nondeterministically a backwards application of rules that takes the input graph to the start symbol. While this parser is correct, its nondeterminism renders it impractical. One of its disadvantages is that it can run into ``dead ends'', situations which can never lead to acceptance, regardless of the remaining input.

\sectionref{s:viable} defines a notion of \emph{viable prefixes} and shows that the naïve shift-reduce parser would avoid running into a dead end if and only if one could make sure that its stack does always contain a viable prefix.

\sectionref{s:nCFA} thus develops a notion of \emph{nondeterministic characteristic finite automaton} (nCFA) and shows that it recognizes (we say \emph{approves}) exactly the viable prefixes. However, since the nCFA is itself nondeterministic, it cannot reasonably be used in order to improve the naïve shift-reduce parser.

\sectionref{s:dCFA}, therefore, shows how the nCFA can be converted into a \emph{deterministic characteristic finite automaton} (dCFA) that is equivalent to the nCFA.

\sectionref{s:ASR} incorporates the dCFA into an improved version of the naïve shift-reduce parser. The \emph{dCFA-assisted shift-reduce parser} makes sure that it can always continue a parse found so far to a successful parse, but possibly not with the current input graph, i.e., it may still run into a dead end. 

Finally, \sectref{s:PSR} discusses how dCFA-assisted shift-reduce parsers 
can be further extended to \emph{predictive shift-reduce parsers} (PSR parsers) which can predict their next move in every situation such that they can never run into dead ends. These parsers, for all practical purposes, run in linear time and space, but they exist only for HR grammars without \emph{conflicts}
and satisfying the so-called \emph{free edge choice property}. 
\sectref{s:PSR} formalizes conflicts as well the free edge choice property, and shows how conflicts can be detected.

Related and future work is discussed in \sectref{s:concl}.

This paper formalizes the concepts developed in~\cite{Drewes-Hoffmann-Minas:17} and provides
detailed correctness proofs for them. We would also like to mention
that the proof of Theorem~1 of~\cite{Drewes-Hoffmann-Minas:17} turned
out to be wrong.

\paragraph*{Acknowledgment} We thank the reviewers for their comments and criticism; we hope to have made good use of them.



\section{Hyperedge Replacement Grammars}\label{s:HRG}
We let $\Nat$ denote the non-negative integers. For set $A$ and $B$,
let $\pow A$ denote the powerset of $A$ and $\pfun(A,B)$ the set of
all partial functions from $A$ to $B$. The domain of a partial function
$f\colon A\pto B$ is denoted by $\dom(f)$, i.e.,
$\dom(f)=\{a\in A\mid\text{$f(a)$ is defined}\}$. For $S\subseteq A$,
we let $f(S)=\{f(a)\mid a\in S\cap\dom(f)\}$. Given two partial
functions $f$ and $g$, we write $f\sqsubseteq g$ if $f\subseteq g$ as
binary relations. The composition $g\circ f$ of (possibly partial)
functions $f\colon A\pto B$ and $g\colon B\pto C$ is defined as usual,
i.e., $(g\circ f)(a)$ equals $g(f(a))$ if both $f(a)$ and $g(f(a))$
are defined, and is undefined otherwise.

$A^*$ denotes
the set of all finite sequences (or strings) over a set $A$; the empty sequence is
denoted by $\emptyseq$ and the length of a sequence $\alpha$ by $|\alpha|$.


For a (total) function $f\colon A \to B$, its extension
$f^*\colon A^* \to B^*$ to sequences is defined by
$f^*(a_1 \cdots a_n) = f(a_1) \cdots f(a_n)$, for all
$a_1, \dots, a_n \in A$, $n\ge 0$.
Given a relation ${\leadsto}\subseteq A \times A$, we denote its $n$-fold composition with itself by $\leadsto^n$ (where $\leadsto^0$ is the identity on $A$), its transitive 
closure by $\leadsto^+$ and its reflexive and transitive closure by 
$\leadsto^*$, as usual.

Throughout the paper, we let $X$ denote a global, countably infinite supply of \emph{nodes} or \emph{vertices}.

\begin{defn}[Graph]%
\label{def:graphs}  
  An \emph{alphabet} is a set $\Lab$ of \emph{symbols} together with an
  \emph{arity function} $\arity\colon \Lab \to \Nat$. Given such an alphabet, a
  \emph{literal $\literal e = a(x_1, \dots, x_k)$ over $\Lab$} consists
  of a symbol $a \in \Lab$ and $k=\arity(a)$ pairwise distinct %
  nodes $x_1,\dots,x_k\in X$. We write $\lab(\literal e) = a$ and denote the set of all literals over $\Lab$ by $\Lit{\Lab}$. 
  
  A \emph{graph $\gamma = \graph(V, \phi)$ over $\Lab$} consists of a finite set 
  $V \subseteq X$ of nodes and a sequence
  $\phi = \literal e_1 \cdots \literal e_n\in\Lit{\Lab}^*$ such that all
  nodes in these literals are in~$V$. $\Gr{\Lab}$ denotes the set of all graphs over ${\Lab}$. %

  We say that two graphs $\gamma = \graph(V, \phi)$ and 
  $\gamma' = \graph(V', \phi')$ are \emph{equivalent}, written
  $\gamma \perm \gamma'$, if $V = V'$ and $\phi$ is a permutation 
  of~$\phi'$.
\end{defn}

Note that graphs are sequences rather than sets of literals, i.e., two graphs 
$\graph(V, \phi)$ and $\graph(V', \phi')$ with the same set of nodes, but 
with different sequences of literals are considered to differ, even if $V=V'$ and $\phi'$ 
is just a permutation of $\phi$. However, such graphs are equivalent, denoted 
by the equivalence relation~$\perm$. In contrast, ``ordinary'' graphs would rather be 
represented using multisets of literals instead of 
sequences. The equivalence classes of graphs, 
therefore, correspond to conventional graphs. The ordering of literals 
is technically convenient for the constructions in this paper. However, input graphs to be
parsed should of course be considered up to equivalence. Thus, we will make
sure that the developed parsers yield identical results on graphs $g,g'$ with $g\perm g'$. 

For a graph $\gamma = \graph(V, \phi)$, we use the notations $X(\gamma) = V$ and $\lit\gamma = \phi$.
An injective function
$\rho\colon X \to X$ is called a \emph{renaming}, and $\gamma^\rho$ denotes the graph obtained by replacing all
nodes in $\gamma$ according to $\rho$. Although renamings are, for technical simplicity, defined as functions on the whole of $X$, in every concrete situation only a finite subset of $X$ will be relevant. The same holds when we, later on in the paper, consider partial functions $\mu\colon X \pto X$.

We define the ``concatenation'' of two graphs 
$\alpha, \beta \in \Gr\Lab$ as $\alpha \beta=\langle X(\alpha) \cup X(\beta), \lit\alpha \, \lit\beta\rangle$. A graph $\gamma$ is a \emph{prefix} of
graph $\alpha$ if there is a graph $\delta$ such that $\alpha = \gamma \delta$. Thus, a prefix is a particular kind of subgraph.
If a graph $\gamma$ is completely determined by its sequence 
$\lit\gamma$ of literals, i.e., if each node in $X(\gamma)$ also occurs 
in some literal in $\lit\gamma$, we simply use $\lit\gamma$ as a shorthand 
for~$\gamma$. In particular, a literal $\literal e \in \Lit\Lab$ is identified with the
graph consisting of just this literal and its nodes.
%


\begin{defn}[HR Grammar]\label{def:hr-grammar}
  Let $\Lab=\NLab\uplus\TLab$ be an alphabet which is partitioned into disjoint subsets
  $\NLab$ and $\TLab$ of \emph{nonterminals} and \emph{terminals}, respectively.
  A \emph{hyperedge replacement rule}
  $r=\qrule{\literal A \to \alpha}$ (a \emph{rule} for short) has a literal
  $\literal A \in \Lit{\NLab}$ as its
  \emph{left-hand side}, and a graph $\alpha \in \Gr{\Lab}$ with
  $X(\literal A) \subseteq X(\alpha)$ as its \emph{right-hand side}.

  Consider a graph $\gamma = \beta \literal{A'} \beta' \in \Gr\Lab$ 
  and a rule $r$ as above.
  A renaming $\mu$ is a \emph{match} (of $r$ to
  $\gamma$) if $\literal A^\mu = \literal{A'}$ and
  $X(\gamma) \cap X(\alpha^\mu) \subseteq X(\literal A^\mu)$.%
  \footnote{\label{fn:cond}This condition makes sure that all nodes that are
    introduced on the right-hand side of a rule are renamed so that
    they are distinct from all nodes that do already occur in the
    graph. See also the discussion in \exref{x:boygirl:HR}, derivation
    \eqref{eq:amr-der}.} %
  If additionally $\beta' \in \Gr{\TLab}$, then a match $\mu$ of $r$ \emph{derives} $\gamma$ to the graph
  $\gamma' = \beta \alpha^\mu \beta'$.
  This is denoted as $\gamma \der_{r,\mu} \gamma'$, or just as
  $\gamma \der_r \gamma'$.
  We write $\gamma \der_\R \gamma'$ if $\gamma \der_r \gamma'$ for some rule $r$ taken
  from a set $\R$ of rules.

  A \emph{hyperedge replacement grammar}
  $\Gamma = (\Lab,\TLab,\R,\startSym)$ (\emph{HR grammar} for short)
  consists of finite alphabets $\Lab,\TLab$ as above,
 a finite set $\R$ of rules over $\Lab$,
  and a \emph{start symbol}~$\startSym \in \NLab$ of arity~0.
  $\Gamma$ generates the language
  \[\L(\Gamma) = \{ g \in \Gr{\TLab} \mid \startSym() \mathrel{\mathop\der\nolimits_\R^*} g \}\]
  of terminal 
  graphs. We call a graph~$g$ \emph{valid} with respect to~$\Gamma$ if
  $\L(\Gamma)$ contains a graph $g'$ with $g \perm g'$.
\end{defn}

Note that the definition of ``derives'' requires that $\beta' \in \Gr{\TLab}$. This
means that we only consider \emph{rightmost} derivations in this paper. As usual, by
the context-freeness of hyperedge replacement this does not
imply any loss of generality. The reader should, however, bear in mind that ``derivation''
always means ``rightmost derivation''.

In the following, $\startLit$ always denotes the literal $\startSym()$ of the start
symbol of $\Gamma$. Moreover, we shall generally omit the subscript in $\der_\R$ and $\der_\R^*$,
thus writing simply $\der$ and $\der^*$ instead because the HR grammar in question will
always be clear from the context.

We call literals in $\Lit \TLab$ \emph{terminal} and denote them as
$\literal a, \literal b, \literal c, \dots$, whereas \emph{nonterminal}
literals from $\Lit \NLab$ are denoted as
$\literal A, \literal B, \literal C, \dots$~.
Terminal graphs, those in $\Gr \TLab$, are denoted as $a, b, c, \dots$, 
whereas graphs in $\Gr\Lab$, i.e., graphs that may contain nonterminal literals, are denoted 
as $\alpha, \beta, \gamma, \dots$~.

The next lemma follows directly from the definition of
derivation steps by a straightforward induction on the length of
the derivation.

\begin{lemma}\label{lemma:image-deriv}%
  $\startLit \To^* \gamma$ implies $\startLit \To^* \gamma^\rho$
  for every renaming $\rho$.%
\end{lemma}

An even more immediate consequence of the definition of the derivation relation
is the following:%
%
\begin{fact}\label{fact:deriv}
  For all graphs
  $\alpha, \alpha', \beta \in \Gr{\Lab}$ and $\beta'\in\Gr{\TLab}$,
  $\alpha \drm^* \alpha'$ implies 
  $\beta \alpha \beta' \drm^* \beta \alpha' \beta'$ if and only if
  $X(\alpha') \cap X(\beta \alpha \beta') \subseteq X(\alpha)$.
\end{fact}

It is a well-known result~\cite[Theorem~IV.4.1.2]{habel:92} that every HR 
grammar can be transformed into an equivalent reduced HR grammar where every nonterminal contributes to its language:

\begin{defn}[Reduced HR Grammar]%
\label{def:reduced-HR-grammar}
  An HR grammar $\Gamma = (\Lab,\TLab,\R,\allowbreak\startSym)$ is \emph{reduced} if, for every nonterminal literal $\literal A \in \Lit{\NLab}$, $\literal A \neq \startLit$, there are graphs $\alpha,\beta \in \Gr\Lab$ and $g \in \Gr \TLab$ such that $\startLit \drm^* \alpha \literal A \beta$ and
  $\literal A \drm^* g$.
\end{defn}

\begin{figure}[hbt]
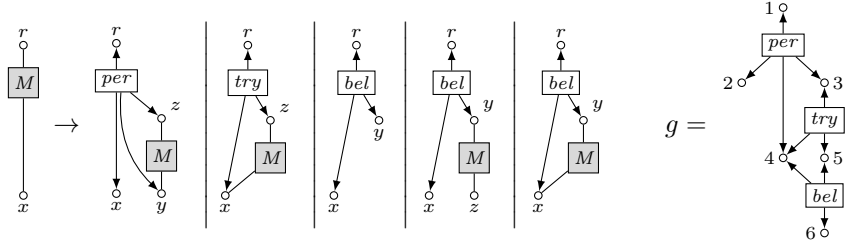

  \def\edgeA#1#2{\Graph[y=5mm,every label/.style={elab}]{%
    \node[o] (r)[label=above:$#1$] at (0,4) {};
    \node[nont] (A) at (0,3) {\nontM};
     \node[o] (x)[label=below:$#2$] at (0,0) {};
   \path (A) edge (r) edge (x);
  }}
  \centering
    $  \edgeA rx \;\to\; 
       \Graph[x=3mm,y=5mm,every label/.style={elab}]{
         \node[o] (r)[label=above:$r$] at (0,4) {};
         \node[term,inner sep=2pt] (B) at (0,3) {\persuade};
         \node[o,label=below:$x$] (x) at (0,0) {};
         \node[o,label=above right:$z$] (y) at (2,2) {};
         \node[nont] (A) at (2,1) {\nontM};
         \node[o,label=below:$y$] (z) at (2,0) {};
         \path[->] (B) edge (r) edge (x) edge (y);
         \path[->] ($(B.south)+(0.2,0)$)  edge[bend angle=25,bend right] (z);
         \path (A) edge (y) edge (z);}
       \;\left|
       \Graph[x=3mm,y=5mm,every label/.style={elab}]{
         \node[o] (r)[label=above:$r$] at (1,4) {};
         \node[term,inner sep=2pt] (B) at (1,3) {\try};
         \node[o,label=below:$x$] (x) at (0,0) {};
         \node[o,label=above right:$z$] (y) at (2,2) {};
         \node[nont] (A) at (2,1) {\nontM};
         \path[->] (B) edge (r) edge (x) edge (y);
         \path (A) edge (y) edge (x);}
       \;\left|
       \Graph[x=3mm,y=5mm,every label/.style={elab}]{
         \node[o] (r)[label=above:$r$] at (1,4) {};
         \node[term,inner sep=2pt] (B) at (1,3) {\believe};
         \node[o,label=below:$x$] (x) at (0,0) {};
         \node[o,label=below:$y$] (y) at (2,2) {};
         \path[->] (B) edge (r) edge (x) edge (y);}
       \;\right|
       \Graph[x=3mm,y=5mm,every label/.style={elab}]{
         \node[o] (r)[label=above:$r$] at (1,4) {};
         \node[term,inner sep=2pt] (B) at (1,3) {\believe};
         \node[o,label=below:$x$] (x) at (0,0) {};
         \node[o,label=above right:$y$] (y) at (2,2) {};
         \node[nont] (A) at (2,1) {\nontM};
         \node[o,label=below:$z$] (z) at (2,0) {};
         \path[->] (B) edge (r) edge (x) edge (y);
         \path (A) edge (y) edge (z);}
       \;\right|
       \Graph[x=3mm,y=5mm,every label/.style={elab}]{
         \node[o] (r)[label=above:$r$] at (1,4) {};
         \node[term,inner sep=2pt] (B) at (1,3) {\believe};
         \node[o,label=below:$x$] (x) at (0,0) {};
         \node[o,label=above right:$y$] (y) at (2,2) {};
         \node[nont] (A) at (2,1) {\nontM};
         \path[->] (B) edge (r) edge (x) edge (y);
         \path (A) edge (y) edge (x);}
       \qquad 
    g =
    \Graph[x=5.5mm,y=5mm,every label/.style={elab}]{
      \node[o] (r)[label=left:$1$] at (2,6) {};
      \node[term,inner sep=2pt] (P) at (2,5) {\persuade};
      \node[o] (d)[label=right:$3$] at (3,4) {};
      \node[term,inner sep=2pt] (T) at (3,3) {\try};
      \node[o] (s)[label=right:$5$] at (3,2) {};
      \node[term,inner sep=2pt] (B) at (3,1) {\believe};
      \node[o] (f)[label=left:$2$] at (1,4) {};
      \node[o] (b)[label=left:$4$] at (2,2) {};
      \node[o] (m)[label=left:$6$] at (3,0) {};
      \path[->] (P) edge (r) edge (f) edge (b) edge (d); 
      \path[->] (T) edge[->] (d) edge (b) edge (s); 
      \path[->] (B) edge[->] (s) edge (b) edge (m); 
    }
  $
  \caption{%
    Diagrams of the rules in~(\ref{eq:pers-rules}) and of the abstract
    meaning representation derived in~\eqref{eq:amr-der}.
    (Circles represent nodes, and boxes represent edges.  The box of
    an edge contains its label, and is connected to the circles of its
    attached nodes by lines; these lines are ordered counter-clockwise
    around the edge, starting to its top.
    Names attached to nodes in rules define the correspondence between
    left-hand side and right-hand side. Vertical bars separate
    the right-hand sides of the rules for the nonterminal $\nontM$.)%
  }
  \label{f:amr-diagrams}
\end{figure}

\begin{example}[Semantic Representation]
  \label{x:boygirl:HR}
  An HR grammar can derive semantic
  representations of sentences of natural language.  The semantic graphs
  in this example are much simplified Abstract Meaning
  Representations~\cite{Banarescu.etAl:13}. As
  in~\cite{Drewes-Jonsson:17} (where the more powerful concept of
  contextual hyperedge replacement~\cite{Drewes-Hoffmann:15} is
  used), we represent the semantics of sentences using the predicates
  (i.e., verbs) `persuade', `try', and `believe'. These yield
  interesting semantic graphs (to the extent such a small example
  reasonably can), because `persuade' is an object control predicate
  (the patient of the persuasion is the agent of whatever she is
  persuaded to do) and `try' is a subject control predicate (the agent
  of the trying is also the agent of whatever is being tried).
  
  The represented patterns are
  \begin{itemize}
  \item ``$x$ \emph{persuades} $y$ to do $z$''
  \item ``$x$ \emph{tries} to do $z$''
  \item ``$x$ \emph{believes} $y$''
  \item ``$x$ \emph{believes} $y$ about $z$''
  \item ``$x$ \emph{believes} $y$ about himself''
  \end{itemize}
  
  The nodes of the graphs represent (anonymous) persons when they
  are leaves, and statements otherwise. Predicates are represented by
  terminal edges with the corresponding label and arity
  (with a further, first tentacle to the root of the statement
  governed by the predicate).
  The rules are as follows
  :
  \begin{equation}\label{eq:pers-rules}
     \begin{array}{r@{\;}c@{\;}l@{\;}l@{\;}lr}
       \startSym()
         & \to & \Mlit rx                    &&&  [s]\\
       \Mlit rx
         & \to & \persuadeL rxyz \: \Mlit zy
         &  |  & \tryL rxz \: \Mlit zx       &  [p,t] \\
         &  |  & \believeL rxy
         &  |  & \believeL rxy \: \Mlit yz   & \; [b_e,b_o] \\
         &  |  & \believeL rxy \: \Mlit yx   &&&  [b_t]
    \end{array}
  \end{equation}
  The graph $g$ representing the phrase ``$1$ \emph{persuades}
  $4$ to \emph{try} to \emph{believe} $6$'' can be derived in four
  steps; the matches $\mu_1$ to $\mu_4$ of the rules are
  given on the right):
  \begin{equation}\label{eq:amr-der}
    \begin{array}{@{}l@{\;}l@{\;}lll@{}}
       \startSym()
         & \Too_s    & \Mlit 12  
         & \mu_1
             = \left\{ \substi{\bm r}1, \substi{\bm x}2 \right\} \\
         & \Too_{p}  & \persuadeL 1243 \: \Mlit 34   
         & \mu_2
              = \left\{ \substi r1, \substi x2,
                        \substi{\bm y}4, \substi{\bm z}3 \right\} \\
         & \Too_t    & \persuadeL 1243 \: \tryL 345 \: \Mlit 54   
         & \mu_3
             = \left\{ \substi r3, \substi x4,
                                 \substi{\bm z}5 \right\} \\
         & \Too_{b_e} & \persuadeL 1243 \: \tryL 345 \: \believeL 546  
         & \mu_4
             = \left\{ \substi r5, \substi x4,
                                 \substi{\bm y}6 \right\} 
     \end{array}     
  \end{equation}
  When a graph is rewritten with a rule, the nodes not occurring on its
  left-hand side (shown in bold in $\mu_1$ to $\mu_4$) have to be
  renamed to nodes that do not occur in that graph. E.g., in the third
  step, the graph $\gamma_2 = \persuadeL 1243 \: \Mlit 34$ is
  rewritten with rule $t$ so that node $z$ is renamed to node
  $5$, which does not occur in $X(\gamma_2) = \{1,2,3,4\}$, or,
  as it is expressed by the condition in \defref{def:hr-grammar},
  $X(\gamma_2) \cap X(\alpha_t^{\mu_3}) \subseteq X(\literal
  A_t^{\mu_3})$,
  i.e., $\{ 1,2,3,4 \} \cap \{ 3,4,5 \} \subseteq \{ 3,4\}$. Mapping
  the ``new'' node $z$ of rule $t$ to node $3$ would falsely match $z$
  to a node in $\gamma_2$.

  Since $g$ can be derived, so can
  $g' = \persuadeL 1623 \: \tryL 325 \: \believeL 524$, i.e., the node
  names in derivations are irrelevant (\lemmaref{lemma:image-deriv}).
  Furthermore, while the
  graph~$h = \tryL 345 \: \persuadeL 1243 \: \believeL 546$ cannot be
  derived, it is valid for this grammar since $g \perm h$.
     \figref{f:amr-diagrams} shows how the rules for $\nontM$ and the
     graph $g$ are drawn as diagrams, a visually convenient notation
     that specifies them up to equivalence.
\end{example}

\section{Shift-Reduce Parsing of Strings}%
\label{s:PSRintro}
The predictive shift-reduce parser for HR grammars borrows concepts
known from the family of context-free \LR(k) parsers for context-free
string grammars \cite{Knuth:65}, and extends them to the parsing of
graphs.  So we recall these concepts first.
As context-free grammars, shift-reduce parsing, and \LR(k) parsing can
be found in every textbook on compiler construction, we discuss these
matters just by means of a small example.

\paragraph{A Context-Free String Grammar for the Dyck Language}
The \emph{Dyck language} of matching nested square brackets
``$\lpar$'' and ``$\rpar$'' is generated by the context-free string
grammar with the nonterminals $\startSym$, $\nontA$, and $\nontB$, and
set of rules %
\[ \D = \{ \startSym \too_0 \nontA,\
           \nontA    \too_1 \lpar\,\nontB\,\rpar,\ 
           \nontB    \too_2 \nontA \nontB,\
           \nontB    \too_3 \emptyseq \},
\]
where $\startSym$ is the start symbol.  A (rightmost) example derivation,
where the replaced nonterminal is underlined in every step, is
\begin{equation}\label{x:rmder}
   \underline{\startSym} \Too_0 \underline{\nontA}
             \Too_1 \lpar \underline{\nontB}             \rpar
             \Too_2 \lpar \nontA \underline{\nontB}      \rpar
             \Too_3 \lpar \underline{\nontA}             \rpar
             \Too_1 \lpar \lpar \underline{\nontB} \rpar \rpar
             \Too_3 \lpar \lpar \,     \rpar \rpar\,.
\end{equation}

\paragraph{A Na\"{\i}ve Shift-Reduce Parser for the Dyck Grammar}
A parser checks whether a string like ``$\lpar\lpar\,\rpar\rpar$''
belongs to the language of a grammar, and constructs a derivation
if this is the case.
A parser is formally defined as a stack automaton that reads
an input string from left to right and uses its stack for remembering
its moves.
A shift-reduce parser is named after its two moves; it constructs a
rightmost derivation 
in reverse.

In a na\"{\i}ve shift-reduce parser, a configuration can be
defined as $\psrbuc(\alpha,w,\bar w)$, where the terminal string
$w\bar w$ is the input (with the vertical bar indicating how far it
has been read), and $\alpha$ is the stack, containing nonterminal and
terminal symbols. The rightmost symbol of~$\alpha$ is the top of the
stack.
Knowing the input, one of $w,\bar w$ can always be reconstructed from
the other, so that we consider simplified configurations
$\alpha \CDOT w$ where the not yet read part of the input is omitted.

The parser performs the following types of moves (where
$\alpha$  and $w$ are as explained above):
\begin{itemize}
\item \emph{Shift} reads the first unread input symbol,%
  \footnote{Since the unread input is omitted in our configurations,
    $a$ seems to come ``out of thin air''.} %
  and pushes it onto the stack. The parser for the Dyck language
  shifts square brackets:
  \[ \alpha \CDOT w \step{} \alpha \lpar \CDOT w \lpar
     \qquad
     \alpha \CDOT w \step{} \alpha \rpar \CDOT w \rpar
  \]
\item \emph{Reduce} pops symbols from the stack if they form the
  right-hand side of a rule, and pushes its left-hand side onto
  it. Thus, in effect, it applies the rule in reverse.  The parser for
  the Dyck language performs the
  following reductions:
  \[ \nontA \CDOT w                         \step 0 \startSym \CDOT w
     \quad
     \alpha \lpar \,\nontB \rpar \, \CDOT w \step 1 \alpha \nontA \CDOT w
     \quad
     \alpha \nontA \nontB           \CDOT w \step 2 \alpha \nontB \CDOT w
     \quad
     \alpha \CDOT w                         \step 3 \alpha \nontB \CDOT w 
  \]
\end{itemize}
A successful \emph{parse} that \emph{accepts} a string $w$ is a sequence of moves starting from the \emph{initial configuration}
$\emptyseq \CDOT \emptyseq$ to an \emph{accepting configuration}
$\startSym \CDOT w$, as below:
\[ \begin{array}
      {l@{\quad}l@{\quad}l@{\quad}l@{\quad}l@{\quad}l@{\quad}l@{\quad}
       l@{\quad}l@{\quad}l@{\quad}l@{\quad}l@{\quad}l@{\quad}l}
          \emptyseq \CDOT \emptyseq & 
   \step{} & \lpar \CDOT  \lpar &
   \step{} & \lpar \lpar \, \CDOT \lpar \lpar &
   \step 3 & \lpar \lpar \nontB   \CDOT  \lpar \lpar &
   \step{} & \lpar \underline{\lpar \nontB \rpar} \CDOT \lpar \lpar \, \rpar &
   \step 1 & \lpar \nontA  \,  \CDOT \lpar \lpar \,\rpar \\  &
   \step 3 & \lpar \underline{\nontA \nontB}   \CDOT \lpar \lpar \, \rpar &
   \step 2 & \lpar  \nontB    \CDOT \lpar \lpar \, \rpar &
   \step{} & \underline{\lpar  \nontB  \rpar} \CDOT \lpar \lpar \, \rpar \rpar & 
   \step 1 & \underline{\nontA} \CDOT \lpar \lpar \, \rpar \rpar &
   \step 0 & \startSym \CDOT \lpar \lpar \, \rpar \rpar
\end{array} \]
(The symbols replaced by reductions are underlined.)
The reductions of a successful parse, read in reverse, yield a
rightmost derivation, in this case the derivation~(\ref{x:rmder})
above.

The na\"{\i}ve shift-reduce parser is correct in the sense that a
string has a successful parse if and only if it has a rightmost
derivation.%

\paragraph{Nondeterminism}
The na\"{\i}ve parser is nondeterministic. E.g., in the
configuration ``$\lpar \nontA \nontB \CDOT \lpar \lpar \, \rpar$'' above, the following moves are possible: %
\begin{enumerate}[(i)]
\item\label{i:redBTB} %
  a reduction by the rule $\nontB \to \nontA\, \nontB$, leading to the
  configuration $\lpar \nontB \CDOT \lpar \lpar \, \rpar$;
\item\label{i:redB} %
  a reduction by the rule $\nontB \to \emptyseq$, leading to the
  configuration
  $\lpar \nontA \nontB \nontB \CDOT \lpar \lpar \, \rpar$; and %
\item\label{i:shiftpar} %
  a shift of an input symbol ``$\lpar$'' or ``$\rpar$'', leading to the configuration
  $\lpar \nontA \nontB \lpar \CDOT \lpar \lpar \, \rpar \lpar$ or $\lpar \nontA \nontB \rpar \CDOT \lpar \lpar \, \rpar \rpar$.
\end{enumerate}
Such a situation is called a conflict.  Only move (\ref{i:redBTB})
will lead to a successful parse, namely the one above.  If the
na\"{\i}ve parser chooses the wrong move in such situations, it will
have to backtrack, i.e., undo shifts and reductions and try
alternative moves until it finds a successful parse, or fails
altogether.

Backtracking makes parsing inefficient. 
To avoid this, the na\"{\i}ve shift-reduce parser can be refined by
gathering information from the grammar that helps to predict which of
its conflicting moves may lead to successful parses:
\begin{itemize}
\item The rules of a grammar allow to predict \emph{viable prefixes}:
  these are
  sequences of terminal and nonterminal symbols
  that occur during rightmost derivations of terminal strings.
  In a successful parse, the stack of the parser always forms
  such a viable prefix.
  For the conflicts discussed above, the sequences
  ``$\lpar \nontA \nontB$'' and ``$\lpar \nontB$'' occurring before and
  after move (\ref{i:redBTB}) are viable prefixes, whereas the
  sequences ``$\lpar \nontA \nontB \nontB$'' and
  ``$\lpar \nontA \nontB \rpar$'' occurring after moves (\ref{i:redB})
  and (\ref{i:shiftpar}) are not. (such stacks cannot be reduced
  further.)
\item A \emph{lookahead} of the $k > 0$ next input symbols 
  may help to decide which move must be taken to make a parse
  successful. In the situation sketched above (where a lookahead of
  $k=1$ suffices), the reductions (\ref{i:redBTB}) and (\ref{i:redB})
  should only be made if the next input symbol is ``$\rpar$'', which
  is the only terminal symbol that may follow $\nontB$ in derivations
  with the grammar.
\end{itemize}
Several ways to determine viable prefixes, and different lengths of
lookahead can be used to construct predictive shift-reduce
parsers. A general one is Knuth's \LR(k) method
\cite{Knuth:65}. Here we just consider the simplest case of DeRemer's
\SLR(k) parser \cite{DeRemer:71}, for a single symbol of
lookahead, i.e., $k=1$.

\paragraph{Nondeterministic Characteristic Finite-State Automata}
The viable prefixes of a context-free grammar form a regular language of
nonterminal and terminal symbols that is generated by an automaton,
known as \emph{characteristic finite-state automaton} (CFA), which can be determined from the grammar as follows:

\begin{itemize}
\item The \emph{states} of the CFA are so-called \emph{items}, rules
  with an additional dot occurring in the
  right-hand side. The dot indicates how far parsing has proceeded. For
  instance, the rule $\nontA \to \lpar\,\nontB \,\rpar$ of the Dyck
  grammar leads to items %
  $\nontA \to \CDOT \lpar \,\nontB \,\rpar$, %
  $\nontA \to \lpar \CDOT \nontB \,\rpar$, %
  $\nontA \to \lpar \,\nontB \CDOT \rpar$, and %
  $\nontA \to \lpar\,\nontB \,\rpar \CDOT $.
\item A state like $\nontA \to \CDOT \lpar \,\nontB \,\rpar$, where
  the dot is before some symbol (terminal or nonterminal), has a
  transition under this symbol to the state where the dot is behind
  that symbol, here a transition under the terminal ``$[$'' to
  $\nontA \to \lpar \CDOT \nontB \, \rpar$.

  A state like $\nontA \to \lpar \CDOT \nontB \, \rpar$, with the dot
  before a nonterminal, does furthermore have transitions
  under the empty string $\emptyseq$ to all items for that nonterminal
  in which the dot is before the first symbol of the right-hand side, e.g. to
  states %
  $\nontB \to \CDOT \nontA \nontB$ and $\nontB \to \CDOT \emptyseq$.
\end{itemize}
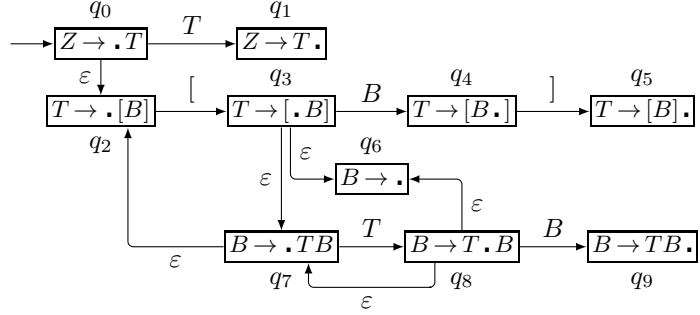
\begin{figure}[bt]
\newcommand{\closT}{\LRo{T}{}{(B)}}
\newcommand{\closB}{\LRo{B}{}{\epsilon}\LRo{B}{}{T B}\closT}
  \centering
  \begin{tikzpicture}[x=24mm,y=-9mm]
    \node[LRq,label=above:$q_0$] (0-0) at (1,0)
      {$\lrQ{\LRo{\startSym}{}{\nontA}}$};
    \node[LRq,label=above:$q_1$] (0-1) at (2,0)
      {$\lrQ{\LRo{\startSym}{\nontA}{}}$};
    \node[LRq,label=below:$q_2$] (1-0) at (1,1)
      {$\lrQ{\LRo{\nontA}{}{\lpar \nontB \rpar}}$};
    \node[LRq,label=above:$q_3$] (1-1) at (2,1)
      {$\lrQ{\LRo{\nontA}{\lpar}{\nontB \rpar}}$};
    \node[LRq,label=above:$q_4$] (1-2) at (3,1)
      {$\lrQ{\LRo{\nontA}{\lpar \nontB}{\rpar}}$};
    \node[LRq,label=above:$q_5$] (1-3) at (4,1)
      {$\lrQ{\LRo{\nontA}{\lpar \nontB \rpar}{}}$};
    \node[LRq,label=above:$q_6$] (3-0) at (2.5,2)
      {$\lrQ{\LRo{\nontB}{}{}}$};
    \node[LRq,label=below:$q_7$] (2-0) at (2,3)
      {$\lrQ{\LRo{\nontB}{}{\nontA \nontB}}$};
    \node[LRq,label=below:$q_8$] (2-1) at (3,3)
      {$\lrQ{\LRo{\nontB}{\nontA}{\nontB}}$};
    \node[LRq,label=below:$q_9$] (2-2) at (4,3)
      {$\lrQ{\LRo{\nontB}{\nontA \nontB}{}}$};
    \path (0.5,0) edge [->] (0-0);
    \path[->] (0-0)
         edge node[above] {$\nontA$} (0-1)
         edge
         node[left] {$\emptyseq$} (1-0);
    \path[->] (1-0)
      edge node[above] {$\lpar$} (1-1);
    \path[->] (1-1)
       edge node[above] {$\nontB$} (1-2)
       edge
       node[left] {$\emptyseq$} (2-0);
    \path[->] (1-2)
       edge node[above] {$\rpar$} (1-3);
    \draw[->,rounded corners=2pt
    ] (2-0.west)
       -| node[below,near start] {$\emptyseq$} ($(1-0.south)+(0.15,0)$) ;
    \draw[->,rounded corners=2pt
    ] ($(1-1.south)+(0.05,0)$)
       |- node[right,near start] {$\emptyseq$} (3-0.west);
    \path[->] (2-0) edge node[above] {$\nontA$} (2-1);
    \path[->] (2-1) edge node[above] {$\nontB$} (2-2);
    \draw[->,rounded corners=2pt
    ] (2-1.north)
       |- node[right,near start] {$\emptyseq$} (3-0.east);
    \draw[->,rounded corners=2pt
    ] ($(2-1.south)-(0.15,0)$)
       |- ($(2-1)+(-0.2,0.6)$)
       -|  node[below,near start] {$\emptyseq$}($(2-0.south)+(0.15,0)$);
  \end{tikzpicture}
  \caption{Nondeterministic characteristic finite-state automaton
           for the Dyck grammar}
  \label{f:nCFA-Dyck}
\end{figure}
\figref{f:nCFA-Dyck} shows the CFA
for the Dyck grammar;
it is nondeterministic, which is caused by its transitions
under the empty string $\emptyseq$.
Its \emph{start state}~$q_0$ (distinguished by the incoming edge
without source node and label) represents the situation
$\startSym \to \CDOT \nontA$ where nothing has been recognized yet.
A path from $q_0$ to some state $q$ in the CFA is an alternating
sequence of states and labels of the transitions connecting them;
the concatenations of the labels along such a path defines a string
generated by the CFA. (Note that a path may contain states and labels
  repeatedly.)

Now a well-known result for shift-reduce parsing reads as follows: a
string is generated by the CFA of a context-free grammar if and only
if it is a viable prefix of a successful parse for that grammar.
E.g., the viable prefixes ``$\lpar \nontA \nontB$'' and
``$\lpar \nontB$'' are generated by the CFA, whereas the sequences
``$\lpar \nontA \nontB \nontB$'', ``$\lpar \nontA \nontB \lpar$'', and ``$\lpar \nontA \nontB \rpar$''
are not.

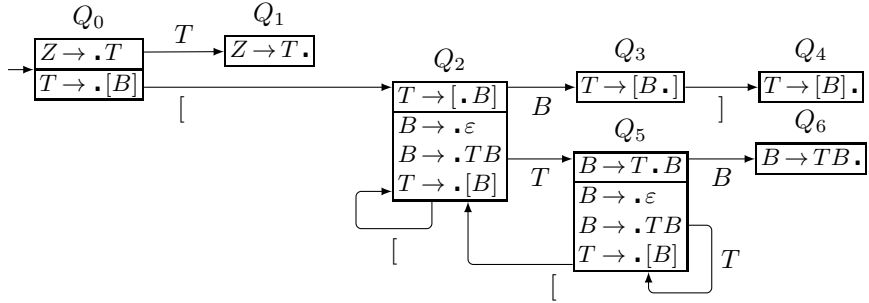
\begin{figure}[b]
\newcommand{\closT}{\LRo{\nontA}{}{\lpar \nontB\rpar}}
\newcommand{\closB}{\LRo{\nontB}{}{\emptyseq}\LRo{\nontB}{}{\nontA \nontB}\closT}
  \centering
  \begin{tikzpicture}[x=24mm,y=-18mm]
    \node[glass] (start) at (-0.5,0.5) {};
    \node[LRq,label=above:$Q_0$] (0) at (0,0.5)
      {$\lrQ{\LRo{\startSym}{}{\nontA}\hline\closT}$};
    \node[LRq,label=above:$Q_1$] (1) at (1,0.365)
      {$\lrQ{\LRo{\startSym}{\nontA}{}}$};
    \node[LRq,label=above:$Q_2$] (2) at (2,1.03)
      {$\lrQ{\LRo{\nontA}{\lpar}{\nontB\rpar}\hline\closB}$};
    \node[LRq,label=above:$Q_3$] (3) at (3,0.63)
      {$\lrQ{\LRo{\nontA}{\lpar \nontB}{\rpar}}$};
    \node[LRq,label=above:$Q_5$] (4) at (3,1.55)
      {$\lrQ{\LRo{\nontB}{\nontA}{\nontB}\hline\closB}$};
    \node[LRq,label=above:$Q_4$] (5) at (4,0.63)
      {$\lrQ{\LRo{\nontA}{\lpar \nontB\rpar}{}}$};
    \node[LRq,label=above:$Q_6$] (6) at (4,1.16)
      {$\lrQ{\LRo{\nontB}{\nontA \nontB}{}}$};
    \path[->] (start) edge  (0);
    \path[->
    ] ($(0.east)+(0,-0.13)$)
         edge node[above
         ] {$\nontA$} (1);
    \path[->] ($(0.east)+(0,+0.13)$)
      edge node[below,pos=.15] {$\lpar$} ($(2.west)-(0,0.40)$);
    \path[->] ($(2.east)-(0,0.4)$)
       edge node[below] {$\nontB$} (3);
    \path[->] ($(2.east)+(0,0.13)$)
       edge node[below] {$\nontA$} ($(4.west)-(0,0.39)$);
    \draw[->,rounded corners=2pt] ($(2.south)-(0.1,0)$)
       |- node[below,near end] {$\lpar$} ($(2.south west)+(-0.2,0.2)$)
       |- ($(2.west)+(0,0.37)$) ;
    \draw[->,rounded corners=2pt] ($(4.west)+(0,0.39)$)
       -| node[below,near start] {\qquad\, $\lpar$} ($(2.south)+(0.1,0)$);
    \path[->] ($(4.east)-(0,0.39)$) edge node[below] {$\nontB$} (6);
    \path[->] (3) edge node[below] {$\rpar$} (5);
    \draw[->,rounded corners=2pt] ($(4.east)+(0,0.1)$)
       -| node[right,near end] {$\nontA$} (3.45,2.15)
       -| ($(4.south)+(0.1,0)$);
  \end{tikzpicture}
\caption{Deterministic characteristic finite-state automaton
         for the Dyck grammar}
\label{f:dCFA-Dyck}
\end{figure}
\paragraph{Deterministic Characteristic Finite-State Automata}
The nondeterministic CFA of a context-free grammar is easy to define,
but of limited practical use for parsing.
Fortunately, it can be turned into a deterministic CFA for the
same language (of viable prefixes). The well-known powerset
construction works as follows: a state set $Q$ joins some state $q$
with all states $q'$ reachable from $q$ by $\emptyseq$-transitions;
$q$ is a \emph{kernel item} of $Q$, whereas the $q'$ are
its \emph{closure items}. Then the non-$\emptyseq$-transitions of the
items in $Q$ have corresponding transitions to successor state sets
$Q'$ that again contain kernel and closure items. Thus state~$q_0$ of
the nondeterministic CFA is joined with state~$q_2$ to form a state
set $Q_0$, and states~$q_3$ and~$q_8$ are both joined with
states~$q_6$, $q_7$, and $q_2$ to form state sets $Q_2$ and $Q_5$,
respectively, while the states $q_1$, $q_4$, $q_5$, and $q_9$ form
singleton state sets $Q_1$, $Q_3$, $Q_4$, and $Q_6$ of the
deterministic CFA.
%
The transition diagram of the deterministic CFA for the Dyck grammar
is shown in \figref{f:dCFA-Dyck}.

The powerset construction 
may let the number of states explode (yielding $2^n$ state sets for
$n$ states of the nondetermistic CFA). However, this rarely occurs in
practice; in our example, the number of states does even decrease.

\paragraph{SLR(1) Parsing}
The stack of the \SLR(1) parser is modified to contain a
sequence like ``$Q_0\lpar Q_2\lpar Q_2\nontA Q_5\nontB Q_6$'', recording a path
$Q_0 \trans{\lpar} Q_2 \trans{\lpar} Q_2 \trans{\nontA} Q_5
\trans{\nontB} Q_6$ in its deterministic CFA, starting in its initial
state. %
The moves of the parser are determined by its current (topmost)
state, and are modified in comparison to those of the nondeterministic
parser as follows:

\begin{itemize}
\item \emph{Shift} reads the next input symbol $a$ if the current
  state is $Q$ and the deterministic CFA contains a transition
  $Q \trans a Q'$. The move pushes $a$ onto the stack, together with
  the successor state~$Q'$.  For our grammar, and state numbers
  $i \in \{0,2,5\}$: %
  \[ \alpha Q_i \CDOT w \step{} \alpha Q_i \, \lpar \, Q_2 \CDOT w \, \lpar
     \qquad
     \alpha Q_3 \CDOT w \step{} \alpha Q_3 \, \rpar \, Q_4 \CDOT w \, \rpar
  \]
\item \emph{Reduce} pops the right-hand side of a rule $A \to \beta$
  (and the intermediate states) off the stack, leaving a state $Q$ on
  top, which has a transition $Q \trans A Q'$. Then $A$ and $Q'$ are
  pushed onto the stack. The \SLR(1) parser performs a reduction only
  if the lookahead---the next input symbol---is a follower symbol of
  $A$.
  We 
  write ``\textbf{if} $\ell=a$'' to indicate that
  the required lookahead symbol is $a$.
  Then the reductions of the example grammar are as follows:
  \def\look{\textbf{if } \ell = \rpar}
  \[ \renewcommand{\arraystretch}{1.1}
  \begin{array}{rcl@{\quad}l}
                  \alpha Q_0 \lpar Q_2 \nontB Q_3 \rpar Q_4 \CDOT w
     & \step 1 & \alpha Q_0 \nontA Q_1 \CDOT w \\
                  \alpha Q_2 \lpar Q_2 \nontB Q_3 \rpar Q_4 \CDOT w
     & \step 1 & \alpha Q_2 \nontA Q_5 \CDOT w \\
                  \alpha Q_5 \lpar Q_2 \nontB Q_3 \rpar Q_4 \CDOT w
     & \step 1 & \alpha Q_5 \nontA Q_5         \CDOT w \\
                  \alpha Q_2 \nontA Q_5 \nontB Q_6 {}\CDOT{}w
     & \step 2 & \alpha Q_2 \nontB Q_3     {}\CDOT{}w & \look \\
                  \alpha Q_5 \nontA Q_5 \nontB Q_6 {}\CDOT{}w
     & \step 2 & \alpha Q_5 \nontB Q_6     {}\CDOT{}w & \look \\
                  \alpha Q_2    {} \CDOT {} w
     & \step 3 & \alpha Q_2 \nontB Q_3 {}\CDOT {} w & \look \\
                  \alpha Q_5     {}\CDOT {} w
     & \step 3 & \alpha Q_5 \nontB Q_6 {}\CDOT{}w & \look 
  \end{array}
  \]
  \item The \SLR(1) parser \emph{accepts} a string $w$ if it reaches 
  a configuration $\alpha Q_1 \CDOT w$.
\end{itemize}
The \SLR(1) parser is correct as well: 
it recognizes the same language as the na\"{\i}ve
shift-reduce parser.

\paragraph{Conflicts} %
The deterministic CFA may reveal conflicts for \SLR(1)  parsing:
\begin{itemize}
\item If a state allows to shift some terminal $a$, and to 
  reduce some rule under the same lookahead symbol $a$, this is a
  \emph{shift-reduce conflict}.
\item If a state allows reductions of different rules under the
  same lookahead symbol, this is a \emph{reduce-reduce conflict}.
\end{itemize}
Whenever the automaton is conflict-free, the \SLR(1) parser
can choose its moves in a deterministic way.

The deterministic CFA for the Dyck grammar is indeed conflict-free: In
states~$Q_2$ and~$Q_5$, rule $\nontB \to \emptyseq$ can be
reduced if the input begins with the only follower symbol $"\rpar"$ of
$\nontB$, which is not in conflict with the shift transitions from
these states under the terminal ``$\lpar$''. Reduce-reduce conflicts do not occur.

A deterministic parse with the \SLR(1) parser is as follows:
\[ \begin{array}{lclcl}
  Q_0 \CDOT \emptyseq & 
  \step{} &
  Q_0 \, \lpar \, Q_2 \CDOT  \lpar & 
  \step{} &
  Q_0 \, \lpar \, Q_2 \, \lpar \, Q_2 \CDOT \lpar \, \lpar
  \\& 
  \step 3 &
  Q_0 \, \lpar \, Q_2 \, \lpar \, Q_2 \nontB \, Q_3 \CDOT \lpar \, \lpar &
  \step{} & 
  Q_0 \, \lpar \, Q_2 \, \lpar \, Q_2 \nontB \, Q_3 \, \rpar \, Q_4 \CDOT \lpar \, \lpar \, \rpar
  \\&
  \step 1 &
  Q_0 \, \lpar \, Q_2 \nontA \, Q_5 \CDOT  \lpar \, \lpar \, \rpar & 
  \step 3 &
  Q_0 \, \lpar \, Q_2 \nontA \, Q_5 \nontB \, Q_6 \CDOT  \lpar \, \lpar \, \rpar
  \\ & 
  \step 2 & 
  Q_0 \, \lpar \, Q_2 \nontB \, Q_3 \CDOT  \lpar \, \lpar \, \rpar & 
  \step{} &
  Q_0 \, \lpar \, Q_2 \nontB \, Q_3 \, \rpar \, Q_4 \CDOT  \lpar \, \lpar \,\rpar \, \rpar
  \\& 
  \step 1 &
  Q_0 \nontA \, Q_1 \CDOT \lpar \, \lpar \, \rpar \, \rpar & 
  &
\end{array} \]
Note that the parser accepts the string
$\lpar \, \lpar \, \rpar \, \rpar$ because it reaches the accepting
configuration
$Q_0 \nontA \, Q_1 \CDOT \lpar \, \lpar \, \rpar \, \rpar$.

Each run of the deterministic parser corresponds to a run of the
corresponding na\"{\i}ve shift-reduce parser (except the last move) when we ignore states and
just consider the symbols on the stack. Thus the deterministic parser
is correct, but it does only apply to grammars that are free of
\SLR(1) conflicts.

\section{A Naïve Shift-Reduce Parser for HR Grammars}\label{s:naive}

We now start to transfer the ideas of shift-reduce string parsing to
HR grammars.  In this section, we describe a naïve nondeterministic
shift-reduce parser, which will be made more practical in the sections
to follow.  We prove the correctness of the naïve parser, i.e., that
it can (nondeterministically) find a derivation for an input graph if
and only if there is one.

\begin{assumption}
  Throughout the rest of the paper, let
  $\Gamma = (\Lab,\TLab,\R,\startSym)$ be the HR grammar for which we
  want to construct a parser.
  Without loss of generality, we assume that $\Gamma$ is reduced.
\end{assumption}
In the remainder of this paper, we will use a HR grammar generating
trees as a running example.
\begin{example}[HR Grammar for Trees]
  \label{x:tree:HR}
  The HR grammar with start symbol $\startSym$ and the
  following rules derives $n$-ary trees. 
  \[  \startSym() \to \startedgeLi x \, \TnontLi x                   \qquad
      \TnontLi y    \to \TnontLi y \, \edgeLi y z \, \TnontLi z \qquad
      \TnontLi y    \to \emptyseq
  \]
  We shall refer to these rules by the number $1,2,3$.
  Note that the unique edge labeled $\startedgeL$ marks the root of the tree,
  which is the unique node at which any derivation of the tree has to start,
  and will thus
  also be the one where parsing has to start. 
  The empty sequence $\emptyseq$ in the last rule is actually a
  short-hand for the graph $\graph(\set{y}, \emptyseq)$ consisting of
  a single node, rather than for the empty graph.
  \figref{f:tree-rmder} shows a derivation of the tree
  $t = \startedgeLi 1 \, \edgeLi 13 \, \edgeLi 12 \, \edgeLi 24$.
  The tree %
  $t' = \edgeLi 24 \, \startedgeLi 1 \, \edgeLi 13 \, \edgeLi 12$
  is valid w.r.t.~the grammar since $t' \perm t$.

\begin{figure}[tbh]
  \[ \begin{array}{r@{\:}c@{\:}l@{\:}c@{\:}l}
    \underline{\startSym()}
      & \drm_1 & \startedgeLi 1 \, \underline{\TnontLi 1} \\
      & \drm_2 & \startedgeLi 1 \,
               \TnontLi 1 \, \edgeLi 12 \, \underline{\TnontLi 2} \\
      & \drm_2 & \startedgeLi 1 \,
               \TnontLi 1 \, \edgeLi 12 \,
               \TnontLi 2 \, \edgeLi 24 \, \underline{\TnontLi 4} \\
      & \drm_3 & \startedgeLi 1 \, \TnontLi 1 \, \edgeLi 12 \,
               \underline{\TnontLi 2} \, \edgeLi 24 \\
      & \drm_3 & \startedgeLi 1 \,
               \underline{\TnontLi 1} \, \edgeLi 12 \, \edgeLi 24 \\
      & \drm_2 & \startedgeLi 1 \,
               \TnontLi 1 \, \edgeLi 13 \,
               \underline{\TnontLi 3} \, \edgeLi 12 \, \edgeLi 24 \\
      & \drm_3 & \startedgeLi 1 \,
               \underline{\TnontLi 1} \, \edgeLi 13 \, \edgeLi 12 \, \edgeLi 24 \\
      & \drm_3 & \startedgeLi 1 \, \edgeLi 13 \, \edgeLi 12 \, \edgeLi 24
  \end{array} \]
  \caption{A derivation of a tree}
  \label{f:tree-rmder}
\end{figure}

  The diagrams of the rules are shown in \figref{f:treerules}, and a
  diagram of the tree~$t$ is shown in \figref{f:tree}.%
\end{example}
\begin{figure}[b]
\def\nvar#1{\Graph[x=6mm]{%
    \node[nont] (x) at (1,0) {$#1$};
}}
\def\uvar#1#2{\Graph[x=6mm]{%
    \node[O] (r)[label=left:\tiny$#2$] at (0,0) {};
    \node[nont] (x) at (1,0) {$#1$};
    \path (r) edge (x);
}}
\def\ee{e}
\def\treefour{\ee (1,2) \, \ee(1,3) \, \ee(2,3) \, \ee(2,4) \,}
  \begin{minipage}[b]{0.74\linewidth}
    \[ \nvar Z \;\to\;
         \Graph[x=6mm,y=6mm]{
           \node[O] (x)[label=below:\tiny$x$] at (1,1) {};
           \node[nont] (t) at (2,1) {$T$};
           \node[nont] (r) at (1,2) {$\mathit{root}$};
           \path (r) edge (x) 
                 (t) edge (x);}
       \quad
       \uvar Ty \;\to\;
          \Graph[x=3mm,y=6mm]{
            \node[O] (r)[label=left:\tiny$y$] at (1,1) {};
            \node[O] (b)[label=left:\tiny$z$] at (1,0) {};
            \node[nont] (x) at (3,1) {$T$};
            \node[nont] (y) at (3,0) {$T$};
            \path (r) edge[->] (b) edge (x); 
            \path (b) edge (y);}
      \quad
      \uvar Ty \;\to\;
                   \Graph{\node[O,label=left:\tiny$y$] at (0,0) {};}
    \] 
    \caption{HR rules deriving trees. The binary terminal edge $\edgeLi yz$
      is drawn as an arrow from node~$y$ to node~$z$.}
    \label{f:treerules}
  \end{minipage} \hfill
  \begin{minipage}[b]{0.2\linewidth}
    \[
        \Graph[x=4mm,y=6mm]{
           \node[op] (r) at (2,3) {$\startedgeL$};
           \node[O] (1) at (2,2) {};
           \node[O] (1-1) at (1,1) {};
           \node[O] (1-2) at (3,1) {};
           \node[O] (1-1-1) at (1,0) {};
           \path[->] (r) edge[->] (1); 
           \path[->] (1) edge[->] (1-1) edge (1-2); 
           \path[->] (1-1) edge[->] (1-1-1) 
           ; 
         }
    \]
    \caption{A tree}
    \label{f:tree}
  \end{minipage}
\end{figure}

A shift-reduce parser of graphs will also 
be modeled by a stack automaton that reads the literals of an input
graph 
and uses a stack for remembering its
moves.
A configuration of the naïve shift-reduce parser could take the
form $\psrbuc(\gamma,g,\bar g)$, where the terminal graph $g\bar g$ is
the input (with the vertical bar indicating how far it has been read),
and $\gamma$ is the stack, containing nonterminal and terminal
literals; the rightmost symbol of~$\gamma$ is the top of the stack.
However, we have to keep in mind that literals of $\bar g$ can be read
in any order, not just from left to right as in string parsing.

In a parse of a certain input graph, the unread part $\bar g$ of the input can always be
reconstructed from the read part $g$, which does not only identify
the edges that have been read, but also that their attached nodes (and only those) have
been read as well.  So we omit the unread input $\bar g$ in
configurations.  %

\begin{defn}[Parser Configuration]\label{def:sr:config}
  A \emph{(shift-reduce parser) configuration} $\buc(\gamma, g)$
  consists of graphs $\gamma \in \Gr\Lab$ and $g \in \Gr \TLab$. The former is
  the stack whereas the latter is the already read subgraph
  of the input graph.
\end{defn}

The parser begins with both the stack and the already read literals
being empty, i.e., its \emph{initial configuration} is
$\buc(\emptyseq, \emptyseq)$. The parser then tries to reach an
accepting configuration using shift and reduce moves, which are
similar to the string case. Shift moves read literals of the input
graph, which are then stored in the parser configuration. The parser
accepts an input graph~$g$ if it reaches a configuration where the
stack consists just of the start graph $\startLit$, and $g$ has been
read completely. This situation is represented by an \emph{accepting
  configuration} $\buc(\startLit, g')$ with $g'\perm g$.
We will show in the following that reaching $\buc(\startLit, g')$
means $\startLit \der^* g'\perm g$, i.e., the parser has identified a
permutation of the input graph literals that proves the validity of
$g$ with respect to the grammar.


\begin{defn}[Shift and Reduce Moves]\label{def:sr-steps}
  A \emph{reduce move} turns a configuration $\buc(\gamma, g)$ into 
  $\buc(\gamma', g)$ if there is a graph 
  $\alpha \in \Gr\Lab$, a rule $\literal A \to \rho$ and a renaming 
  $\mu$ such that 
  $\gamma = \alpha \rho^\mu$, 
  $\gamma' = \alpha \literal A^\mu$,
  and
  $X(\alpha) \cap X(\rho^\mu) \subseteq X(\literal A^\mu)$. We write 
  $\buc(\alpha \rho^\mu, g) 
   \lstep{\literal A^\mu \To \rho^\mu} 
   \buc(\alpha \literal A^\mu, g)$.

  A \emph{shift move} turns a configuration $\buc(\gamma, g)$ into 
  $\buc(\gamma\literal a, g \literal a)$ for a literal 
  $\literal a \in \Lit \TLab$ 
  if $X(\literal a) \cap X(g) \subseteq X(\gamma)$. We write 
  $\buc(\gamma, g) 
   \shift 
   \buc(\gamma \literal a, g \literal a)$.

We write  
$\buc(\gamma, g) \cstep \buc(\gamma', g')$ and call it a \emph{move} of 
the parser if 
$\buc(\gamma, g) \shift \buc(\gamma', g')$ or 
$\buc(\gamma, g) \lstep{\literal A^\mu \To \rho^\mu} \buc(\gamma', g')$.
\end{defn} 

Let us briefly discuss the difference between these shift and reduce
moves on the one hand and their counterparts in string parsing on the other hand.

A shift move in string parsing always reads the first symbol
of the remaining input; the string parser cannot choose the symbol to
be shifted. The graph parser, in contrast, can pick any of the
remaining (terminal) literals for a shift move, as long as the requirement $X(\literal a) \cap X(g) \subseteq X(\gamma)$ is satisfied. This adds another dimension of
nondeterminism  to the parsing of graphs.

A reduce move in string parsing replaces the right-hand side of
a rule on the stack by its left hand side without further
consideration. The graph parser, in contrast, must first rename the nodes in
the rule (cf.{} \defref{def:hr-grammar}). The
condition $X(\alpha) \cap X(\rho^\mu) \subseteq X(\literal A^\mu)$
makes sure that
$\gamma' = \alpha \literal A^\mu \der \alpha \rho^\mu = \gamma$, i.e.,
the application of rule $\literal A \to
\rho$ to~$\gamma'$ indeed yields $\gamma$, and so the move is a correct
backwards application of the rule. In other words, replacing the
literals of~$\rho^\mu$ by $\literal A^\mu$ also removes all nodes 
from $\gamma$ that are
generated by the derivation step $\gamma' \der \gamma$. The requirement
$X(\literal a) \cap X(g) \subseteq X(\gamma)$ of shift moves
eventually checks that these nodes do not occur in
literals still to be processed. Note that if a literal
violates the condition for a shift move once, it will never satisfy
this condition, and will thus never be shifted. Once a condition for a
shift move fails for any of the remaining literals, the parse will
eventually fail altogether.

\newcommand{\Bx}{\textbf{\itshape{x}}}%
\newcommand{\By}{\textbf{\itshape{y}}}%
\newcommand{\Bz}{\textbf{\itshape{z}}}%
\begin{example}[Naïve Shift-Reduce Parser for Trees\label{x:nSR-tree-parser}]
  The naïve shift-re\-duce parser for the tree-generating grammar of
  \exref{x:tree:HR} has the following operations:
  \begin{itemize}
  \item shift operations for the edges labeled with
    $\startedgeL$ and $\edgeL$, and
  \item reductions for the tree-generating rules.
  \end{itemize}
  \figref{f:nsrmoves} shows the moves of the naïve
  shift-reduce parser recognizing the tree $t$ with
  $t \perm \startedgeLi 1 \, \edgeLi 12 \, \edgeLi 13 \, \edgeLi 24$. 
  In many steps of this parse, the parser could
  make alternative decisions that may lead into a dead end:
  \begin{enumerate}
  \item In the first step, the parser could have shifted edge
    $\edgeLi 13$ instead of $\startedgeLi 1$; however, this would lead
    to a dead end.
  \item In the third step, the parser shifts the edge $\edgeLi 12$; it
    could have chosen $\edgeLi 13$ instead, which is another match of
    the edge pattern $\edgeLi yz$. In this case the parser
    could succeed anyway, arriving at the graph
    $g' = \startedgeLi 1 \, \edgeLi 13 \, \edgeLi 12 \, \edgeLi
    24$.
    However, it holds that $g \perm g'$.

    Shifting $\edgeLi 24$ is
    also possible in this step, but would lead to failure.
  \item In step four, the parser could shift edge $\edgeLi 24$
    instead of reducing rule~3. This choice of a shift instead of a
    reduction would also cause failure.
  \item Instead of reducing $\TnontLi 2 \, \edgeLi 24 \, \TnontLi 4$
    to $\TnontLi 2$ in the seventh step, the parser could reduce by
    rule~3. Reduction of a rule like $\TnontLi y \to \emptyseq$ is
    possible in every step. Here it would also lead to a dead end.
  \end{enumerate}
\end{example}
\begin{figure}[bht]
\[ \renewcommand{\arraystretch}{1.1}
   \begin{array}{lr@{}c@{}l@{\quad}l}
           & \emptyseq 
             & \CDOT & \emptyseq & \\
  \step{}   & \startedgeLi 1
             & \CDOT & \startedgeLi 1 &   \\
  \step 3 & \startedgeLi 1 \, \TnontLi 1 
             & \CDOT & \startedgeLi 1  & y/1 \\
  \step{}   & \startedgeLi 1 \, \TnontLi 1 \, \edgeLi 12  
             & \CDOT & \startedgeLi 1 \, \edgeLi 12  &  \\
  \step 3 & \startedgeLi 1 \, \TnontLi 1 \, \edgeLi 12 \, \TnontLi 2   
             & \CDOT & \startedgeLi 1 \, \edgeLi 12  & y/2 \\
  \step{}   & \startedgeLi 1 \, \TnontLi 1 \, \edgeLi 12 \, \TnontLi 2 \, \edgeLi 24
             & \CDOT & \startedgeLi 1 \, \edgeLi 12 \, \edgeLi 24 &  \\
  \step 3 & \startedgeLi 1 \, \TnontLi 1 \, \edgeLi 12 \, \underline{ \TnontLi 2 \, \edgeLi 24 \, \TnontLi 4 }
             & \CDOT & \startedgeLi 1 \, \edgeLi 12 \, \edgeLi 24 & y/4 \\
  \step 2 & \startedgeLi 1 \, \underline{\TnontLi 1 \, \edgeLi 12 \, \TnontLi 2}   
             & \CDOT & \startedgeLi 1 \, \edgeLi 12 \, \edgeLi 24 & y/2, z/4 \\
  \step 2 & \startedgeLi 1 \, \TnontLi 1    
             & \CDOT & \startedgeLi 1 \, \edgeLi 12 \, \edgeLi 24 & y/1, z/2 \\
  \step{}   & \startedgeLi 1 \, \TnontLi 1 \, \edgeLi 13
             & \CDOT & \startedgeLi 1 \, \edgeLi 12 \, \edgeLi 24 \, \edgeLi 13
                                                            &  \\
  \step 3   & \startedgeLi 1 \, \underline{\TnontLi 1 \, \edgeLi 13 \, \TnontLi 3}
             & \CDOT & \startedgeLi 1 \, \edgeLi 12 \, \edgeLi 24 \, \edgeLi 13 
                                                            & y/3 \\
  \step 2   & \underline{\startedgeLi 1 \, \TnontLi 1}
             & \CDOT & \startedgeLi 1 \, \edgeLi 12 \, \edgeLi 24 \, \edgeLi 13
                                                            & y/1, z/3 \\
  \step 1   & \startSym()
             & \CDOT & \startedgeLi 1 \, \edgeLi 12 \, \edgeLi 24 \, \edgeLi 13
                                                            & x/1
\end{array} \]
\caption{Moves of the naïve shift-reduce parser when recognizing the tree
  in \exref{x:tree:HR}. 
  Places on the stack where reductions occur are underlined. 
  Matches for rules in reductions appear in the rightmost column.}
  \label{f:nsrmoves}
\end{figure}

%

We now show that a parse consisting of shift and reduce moves
corresponds to a derivation and vice versa. We first show that each parse yields a derivation 
(\lemmaref{lemma:step-to-deriv}) and then that each derivation 
yields a parse (\lemmaref{lemma:deriv-to-step}).

\begin{lemma}\label{lemma:step-to-deriv}
  For every sequence $\buc(\gamma, g) \cstep^* \buc(\gamma', g')$ 
  of moves with $X(\gamma) \subseteq X(g)$, there is a graph 
  $u \in \Gr{\TLab}$ such that
  $g' = g u$ and $\gamma' \drm^* \gamma u$.
  Moreover, $X(\gamma') \subseteq X(g')$.
\end{lemma}

\begin{proof}
  Let $\buc(\gamma, g) \cstep^n \buc(\gamma', g')$ be any sequence of
  moves with $X(\gamma) \subseteq X(g)$. 
  We prove the statement by induction on~$n$.
  For $n = 0$ it holds with
  $u = \emptyseq$. 
  
  For $n > 0$ and the last move being a shift move, the sequence has the form
  $$
  \buc(\gamma, g) \cstep^{n-1} 
  \buc(\gamma'', g'') \shift 
  \buc(\gamma'' \literal a, g'' \literal a) = \buc(\gamma', g')
  $$
  for some $\literal a \in \Lit\TLab$, $g'' \in \Gr{\TLab}$, and 
  $\gamma'' \in \Gr{\Lab}$. 
  By the induction hypothesis, 
  there is a graph $u' \in \Gr{\TLab}$ such that $g'' = g u'$ and
  $\gamma'' \drm^* \gamma u'$. Let $u=u'\literal a$. Since 
  $X(\gamma) \subseteq X(g)$, the definition of shift moves yields 
  $X(\literal a) \cap X(\gamma u') 
   \subseteq X(\literal a) \cap X(g u') 
   = X(\literal a) \cap X(g'')
   \subseteq X(\gamma'')$.
  Therefore, by \factref{fact:deriv},
  $\gamma' = \gamma'' \literal a \drm^* \gamma u' \literal a=\gamma u$
  and $g' = g'' \literal a = g u' \literal a=gu$.
 
  For $n > 0$ and the last move being a reduce move, the sequence has the form
  $$
  \buc(\gamma, g) \cstep^{n-1} 
  \buc(\alpha \rho^\mu, g') \lstep{\literal A^\mu \To \rho^\mu}
  \buc(\alpha \literal A^\mu, g') = \buc(\gamma', g')
  $$
  for a rule $r = \literal A \to \rho $ and a renaming 
  $\mu$. By the induction hypothesis, there is a graph 
  $u \in \Gr{\TLab}$ such that $g' = g u$ and
  $\alpha \rho^\mu \drm^* \gamma u$, and by the definition of reduce 
  moves, $X(\alpha) \cap X(\rho^\mu) \subseteq X(\literal A^\mu)$. Therefore, 
  $\gamma' = \alpha \literal A^\mu \drm \alpha \rho^\mu \drm^* \gamma u$.

  Finally,  
  $X(\gamma') \subseteq X(\gamma u) \subseteq X(g u) = X(g')$ as
  $\gamma' \drm^* \gamma u$ and $X(\gamma) \subseteq X(g)$.
  \qed
\end{proof}

The following lemma is needed in the proof of \lemmaref{lemma:deriv-to-step};
it generalizes the condition for applying one shift move to sequences of 
shift moves:

\begin{lemma}\label{lemma:n-shifts}
  $X(g) \cap X(u) \subseteq X(\gamma)$ implies
  $\buc(\gamma, g) \shift^* \buc(\gamma u, g u)$ 
  for all graphs $\gamma \in \Gr\Lab$ and $g, u \in \Gr\TLab$.
\end{lemma}

\begin{proof}
  We prove the lemma by induction on $n = |u|$.
  For $n = 0$ it follows from $u = \emptyseq$.
  For $n > 0$, let $g, u, \gamma$ as in the lemma, $u = \literal a u'$ 
  for some $\literal a \in \Lit\TLab$ and $u' \in \Gr\TLab$. Then, 
  $X(g) \cap X(\literal a) \subseteq X(g) \cap X(u) \subseteq X(\gamma)$, and 
  therefore $\buc(\gamma, g) \shift \buc(\gamma \literal a, g \literal a)$. 
  Further,
  $
   X(g \literal a) \cap X(u') = (X(g) \cap X(u')) \cup (X(\literal a) \cap X(u')) 
                  \subseteq X(\gamma) \cup X(\literal a) 
                  = X(\gamma \literal a),
  $
  which satisfies the condition of the lemma, hence 
  $\buc(\gamma \literal a, g \literal a) 
    \shift^* 
    \buc(\gamma \literal a u', g \literal a u') = \buc(\gamma u, g u)$
  by the induction hypothesis.
  \qed
\end{proof}

\begin{lemma}\label{lemma:deriv-to-step}
  $\gamma \drm^* g$ implies 
  $\buc(\emptyseq, \emptyseq) \cstep^* \buc(\gamma, g)$
  for all graphs $\gamma\in\Gr\Lab$ and $g \in \Gr\TLab$.
\end{lemma}

\begin{proof}
  Let  $\gamma \drm^n g$ be any derivation as in the lemma.
  We proceed by induction on $n$.
  
  For $n = 0$, we have $\gamma = g$ and, by \lemmaref{lemma:n-shifts}, 
  $\buc(\emptyseq, \emptyseq) \shift^* 
   \buc(g, g) = \buc(\gamma, g)$.
  
  For $n > 0$, the derivation must be of the form 
  \begin{equation}\label{eq:deriv-to-step-a}
   \gamma = \alpha \literal A^\mu v 
          \drm \alpha \rho^\mu v 
          \drm^{n-1} u v
          = g
  \end{equation}
  for some $u, v \in \Gr \TLab$, $\alpha \in \Gr{\Lab}$, rule $\literal A \to \rho$,
   and renaming $\mu$.
  By $\alpha \rho^\mu \drm^{n-1} u$ and the induction hypothesis, 
  $$
    \buc(\emptyseq, \emptyseq) 
    \cstep^* 
    \buc(\alpha \rho^\mu, u).
  $$
  The first derivation step in \eqref{eq:deriv-to-step-a} implies 
  $X(\alpha) \cap X(\rho^\mu) \subseteq X(\literal A^\mu)$, 
  which enables the following reduce move:
  $$
    \buc(\alpha \rho^\mu, u) 
    \lstep{\literal A^\mu \To \rho^\mu}
    \buc(\alpha \literal A^\mu, u).
  $$
  The inclusion $X(u) \cap X(v) \subseteq X(\alpha \literal A^\mu)$ holds by 
  \eqref{eq:deriv-to-step-a} and \factref{fact:deriv}, and thus
  $$
    \buc(\alpha \literal A^\mu, u) \shift^*
    \buc(\alpha \literal A^\mu v, u v) =
    \buc(\gamma, g)
  $$
  by \lemmaref{lemma:n-shifts}.
  \qed
\end{proof}


We have thus proved the correctness of the naïve shift-reduce parser:

\begin{theorem}\label{the:sr:correct}
  For each graph $h \in \Gr{\TLab}$,
  $\buc(\emptyseq, \emptyseq) \cstep^* \buc(\startLit, h)$
  if and only if $\startLit \drm^* h$. 
\end{theorem}

\begin{proof}
For the \emph{only-if} direction, set $\gamma = g = \emptyseq$, 
$\gamma' = \literal Z$, and $g' = h$
in \lemmaref{lemma:step-to-deriv}, and for the \emph{if} direction, set 
$\gamma = \literal Z$ in \lemmaref{lemma:deriv-to-step}.
\qed
\end{proof}
Although the naïve parser is correct, it is not practically useful,
as it is highly nondeterministic, which requires inefficient backtracking.
In particular, it does not take into account the unread part of the input 
graph when selecting its next move.
%
The following five sections are devoted to enhancements that make
the naïve parser less nondeterministic and more efficient. We
follow the ideas for shift-reduce string parsing described in
\sectref{s:PSRintro}:
\begin{enumerate}
\item In \sectref{s:viable}, we define viable prefixes of derivations,
  which are those graphs that may occur on the stack of successful
  parses.
\item In \sectref{s:nCFA}, we construct the nondeterministic
  characteristic finite automaton (CFA) of a grammar. The CFA generates
  all viable prefixes of the grammar; we say it \emph{approves} them.
\item The nondeterministic CFA is then turned into an equivalent
  deterministic CFA, in \sectref{s:dCFA}.

\item In \sectref{s:ASR}, the deterministic CFA is used to assist the
  naïve SR parser so that it only constructs viable prefixes of
  derivations. This makes sure that the parser can always continue a
  parse found so far, by some remaining input, but not necessarily by the remaining input of
  the current input graph. This is so because it does not ``look ahead''
  into the unread part of the input graph when selecting its next move.
  An implementation would have to use backtracking when it runs into a dead end.

\item Finally, in \sectref{s:PSR}, we describe how the parser can
  avoid backtracking by selecting its next move based on information
  from the unread part of the input graph. This corresponds to the
  lookahead in string parsing. And, likewise, this is not possible if
  the deterministic CFA has \emph{conflicts}. For the HR grammar meeting these
  conditions, we can construct a predictive SR parser that ``knows''
  the appropriate move in each situation, so that it does not need
  backtracking, and is efficient.
 
\end{enumerate}



\section{Viable Prefixes of Graphs\label{s:viable}}

The na\"{\i}ve shift-reduce parser can find a successful parse
for every valid graph (and only for those), but it must choose the
right moves to avoid backtracking. Bear in mind that the parser can
always perform a shift move 
as long as the input graph has not yet been read in its entirety. In
particular, all literals can be shifted right away.  Also, an
$\emptyseq$-rule (like $T(y) \to \emptyseq$ in \exref{x:tree:HR}) can
always be reduced. This will typically lead into a dead end. We shall
now distinguish stacks that may occur in successful parses from those
that do not. This will eventually result in the characteristic finite
automaton that ``assists'' the parser. We follow a similar line of
argument as for string parsing and define so-called \emph{viable
  prefixes} first~\cite[Sect.~5.3.2]{Aho:1972}.


\begin{assumption}
  For the remainder of the paper, we add to $\Gamma$ a new nonterminal
  $\pseudoSym$ of arity zero, and the rule
  $\pseudoLit \to \startLit$ with $\pseudoLit=\pseudoSym()$.
  Thus, the derivations starting with $\pseudoLit$ are just those in
  the original grammar, but with an additional first step
  $\pseudoLit\drm\startLit$. Clearly, the generated language is
  independent of whether $Z$ or $\pseudoSym$ is
  considered to be the initial nonterminal.
\end{assumption}

A viable prefix is a prefix of a graph derivable from $\pseudoLit$ by a nonempty
derivation, provided that this prefix does
not extend past the right-hand side of the most recently applied rule.
More formally:

\begin{defn}[Viable Prefix]\label{def:viable-prefix}
  A graph $\gamma \in \Gr\Lab$ is called a \emph{viable prefix} 
  if there are graphs $\alpha, \beta \in \Gr\Lab$ as 
  well as $z \in \Gr{\TLab}$ and a literal $\literal A \in \Lit \NLab$ such that 
  $
  \pseudoLit \drm^* \alpha \literal A z \drm \alpha \beta z
  $ 
  and $\gamma$ is a prefix of $\alpha \beta$.
\end{defn}

\begin{example}
  We illustrate \defref{def:viable-prefix} by using an initial segment
  of the derivation in \figref{f:tree-rmder} (though now beginning
  with $\pseudoLi$):
  \begin{align*}
    \pseudoLi
     & \drm^*\ \overbrace{\startedgeLi 1}^\alpha \, 
             \overbrace{\TnontLi 1}^{\literal A} \, 
             \overbrace{\edgeLi 12 \, \edgeLi 24}^z \\[-2mm] &
      \drm_2\ \underbrace{
                 \overbrace{\startedgeLi 1}^\alpha \,
                 \overbrace{\TnontLi 1 \, \edgeLi 13 \, \TnontLi 3}^\beta}_\gamma \, 
                 \overbrace{\edgeLi 12 \, \edgeLi 24}^z \,.
  \end{align*}
  The graph
  $\gamma = \startedgeLi 1 \, \TnontLi 1 \, \edgeLi 13 \, \TnontLi 3$
  is the longest viable prefix of the derived graph; all
  prefixes of $\gamma$ are viable as well. In contrast to that, the graph $\startedgeLi 1 \, \TnontLi 1 \, \edgeLi 13 \, \TnontLi 3 \, \edgeLi 12$ is not a viable prefix, because the edge $\edgeLi 12$ is ``beyond'' the part of the graph that is generated from nonterminal $\TnontLi 1$. 
\end{example}

Before we show that the set of viable prefixes is just the set of all
stacks occurring in successful parses, we need
two technical lemmata. \lemmaref{lemma:viable-prefix} states that the
set of viable prefixes does not change if we add to
\defref{def:viable-prefix} the additional requirement that the
suffix~$v$ with $ \gamma v =\alpha \beta$ is a terminal graph. 

\begin{lemma}\label{lemma:viable-prefix}
  A graph $\gamma \in \Gr\Lab$ is a viable prefix if and only if there are 
  graphs $\alpha, \beta \in \Gr\Lab$ as well as $v, z \in \Gr{\TLab}$ and a 
  literal $\literal A \in \Lit \NLab$ such that 
  $
  \pseudoLit \drm^* \alpha \literal A z \drm \alpha \beta z = \gamma v z.
  $ 
\end{lemma}

\begin{proof}
  The \emph{if} direction follows immediately from the definition of viable 
  prefixes. For the \emph{only-if} direction, let $\gamma$ be any viable prefix.
  Hence, there is a derivation  
  \begin{equation*}
    \pseudoLit \drm^n \alpha \literal A z \drm \alpha \beta z\drm^*g
  \end{equation*}
  and $\alpha \beta = \gamma \delta$, for some $\delta\in\Gr\Lab$ and 
  $g\in\Gr\TLab$ (because we assume that $\Gamma$ is reduced).
  Without loss of generality, assume that the step $\alpha \literal A z \drm \alpha \beta z$
  is the last one in this derivation such that $\gamma$ is a prefix of $\alpha \beta$.
  Then the next step (if there is one) replaces a nonterminal literal in $\gamma$,
  which means that $\delta z\in\Gr\TLab$ (since all derivations are rightmost).
  Thus, the statement holds with $v=\delta$.
  \qed
 \end{proof}




We now show that the set of viable prefixes is just the set of all
stacks occurring in successful parses. 

\begin{lemma}\label{lemma:prefix-to-parse}
  For every sequence  
  $\buc(\emptyseq, \emptyseq) \cstep^* \buc(\gamma, g)$ of 
  moves such that $\gamma$ is a viable prefix, there is a graph 
  $g' \in \Gr \TLab$ with
  $\buc(\gamma, g) \cstep^* \buc(\startLit, g g')$.
\end{lemma}

\begin{proof}
  Let $\buc(\emptyseq, \emptyseq) \cstep^* \buc(\gamma, g)$
  be a sequence of moves as in the lemma. By 
  \lemmaref{lemma:step-to-deriv}, $\gamma \drm^* g$. 
  We first show that there is a derivation 
  \begin{equation}\label{eq:ptp-deriv-form}
    \pseudoLit \drm^n \alpha \literal A z \drm \alpha \beta z
              =\gamma v z \drm^* g v z
  \end{equation}
  where $v \in \Gr\TLab$.
  Since $\gamma$ is a 
  viable prefix and according to \lemmaref{lemma:viable-prefix}, there is a 
  derivation 
  $\pseudoLit \drm^n {\hat\alpha} \hat{\literal A} {\hat z} 
              \drm {\hat\alpha} {\hat\beta} {\hat z}$
  for some $n\in\Nat$ such that $\gamma {\hat v}={\hat\alpha} {\hat\beta}$ for 
  a terminal graph ${\hat v}\in\Gr\TLab$. However, one cannot conclude 
  $\gamma {\hat v} {\hat z} \drm^* g {\hat v} {\hat z}$ because of possible 
  name clashes. To circumvent this, we rename nodes that may 
  cause such problems. For this purpose, choose any renaming 
  $\mu$ with $\mu(x) = x$ if $x \in X(\gamma)$ and 
  $\mu(x) \notin X(g)$ otherwise, and let $\alpha = \hat\alpha^\mu$, 
  $\beta = \hat\beta^\mu$, $v = \hat v^\mu$, $z = \hat z^\mu$, and
  $\literal A = \hat{\literal A}$. By the choice of $\mu$, $\gamma = \gamma^\mu$ and
  $\gamma v = \alpha \beta$, as well as
  \begin{equation}\label{eq:ptp-subset}
    X(v z) \cap X(g) \subseteq X(\gamma).
  \end{equation}
  Hence,
  $\gamma v z \drm^* g v z$ by \factref{fact:deriv}. 
  By \lemmaref{lemma:image-deriv}, we thus have
  a derivation as in~\eqref{eq:ptp-deriv-form}.
  Note that
  \begin{equation}\label{eq:ptp-subset-b}
    X(\alpha) \cap X(\beta) \subseteq X(\literal A)
  \end{equation}
  follows from $\alpha \literal A z \drm \alpha \beta z$.

  We now show that, for every sequence  
  $\buc(\emptyseq, \emptyseq) \cstep^* \buc(\gamma, g)$ 
  of moves and every 
  derivation~\eqref{eq:ptp-deriv-form},
  there is a sequence
  $\buc(\gamma, g) \cstep^* \buc(\startLit, g v z)$ 
  by induction on the length of the derivation in~\eqref{eq:ptp-deriv-form}.
    
  For $n = 0$, we have $\alpha = z = \emptyseq$ and 
  $\pseudoLit \drm \beta = \startLit = \gamma v$, i.e., $\gamma=\startLit$ and $v=\emptyseq$. 
  Hence,
  $
    \buc(\gamma, g) 
    = 
    \buc(\startLit, g vz)
    \cstep^0 
    \buc(\startLit, g vz).
  $
  
  For $n > 0$, we distinguish between two cases.
  
  \begin{enumerate}[(1)]
  \item The derivation~\eqref{eq:ptp-deriv-form} has the form 
  \begin{equation}\label{eq:ptp-case-a}
    \pseudoLit \drm^{n-1} 
    \delta \literal B u \drm 
    \delta \phi \literal A w u \drm
    \alpha \beta z = \gamma v z \drm^* g v z
  \end{equation}
  where $\alpha = \delta \phi$, $z = w u$, and $\alpha \beta = \gamma v$. 
  By~\eqref{eq:ptp-subset}, \eqref{eq:ptp-subset-b}, and 
  \lemmaref{lemma:n-shifts},
  \begin{equation}\label{eq:ptp-case-a-steps}
    \buc(\gamma, g) 
    \shift\nolimits^* 
    \buc(\gamma v, g v) 
    =
    \buc(\alpha \beta, g v) 
    \lstep{\literal A \der \beta} 
    \buc(\alpha \literal A, g v) = \buc(\delta \phi \literal A, g v).
  \end{equation}
  Note that~\eqref{eq:ptp-case-a} has the form of~\eqref{eq:ptp-deriv-form} 
  where $\literal B$ plays the role of~$\literal A$, 
  $\delta \phi \literal A$ the role 
  of~$\gamma$, $w$~the role of~$v$, and $g v$ the role of~$g$. Because 
  of~\eqref{eq:ptp-case-a-steps},
  we can make use of the induction hypothesis to conclude that
  $\buc(\delta \phi \literal A, g v) 
   \cstep^*
   \buc(\startLit, g v z)$.
  
  \item The given derivation~\eqref{eq:ptp-deriv-form} has the form
  \begin{equation}\label{eq:ptp-case-b}
    \pseudoLit \drm^{n-1} 
    \alpha \literal A u \literal B w \drm 
    \alpha \literal A u v' w \drm
    \alpha \beta z  = \gamma v z \drm^* g v z
  \end{equation}
  with $\literal B \der v' \in \Gr\TLab$ and $z = u v' w$.
  By~\eqref{eq:ptp-subset}, \eqref{eq:ptp-subset-b}, and 
  \lemmaref{lemma:n-shifts}, 
  \begin{equation}\label{eq:ptp-case-b-steps}
    \buc(\gamma, g) 
    \shift\nolimits^* 
    \buc(\gamma v, g v) 
    =
    \buc(\alpha \beta, g v) 
    \lstep{\literal A \der \beta} 
    \buc(\alpha \literal A, g v).
  \end{equation}
  Note that~\eqref{eq:ptp-case-b} has the form of~\eqref{eq:ptp-deriv-form} 
  where $\literal B$ plays the role of~$\literal A$, $\alpha \literal A$ 
  the role of~$\gamma$, $u v'$ the role of~$v$, and $g v$ the role of~$g$. 
  Because of~\eqref{eq:ptp-case-b-steps}, we can 
  use the induction hypothesis to conclude that
  $\buc(\alpha \literal A, g v) 
   \cstep^* 
   \buc(\startLit, g v z)$.
  \qed
  \end{enumerate}
\end{proof}

\begin{lemma}\label{lemma:parse-to-prefix}
  For every sequence  
  $
    \buc(\emptyseq, \emptyseq)
    \cstep^*
    \buc(\gamma, u) 
    \cstep^*
    \buc(\startLit, u v)
  $ 
  of moves, $\gamma$~is a viable prefix.
\end{lemma}

\begin{proof}
  If $\gamma=\startLit$, then it is
  a viable prefix since $\pseudoLit\drm\startLit$. Otherwise, the sequence of
  moves has the form
  $
    \buc(\emptyseq, \emptyseq)
    \cstep^*
    \buc(\gamma, u) 
    \cstep^+
    \buc(\startLit, u v)
  $.
  As the last move is a reduce move,
  the sequence can be written as
  $$
    \buc(\emptyseq, \emptyseq)
    \cstep^*
    \buc(\gamma, u) 
    \shift\nolimits^*
    \buc(\gamma v', u v') = \buc(\alpha \beta, u v')
    \lstep{\literal A \der \beta}
    \buc(\alpha \literal A, u v')
    \cstep^* 
    \buc(\startLit, u v' v'')
  $$
  for graphs $v', v'' \in \Gr\TLab$ with $v = v' v''$. 
  Applying 
  \lemmaref{lemma:step-to-deriv} to each subsequence yields
  $
    \pseudoLit\drm\startLit \drm^* \alpha \literal A v'' \drm \alpha \beta v'' = \gamma v' v'',
  $
  i.e., $\gamma$~is a viable prefix.
  \qed
\end{proof}

An immediate consequence of \lemmaref{lemma:prefix-to-parse} and
\lemmaref{lemma:parse-to-prefix} is the following:

\begin{theorem}\label{thm:prefix}
  For every sequence  
  $\buc(\emptyseq, \emptyseq) \cstep^* \buc(\gamma, u)$ of 
  moves, there is a graph 
  $v \in \Gr \TLab$ with
  $\buc(\gamma, u) \cstep^* \buc(\startLit, u v)$
  if and only if $\gamma$~is a viable prefix.
\end{theorem}

In other words, the stack of a reachable configuration is a
viable prefix if and only if the naïve parser can reach the
accepting configuration for \emph{some} possible remaining sequence
of literals.



\section{Nondeterministic Characteristic Finite-State Automata}\label{s:nCFA}

Let us now start to develop the means to ``assist'' the shift-reduce
parser to restrict its moves to promising ones. The first step towards
this goal is the construction of nondeterministic characteristic
finite automata, defined next.
\begin{defn}[Nondeterministic  CFA]\label{def:auto-gamma}
  The 
  \emph{nondeterministic characteristic finite automaton} (nCFA) for $\Gamma$
  is the tuple $\nCFA = (Q, q_0, \Delta)$ consisting of the following components:
  \begin{enumerate}
    \item $Q = \set{ \literal A \to \alpha \CDOT \beta \mid 
                    \qrule{\literal A \to \alpha \beta} \in \R}$ is the finite set of \emph{states}.

    \item 
      $q_0 = \qrule{\pseudoLit \to \CDOT \startLit}$ is the \emph{initial state}.

  \item
    $\Delta \subseteq Q \times (\Lit\Lab \cup \set{\emptyseq}) \times
    Q$
    is a ternary \emph{transition relation}. Writing $p \atrans{x} q$
    if $(p,x,q) \in \Delta$, the transitions constituting $\Delta$
    are:
    \begin{enumerate}
      \item
        $ \qrule{\literal A \to \alpha \CDOT \literal l \beta}
        \atrans{\literal l} \qrule{\literal A \to \alpha \literal l \CDOT
          \beta} $
        for every state
        $\qrule{\literal A \to \alpha \CDOT \literal l \beta}\in Q$,
        where $\literal l \in \Lit\Lab$;
        such a transition is called \emph{goto transition}.
      \item $q \atrans{\emptyseq} p$ for all states 
      $q = \qrule{\literal A \to \alpha \CDOT \literal B \beta} \in Q$ 
      and 
      $p = \qrule{\literal C \to \CDOT \gamma} \in Q$
      such that $\lab(\literal B) = \lab(\literal C) \in \NLab$;
      such a transition is called \emph{closure transition}.
    \end{enumerate}
  \end{enumerate}
\end{defn}

\begin{assumption}
  Since we assume a fixed HR grammar $\Gamma$, we assume a fixed
  nCFA $\nCFA = (Q, q_0, \Delta)$ obtained from $\Gamma$ from now on.
\end{assumption}

Following the ideas discussed earlier, each item is a rule with a dot somewhere between
literals in its right-hand side, indicating the division between literals that have already 
been processed and those which have not. Accordingly, the dot is moved across a literal when
a corresponding literal is processed.
We now start to formalize
how an nCFA ``approves'' graphs
(which will later turn out to be viable prefixes).

Intuitively, an nCFA \emph{approves} a graph $\phi$ if the sequence $\lit\phi$ of literals
corresponds to a sequence of state transitions, starting at the initial
state. We define the notion of \emph{nCFA configurations} (or just
\emph{configurations} if it is clear from the context) to formalize this:

\begin{defn}[nCFA Configuration]\label{def:auto-configuration}
  An \emph{nCFA configuration} $\autoconf(q, \mu, \phi)$ 
  consists of
  \begin{itemize}
  \item a graph $\phi \in \Gr\Lab$,
  \item a state $q = \qrule{\literal A \to \alpha \CDOT \beta} \in Q$, and
  \item an injective partial function $\mu \colon X \pto X$ with \\
  $X(\alpha) \subseteq \dom(\mu)
             \subseteq X(\alpha) \cup X(\literal A)$.
  \end{itemize}
\end{defn}

The function~$\mu$ in an nCFA configuration $\autoconf(q, \mu, \phi)$
corresponds to the match defined in \defref{def:hr-grammar}; it
maps rule nodes to nodes of the graph read so far. In an nCFA
configuration, this match is not yet completely 
determined in general; the mapping of nodes that have not yet been read is still undefined.
The mapping $\mu$ is extended when a literal is read,
which means that all its attached nodes, if they have not been read as
nodes of other literals earlier, are now read as well.
As a consequence, nodes of state~$q$ must be mapped by~$\mu$
to nodes in~$\phi$---unless they have not been read yet, in which case they
are not in~$\dom(\mu)$. Such nodes may only occur
in literals behind the dot in~$q$, which is reflected by the requirement that 
$X(\alpha) \subseteq \dom(\mu) \subseteq X(\alpha) \cup X(\literal A)$.

To compare literals that have only partially been matched, let $\unknown \notin X$ be a special
value denoting `undefined'. Given a partial injective function
$\mu\colon X \pto X$ and a literal $\literal l=a(x_1,\dots,x_k)$, we
let $\literal l^\mu=a(y_1,\dots,y_k)$ where, for all $1\le i\le k$,
$y_i=\mu(x_i)$ if $x_i\in\dom(\mu)$ and $y_i=\unknown$ otherwise. Note
that $\literal l^\mu$ is a literal if (and only if)
$x_1,\dots,x_k\in\dom(\mu)$. ``Literals'' which may contain `$\unknown$' are called
\emph{pseudo-literals}. We let $X(\literal l^\mu)$ denote 
$\mu(X(\literal l))$.
A slightly more general form of pseudo-literals will play an important
role in \sectionref{s:PSR}.

An nCFA works by processing literals step
by step while moving from state to state, represented by a
corresponding sequence of nCFA configurations, starting at
$\autoconf(q_0, \iota, \emptyseq)$, the initial
configuration. Here, $\iota \colon X \pto X$ is the
totally undefined function with $\dom(\iota) =
\emptyset$. Intuitively, $\autoconf(q_0, \iota, \emptyseq)$
being the initial configuration means that the nCFA starts with the
empty graph~$\emptyseq$ in $q_0 = \qrule{\pseudoLit \to \CDOT \startLit}$
and with no nodes mapped yet,
the latter being indicated by the empty domain of $\iota$.

Each step of the nCFA is modeled by a move, defined as follows:

\begin{defn}[nCFA Move]\label{def:auto-conf}
  Let $\autoconf(q, \mu, \phi)$ be a configuration.  
  A goto transition $q \atrans{\literal l} q'$ induces a \emph{goto move} 
  $$
  \autoconf(q, \mu, \phi) \consumestep \autoconf(q', \nu, \phi \literal l^\nu)
  $$
  where $\mu \sqsubseteq \nu$,
  $\dom(\nu) = \dom(\mu) \cup X(\literal l)$,
  and $X(\literal l^\nu) \cap X(\phi) \subseteq X(\literal l^\mu)$.
  
  A closure transition $q \atrans{\emptyseq} q'$ with $q = \qrule{\literal A \to \alpha \CDOT \literal B \beta}$ and
  $q' = \qrule{\literal C \to \CDOT \delta}$
  induces a \emph{closure move}
  $$
  \autoconf(q, \mu, \phi) \expandstep \autoconf(q', \nu, \phi)
  $$
  where $\literal B^\mu = \literal C^\nu$ and $\dom(\nu) \subseteq X(\literal C)$. 
  
  We write $C \autostep C'$ if either $C \consumestep C'$ or $C \expandstep C'$, and call this a \emph{move}.
\end{defn}

A goto move applies a goto transition triggered by a (possibly nonterminal)
literal~$\literal l$. When the corresponding literal is processed,
which also means that all its nodes have been read,
all nodes of~$\literal l$ are mapped by the resulting
node mapping~$\nu$. The processed literal is hence $\literal l^\nu$,
which is added to the end of the approved graph, resulting in $\phi
\literal l^\nu$. The first two conditions, $\mu \sqsubseteq \nu$ and
$\dom(\nu) = \dom(\mu) \cup X(\literal l)$, state that the mapping $\nu$
extends the previous mapping $\mu$ so as to map the entire literal $\literal l$.
The remaining condition $X(\literal l^\nu) \cap X(\phi) \subseteq X(\literal l^\mu)$
ensures that nodes that have already been read (i.e., those in $\phi$) are not matched another time by extending $\mu$ to $\nu$.

A closure move applies a closure transition and corresponds to a
derivation step, i.e., the
mapping~$\mu$ of nodes in~$\literal B$ is translated into a
mapping~$\nu$ of the corresponding nodes in~$\literal C$. Note
that~$\dom(\mu)$ and~$\dom(\nu)$ are unrelated because the nodes in
$\literal B$ and $\literal C$ may differ. Only nodes appearing
in~$\literal C$---but not necessarily all of them---are mapped by~$\nu$; 
other nodes of state~$q'$ are
not mapped because their corresponding nodes have not yet
been read.

\begin{defn}\label{def:auto-approve}
  The nCFA \emph{approves} a graph $\phi \in \Gr\Lab$ if there is a configuration 
  $C = \autoconf(q, \mu, \phi)$ such that
  $\autoconf(q_0, \iota, \emptyseq) \autostep^* C$.
\end{defn}
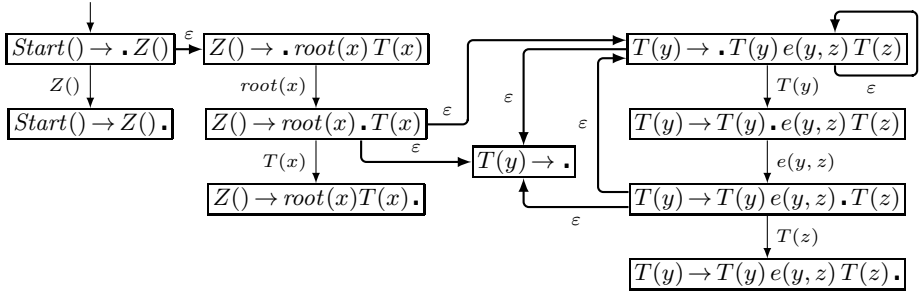
\begin{figure}[t]
  \centering
  \begin{tikzpicture}[x=30mm,y=-10mm]
    \node (0) at (1,.25) {};
    \node[LRq] (0-0) at (1,1)
      {$\lrQ{\LRo{\pseudoLi}{}{\startLi}}$};
    \node[LRq] (0-1) at (1,2)
      {$\lrQ{\LRo{\pseudoLi}{\startLi}{}}$};
    \node[LRq] (1-0) at (2,1)
      {$\lrQ{\LRo{\startLi}{}{\startedgeLi x \, \TnontLi x}}$};
    \node[LRq] (1-1) at (2,2)
      {$\lrQ{\LRo{\startLi}{\startedgeLi x}{\TnontLi x}}$};
    \node[LRq] (1-2) at (2,3)
      {$\lrQ{\LRo{\startLi}{\startedgeLi x \TnontLi x}{}}$};
    \node[LRq] (2-0) at (2.92,2.5)
      {$\lrQ{\LRo{\TnontLi y}{}{}}$};
    \node[LRq] (3-0) at (4,1)
      {$\lrQ{\LRo{\TnontLi y}{}{\TnontLi y \,\edgeLi yz \,\TnontLi z}}$};
    \node[LRq] (3-1) at (4,2)
      {$\lrQ{\LRo{\TnontLi y}{\TnontLi y}{\edgeLi yz \,\TnontLi z}}$};
    \node[LRq] (3-2) at (4,3)
      {$\lrQ{\LRo{\TnontLi y}{\TnontLi y \, \edgeLi yz}{\TnontLi z}}$};
    \node[LRq] (3-3) at (4,4)
      {$\lrQ{\LRo{\TnontLi y}{\TnontLi y \, \edgeLi yz \, \TnontLi z}{}}$};
    \path[->]
      (0)   edge (0-0)
      (0-0) edge node[left] {\scriptsize$\startLi$} (0-1)
            edge[thick] node[above] {\scriptsize$\emptyseq$} (1-0)
      (1-0) edge node[left] {\scriptsize$\startedgeLi x$} (1-1)
      (1-1) edge node[left] {\scriptsize$\TnontLi x$} (1-2)
      (3-0) edge node[right] {\scriptsize$\TnontLi y$} (3-1)
      (3-1) edge node[right] {\scriptsize$\edgeLi yz$} (3-2)
      (3-2) edge node[right] {\scriptsize$\TnontLi z$} (3-3);
    \draw[->,rounded corners=2pt,thick]
      ($(1-1.south)+(0.2,0)$)
        |- node[above,near end] {\scriptsize$\emptyseq$}
      ($(2-0.west)$);
    \draw[->,rounded corners=2pt,
          thick]
      ($(1-1.east)$)
        -| node[above,near start] {\scriptsize$\emptyseq$} 
          ($(1-1.east)+(0.18,0)$)
        |- ($(3-0.west)+(0,-0.12)$);
    \draw[->,rounded corners=2pt,thick]
      ($(3-0.west)$)
         -| node[left,near end] {\scriptsize$\emptyseq$}
            ($(2-0.north)$);
    \draw[->,rounded corners=2pt,thick]
      ($(3-0.south)+(0.3,0)$)
         |- node[below,near end] {\scriptsize$\emptyseq$}
            ($(3-0.east)+(0,0.35)$)
         -| ($(3-0.east)+(0.04,0)$)
         |- ($(3-0.east)+(0,-0.5)$)
         -| ($(3-0.north)+(0.3,-0)$);
    \draw[->,rounded corners=2pt,thick]
      ($(3-2.west)+(0,+0.1)$)
         -| node[below,near start] {\scriptsize$\emptyseq$}
      (2-0.south);
    \draw[->,rounded corners=2pt,thick]
      ($(3-2.west)+(0,-0.1)$)
        -| node[left,at end] {\scriptsize$\emptyseq$} ($(3-1)+(-0.75,0)$)
        |- ($(3-0.west)+(0,+0.12)$);
  \end{tikzpicture}
  \caption{Nondeterministic CFA for the tree-generating grammar in
    \exref{x:tree:HR}. The initial state appears in the upper
    left. Closure transitions are drawn with thicker lines.}
    \label{fig:nCFA}
\end{figure}
\begin{example}[The nCFA for the Tree-Generating
  Grammar\label{ex:nCFA}]\hspace*{0pt plus 1em}%
  \figref{fig:nCFA} shows the transition diagram of the
  nondeterministic CFA for the tree-generating grammar in
  \exref{x:tree:HR}. %
  In \figref{f:nCFA-moves} we show moves of the nondeterministic CFA.

  Every graph approved by the automaton is a viable prefix occurring
  in the derivation in \figref{f:tree-rmder} of
  \exref{x:tree:HR}, and in the parse shown in \figref{f:nsrmoves} of
  \exref{x:nSR-tree-parser}.
  We will show in the sequel that this is not a coincidence.  \qed
\end{example}%
\begin{figure}[b!]
  \[ \begin{array}{c@{\;}r@{\,}c@{\,}l}
                 & \Autoconf(\lro{\pseudoLi}{}{\startLi}, , \emptyseq) \\
     \expandstep & 
     \Autoconf(\lro{\startLi}{}{\startedgeLi x \TnontLi x}, , \emptyseq) 
     \\ \consumestep & 
     \Autoconf(\lro{\startLi}{\startedgeLi x}{\TnontLi x}, x/1, \startedgeLi 1) 
     \\ \expandstep & 
     \Autoconf(\lro{\TnontLi y}{}{\TnontLi y \, \edgeLi yz \, \TnontLi z}, {y/1}, \startedgeLi 1) 
     \\ \consumestep & 
     \Autoconf(\lro{\TnontLi y}{\TnontLi y}{ \edgeLi yz \, \TnontLi z}, {y/1}, \startedgeLi 1 \, \TnontLi 1) 
     \\ \consumestep & 
     \Autoconf(\lro{\TnontLi y}{\TnontLi y \, \edgeLi yz}{ \TnontLi z}, {y/1,z/2}, \startedgeLi 1 \, \TnontLi 1 \, \edgeLi 12) 
     \\ \expandstep & 
     \Autoconf(\lro{\TnontLi y}{}{\TnontLi y \, \edgeLi yz \, \TnontLi z}, {y/2}, \startedgeLi 1 \, \TnontLi 1 \, \edgeLi 12) 
     \\ \consumestep & 
     \Autoconf(\lro{\TnontLi y}{\TnontLi y }{\edgeLi yz \, \TnontLi z}, {y/2}, \startedgeLi 1 \, \TnontLi 1 \, \edgeLi 12 \, \TnontLi 2) 
     \\ \consumestep & 
     \Autoconf(\lro{\TnontLi y}{\TnontLi y \, \edgeLi yz }{\TnontLi z}, {y/2,z/4}, \startedgeLi 1 \, \TnontLi 1 \, \edgeLi 12 \, \TnontLi 2 \, \edgeLi 24) 
     \\ \consumestep & 
     \Autoconf(\lro{\TnontLi y}{\TnontLi y \, \edgeLi yz \, \TnontLi z}{}, {y/2,z/4}, \startedgeLi 1 \, \TnontLi 1 \, \edgeLi 12 \, \TnontLi 2 \, \edgeLi 24 \, \TnontLi 4) 
  \end{array}
  \]
  \caption{Approval of a graph with a nondeterministic CFA. Renamings
    $\mu$ of states $q_i$ with
    $\mu(x_1) = y_1, \dots, \mu(x_k) = y_k$ are represented by exponents
    $x_1/y_1, \dots, x_k/y_k$}
  \label{f:nCFA-moves}
\end{figure}

It is rather obvious that one can arbitrarily rename input graph nodes
without affecting approval by the nCFA:

\begin{fact}\label{fact:step-inj}
  $\autoconf(q_0, \iota, \emptyseq) \autostep^n \autoconf(q, \mu, \phi)$
  implies
  $
    \autoconf(q_0, \iota, \emptyseq) 
    \autostep^n 
    \autoconf(q, \rho \circ \mu, \phi^\rho)
  $
  for every renaming~$\rho$.  
\end{fact}

Moreover, properties of goto moves can be generalized to goto sequences:

\begin{lemma}\label{lemma:n-gotos}
  $
    \autoconf(\literal A \to \alpha \CDOT \beta \gamma, \mu, \phi)
    \consumestep^*
    \autoconf(\literal A \to \alpha \beta \CDOT \gamma, \nu, \phi \psi)
  $ 
  implies
  $\mu \sqsubseteq \nu$, 
  $\dom(\nu) = \dom(\mu) \cup X(\beta)$, and
  $X(\phi) \cap X(\alpha^\nu \beta^\nu \gamma^\nu) \subseteq
  X(\alpha^\mu \beta^\mu \gamma^\mu)$.
\end{lemma}

\begin{proof}
  Consider any sequence 
  $
    \autoconf(\literal A \to \alpha \CDOT \beta \gamma, \mu, \phi)
    \consumestep^n
    \autoconf(\literal A \to \alpha \beta \CDOT \gamma, \nu, \phi \psi).
  $ 
  We prove the lemma by induction on~$n$. For
  $n = 0$ it follows from $\mu = \nu$.

  For $n > 0$, we have the sequence
  $$
  \autoconf(\literal A \to \alpha \CDOT \beta \gamma, \mu, \phi)
  \consumestep^{n-1}
  \autoconf(\literal A \to \alpha \beta' \CDOT \literal e \gamma, \nu', \phi \psi')
  \consumestep
  \autoconf(\literal A \to \alpha \beta \CDOT \gamma, \nu, \phi \psi) 
  $$
  with $\beta = \beta' \literal e$ and $\psi = \psi' \literal e^\nu$. By the
  induction hypothesis and the definition of goto moves, 
  \begin{align}
    \mu &\sqsubseteq \nu' \sqsubseteq \nu \label{eq:n-gotos-c}  \\
    X(\phi) \cap X(\alpha^{\nu'} \beta^{\nu'} \gamma^{\nu'}) &\subseteq
      X(\alpha^\mu \beta^\mu \gamma^\mu) \label{eq:n-gotos-b} \\
    \dom(\nu) & = \dom(\nu') \cup X(\literal e) 
                = \dom(\mu) \cup X(\beta') \cup X(\literal e) \notag \\
              & = \dom(\mu) \cup X(\beta) \label{eq:n-gotos-d}  \\
    X(\literal e^\nu) \cap X(\phi \psi') & \subseteq X(\literal e^{\nu'}) \label{eq:n-gotos-a}
  \end{align}
  By~\defref{def:auto-configuration} and \eqref{eq:n-gotos-c}, 
  $X(\alpha^\nu) = X(\alpha^{\nu'})$ and $X({\beta'}^\nu) = X({\beta'}^{\nu'})$. 
  Moreover, 
  $X(\gamma^\nu) \subseteq X(\gamma^{\nu'}) \cup X(\literal e^\nu)$ 
  using \eqref{eq:n-gotos-c} and~\eqref{eq:n-gotos-d}, and 
  $X(\literal e^\nu) \cap X(\phi) \subseteq X(\literal e^{\nu'}) \cap X(\phi)$ 
  using~\eqref{eq:n-gotos-a}.
  Therefore, 
  $X(\phi) \cap X(\alpha^\nu \beta^\nu \gamma^\nu) \subseteq 
  X(\phi) \cap X(\alpha^{\nu'} \beta^{\nu'} \gamma^{\nu'}) \subseteq 
  X(\alpha^\mu \beta^\mu \gamma^\mu)$ using~\eqref{eq:n-gotos-b}.
  \qed
\end{proof}

The following lemma shows that all images of nodes of the current nCFA
state have been read already, i.e., occur in processed literals.

\begin{lemma}\label{lemma:read-nodes}
  $
  \autoconf(q_0, \iota, \emptyseq)
  \autostep^*
  \autoconf(\literal A \to \alpha \CDOT \beta , \mu, \phi)
  $
  implies
  $X(\alpha^\mu \beta^\mu) \subseteq X(\phi)$.
\end{lemma}

\begin{proof}
  We prove the lemma by induction on the length~$n$ of the sequence 
  of moves. 
  For $n=0$ the inclusion follows 
  from $\literal A = \pseudoLit$, $\alpha = \emptyset$, 
  $\beta = \startLit$, and $\mu = \iota$.
  
  For $n > 0$ and the last move being a closure move, the sequence has
  the form
  $$
    \autoconf(q_0, \iota, \emptyseq)
    \autostep^{n-1}
    \autoconf(\literal B \to \delta \CDOT \literal C \gamma, \nu, \phi)
    \expandstep
    \autoconf(\literal A \to \alpha \CDOT \beta , \mu, \phi).
  $$
  with $\alpha = \emptyseq$, $\literal C^\nu = \literal A^\mu$, and
  $\dom(\mu) \subseteq X(\literal A)$. Therefore, 
  $X(\alpha^\mu \beta^\mu) = X(\literal A^\mu) = X(\literal C^\mu)
   \subseteq X(\delta^\nu \literal C^\nu \gamma^\nu) \subseteq 
   X(\phi)$
  using the induction hypothesis.

  For $n > 0$ and the last being a goto move, the sequence has the form
  $$
    \autoconf(q_0, \iota, \emptyseq)
    \autostep^{n-1}
    \autoconf(\literal A \to \alpha' \CDOT \literal e \beta , \nu, \phi')
    \consumestep
    \autoconf(\literal A \to \alpha \CDOT \beta , \mu, \phi)
  $$
  with $\alpha = \alpha' \literal e$, $\phi = \phi' \literal e^\mu$,
  $\nu \sqsubseteq \mu$, $\dom(\mu) = \dom(\nu) \cup X(\literal e)$, and
  $X(\literal e^\mu) \cap X(\phi') \subseteq X(\literal e^\nu)$. Therefore,
  $X(\beta^\mu) \subseteq X(\beta^\nu) \cup X(\literal e^\mu)$ and
  \begin{align*}
    X(\alpha^\mu \beta^\mu) 
    & \subseteq X({\alpha'}^\nu) \cup X(\literal e^\mu) \cup X(\beta^\nu) \\
    & \subseteq X({\alpha'}^\nu \literal e^\nu \beta^\nu) \cup X(\literal e^\mu) \\
    & \subseteq X(\phi') \cup X(\literal e^\mu)\\
    & = X(\phi)
  \end{align*}
  using the induction hypothesis.
  \qed
\end{proof}

We now show that the graphs approved by the nCFA are
viable prefixes (\lemmaref{lemma:auto-to-vp}) and vice versa
(\lemmaref{lemma:auto-from-vp}).

\begin{lemma}\label{lemma:auto-to-vp}
  For every sequence
  $$\autoconf(q_0, \iota, \emptyseq) \autostep^* 
   \autoconf(\literal A \to \alpha \CDOT \beta, \mu, \phi)$$
  of moves and every injective function $\tau \colon X(\alpha\beta) \to X$ with
  $\mu \sqsubseteq \tau$ and $X(\beta^\tau) \cap X(\phi) \subseteq X(\beta^\mu)$,
  there exist $\psi \in \Gr\Lab$ and $z \in \Gr{\TLab}$ such that
  $$ \pseudoLit \drm^* \psi \literal A^\tau z \drm 
    \psi \alpha^\tau \beta^\tau z = \phi \beta^\tau z.$$ 
\end{lemma}

\begin{proof}
  We proceed by induction on the number $n$ of moves.
  If $n = 0$ the statement follows from the definition of initial nCFA 
  configurations (with $\tau = \iota$, $\literal A=\pseudoLit=\pseudoLit^\tau$, $\phi=\alpha=z=\emptyseq$, and $\beta=\pseudoLit$).
  
  If $n > 1$ and the last move is a goto move, we have
  $$
  \autoconf(q_0, \iota, \emptyseq)
  \autostep^{n-1}
  \autoconf(\literal A \to \alpha' \CDOT \literal e \beta, \nu, \phi')
  \consumestep
  \autoconf(\literal A \to \alpha' \literal e \CDOT \beta, \mu, \phi) 
  =
  \autoconf(\literal A \to \alpha \CDOT \beta, \mu, \phi)
  $$
  where     
  \begin{align}
    \phi & = \phi' \literal e^\mu\\
    \nu & \sqsubseteq \mu\label{eq:auto-to-v-a}\\
    \dom(\mu) & = \dom(\nu) \cup X(\literal e) 
    \label{eq:auto-to-v-b}\\
    X(\literal e^\mu) \cap X(\phi') & \subseteq X(\literal e^\nu),
    \label{eq:auto-to-v-c}
  \end{align}
  Let $\tau$ be as in the lemma.
  Then $\nu\sqsubseteq\mu \sqsubseteq \tau$.
  To make use of the induction hypothesis, we additionally need that
  $X(\literal e^\tau\beta^\tau) \cap X(\phi')\subseteq X(\literal e^\nu\beta^\nu)$.
  Indeed,
  $$
  \begin{aligned}
  X(\literal e^\tau\beta^\tau) \cap X(\phi') &= (X(\literal e^\tau)\cap X(\phi'))\cup(X(\beta^\tau) \cap X(\phi'))\\
  &= (X(\literal e^\mu)\cap X(\phi'))\cup(X(\beta^\tau) \cap X(\phi) \cap X(\phi'))\\
  &\subseteq X(\literal e^\nu)\cup(X(\beta^\mu) \cap X(\phi'))\\
  &\subseteq X(\literal e^\nu)\cup X(\beta^\nu)\\
  &= X(\literal e^\nu\beta^\nu).
  \end{aligned}
  $$
  Hence the induction hypothesis applies, yielding $\psi \in \Gr\Lab$ and $z \in \Gr{\TLab}$ such that
  $ \pseudoLit \drm^* \psi \literal A^\tau z \drm 
  \psi \alpha'^\tau \literal e^\tau \beta^\tau z = \phi \beta^\tau z$,
  which proves the proposition.

  If $n > 0$ and the last move is a closure move, we have
  $$
  \autoconf(q_0, \iota, \emptyseq)
  \autostep^{n-1}
  \autoconf(\literal B \to \gamma \CDOT \literal C \delta, \nu, \phi)
  \autostep
  \autoconf(\literal A \to \CDOT \beta, \mu, \phi) 
  $$
  where
  $\literal C^\nu = \literal A^\mu$,
  $\dom(\mu) \subseteq X(\literal A)$, and $\alpha=\emptyseq$.
  Again, let
  $\tau \colon X(\alpha \beta) \to X$ be as in the statement of the lemma. To be able to 
  use the induction hypothesis we need an injective function 
  $\eta \colon X(\gamma\literal C\delta) \to X$ such that $\nu\sqsubseteq\eta$ and
  $X(\literal C^\eta\delta^\eta)\cap X(\phi)\subseteq X(\literal C^\nu\delta^\nu)$. But we also need $\literal C^\eta = \literal A^\tau$   
  in order to conclude a derivation
  $\pseudoLit \drm^* \psi \literal B^\eta w \drm 
   \psi \gamma^\eta \literal C^\eta \delta^\eta w = 
   \phi \literal A^\tau \delta^\eta w$. 
  However, this may be impossible because of name clashes. 
  
  To solve this problem, we rename all nodes that may cause 
  name clashes and use \factref{fact:step-inj}. Choose any renaming 
  $f$ with $f(x) = x$ for $x \in X(\phi)$ and
  $f(x) \notin X(\beta^\tau)$ for $x \in X(\beta^\tau) \setminus X(\phi)$.
  Then $X(\literal C^\nu) \subseteq X(\phi)$ because of 
  \lemmaref{lemma:read-nodes}, and hence
  $\literal C^{f \circ \nu} = (\literal C^\nu)^f = \literal C^\nu 
                            = \literal A^\mu$
  as well as $\phi^f = \phi$, and therefore 
  $$
  \autoconf(q_0, \iota, \emptyseq)
  \autostep^{n-1}
  \autoconf(\literal B \to \gamma \CDOT \literal C \delta, f \circ \nu, \phi)
  \autostep
  \autoconf(\literal A \to \CDOT \beta, \mu, \phi). 
  $$
  We can now choose a renaming $\eta$ with
  $f \circ \nu\sqsubseteq\eta$,
    $X(\literal C^\eta \delta^\eta) \cap X(\phi) \subseteq 
   X(\literal C^{f \circ \nu} \delta^{f \circ \nu})$, and 
  $\literal C^\eta = \literal A^\tau$, and by the induction hypothesis,
  $\pseudoLit \drm^* \psi \literal B^\eta w \drm 
   \psi \gamma^\eta \literal C^\eta \delta^\eta w = 
   \phi \literal A^\tau \delta^\eta w$.
  Although $\literal A^\tau \der \beta^\tau$, we cannot conclude  
  $\phi \literal A^\tau \delta^\eta w \der \phi \beta^\tau \delta^\eta w$
  because $\delta^\eta w$ may contain nodes that are created by 
  the derivation $\literal A^\tau \der \beta^\tau$. Again, we solve this problem 
  by renaming the nodes in $\delta^\eta w$ to new nodes. For this 
  purpose, let $Y = X(\delta^\eta w) \setminus X(\phi \literal A^\tau)$ and 
  choose, for each $y \in Y$, a new node
  $n_y \in X \setminus X(\phi \beta^\tau \delta^\eta w)$. Let 
  $h$ be a renaming with $h(x) = n_x$ if $x \in Y$ 
  and $h(x) = x$ for $x \in X(\phi \beta^\tau)$.
  By \lemmaref{lemma:image-deriv}, 
  $
  \pseudoLit 
  \drm^* 
  (\phi \literal A^\tau \delta^\eta w)^h 
  = 
  \phi \literal A^\tau \delta^{h \circ \eta} w^h.
  $
  By the definition of~$h$, and because $\Gamma$ is 
  reduced, there is a graph $u \in \Gr{\TLab}$ such that
  $$
  \phi \literal A^\tau \delta^{h \circ \eta} w^h 
  \der 
  \phi \beta^\tau \delta^{h \circ \eta} w^h
  \mathrel{\mathop\der\nolimits^*}
  \phi \beta^\tau u w^h.
  $$
  Therefore, $\pseudoLit \drm^* \phi \beta^\tau z$ for $z = u w^h$,
  which completes the proof.
  \qed
\end{proof}

\begin{lemma}\label{lemma:auto-from-vp}
  For every derivation
  $\pseudoLit \drm^* \alpha \literal A z \drm \alpha \beta z$ 
  and each prefix~$\phi$ of $\alpha \beta$, there is a sequence 
  $
  \autoconf(q_0, \iota, \emptyseq)
  \autostep^*
  \autoconf(p, \nu, \phi)
  $
  of moves (for a suitable state $[p]^\nu$).
\end{lemma}

\begin{proof}
  We prove by induction on~$n$ that 
  $\pseudoLit \drm^n \alpha \literal A z \drm \alpha \beta z$ implies 
  $\autoconf(q_0, \iota, \emptyseq) \autostep^* \autoconf(p, \nu, \phi)$ 
  for every prefix $\phi$ of $\alpha \beta$.
  
  For $n = 0$, the derivation is
  of the form $\pseudoLit \drm \startLit = \beta$ and we have
  $\phi=\emptyseq$ or $\phi=\startLit$. Hence, $\phi\in\set{\emptyseq,\startLit}$,
  and the proposition follows by making no move at all, or by making the goto move
  $$
    \autoconf(q_0, \iota, \emptyseq)
    \consumestep
    \autoconf(\pseudoLit \to \startLit \CDOT, \iota, \startLit).
  $$

  For $n > 0$, the initial part of the derivation up to $\alpha\literal Az$ has the form 
  $$
  \pseudoLit \drm^{n-1} \theta \literal X w \drm \theta \rho w 
   = \alpha \literal Az.
  $$
  There are two cases:
  \begin{enumerate}[(1)]
    \item $\rho \in \Gr{\TLab}$. 

    Then $\theta = \alpha \literal A u$ for some $u\in\Gr\TLab$, 
    $
    \pseudoLit 
    \drm^{n-1} 
    \alpha \literal A u \literal X w \drm \alpha \literal A u \rho w 
    \drm 
    \alpha \beta u \rho w $. We distinguish two sub-cases:

    \begin{enumerate}[({1}a)]
      \item \label{item:case-a}$\phi$ is a prefix of $\alpha$.
      
      Then $\phi$ is a prefix of $\alpha \literal A u \rho$ and hence the claimed
      sequence is obtained directly from the induction hypothesis.

      \item \label{item:case-b}$\phi = \alpha \tau$ for a prefix $\tau$ of $\beta$.

      By the induction hypothesis, there is a sequence
      $
      \autoconf(q_0, \iota, \emptyseq)
      \autostep^m
      \autoconf(p, \nu, \alpha \literal A)
      $
      of moves. W.l.o.g, let $m$ be the minimum number of such moves. 
      By \defref{def:auto-conf}, this sequence must be of the form
      $$
      \autoconf(q_0, \iota, \emptyseq)
      \autostep^{m-1}
      \autoconf(\literal B \to \delta \CDOT \literal C \delta', \mu, \alpha )
      \consumestep
      \autoconf(\literal B \to \delta \literal C \CDOT \delta', \nu, \alpha \literal A)
      $$
      with $\literal A = \literal C^\nu$. Suppose that 
      $\literal A \der \beta$ uses rule $\literal D \to \psi \bar\psi$ with
      $|\tau| = |\psi|$.  By making a closure move instead of 
      the goto move at the end of the sequence above, we obtain
      $$
      \autoconf(q_0, \iota, \emptyseq)
      \autostep^{m-1}
      \autoconf(\literal B \to \delta \CDOT \literal C \delta', \mu, \alpha )
      \expandstep
      \autoconf(\literal D \to \CDOT \psi \bar\psi, \sigma, \alpha )
      $$
      with $\literal C^\mu = \literal D^\sigma$. The claimed sequence is obtained by 
      applying the appropriate number of goto moves:
      $$
      \autoconf(\literal D \to \CDOT \psi \bar\psi, \sigma, \alpha )
      \consumestep^*
      \autoconf(\literal D \to \psi \CDOT \bar\psi, \sigma', \alpha \tau)
      =
      \autoconf(\literal D \to \psi \CDOT \bar\psi, \sigma', \phi).
      $$
    \end{enumerate}

    \item $\rho \notin \Gr{\TLab}$.

    Then $\literal A$ is a literal in $\rho$ and the given derivation has the form
    $
    \pseudoLit \drm^{n-1} \theta \literal X w 
    \drm \theta \gamma \literal A u w \drm \theta \gamma \beta u w
    $
    where $\theta\gamma=\alpha$, and thus $\phi$ is a prefix of $\theta \gamma \beta$. We distinguish two sub-cases:
    \begin{enumerate}[({2}a)]
      \item $\phi$ is prefix of $\theta\gamma$.
      
      As in case~\ref{item:case-a}, the proposition follows directly from the induction hypothesis because $\phi$ is a prefix of $\theta \gamma \literal A u$.
%
%
%
%
      \item $\phi = \theta \gamma \tau$ for a prefix $\tau$ of $\beta$.

      By the induction hypothesis, there is a sequence
      $
      \autoconf(q_0, \iota, \emptyseq)
      \autostep^*
      \autoconf(p, \nu, \theta \gamma \literal A)
      $, 
      and with a similar argument as in case~\ref{item:case-b},
      $
      \autoconf(q_0, \iota, \emptyseq)
      \autostep^*
      \autoconf(p', \sigma, \theta \gamma \tau).
      $
      \qed
    \end{enumerate}
  \end{enumerate}
\end{proof}

An immediate consequence of \lemmaref{lemma:auto-to-vp} and \lemmaref{lemma:auto-from-vp} is the following:

\begin{theorem}\label{thm:nCFA-prefix}
A graph $\phi \in \Gr\Lab$ is a viable prefix if and only if the nCFA approves~$\phi$.
\end{theorem}

On the one hand, we have shown at the end of \sectref{s:naive} that the
na\"{\i}ve nondeterministic parser can reach the accepting configuration 
(with some appropriate remaining input)
if and only if the current stack content is a viable prefix
(\thmref{thm:prefix}). On the other hand, \thmref{thm:nCFA-prefix}
shows that the nCFA approves precisely the viable prefixes. In other words,
the naïve parser can avoid running into a situation in which no remaining input
could ever make it accept, if moves are restricted to those which
produce stacks approved by the nCFA.



\section{Deterministic Characteristic Finite-State Automata}\label{s:dCFA}

Because of its spontaneously acting closure transitions, the nCFA cannot
efficiently be used to improve the naïve shift-reduce parser by making sure that the
stack of the parser is always a viable prefix.
This is so because the nCFA, whenever it reaches a configuration, may
also be in any configuration reachable by closure moves. In a deterministic
implementation, all these
configurations must be maintained simultaneously when the next goto
move shall be made. To avoid this, we preprocess the nCFA and create the
deterministic CFA (dCFA) by combining such simultaneously reachable states
into new states, using a procedure similar to the classical powerset
construction.

Literals of prefixes approved by an nCFA are images of literals
of transitions and nCFA states under node mappings. The idea behind our
adaptation of the traditional powerset construction to CFAs
is to split such a node
mapping into a composition of two mappings; the so-called
\emph{parameter mapping} is applied first, intuitively providing the node with a formal parameter name under
which the node can be addressed. Later, the parameters are bound
to nodes of the input graph by an \emph{\inputbinding}. Thus, the \inputbinding\ maps parameters to
nodes of the actual input graph during an actual run of the dCFA,
whereas the parameter mapping is chosen when constructing the
dCFA. Different nodes that are always mapped to the same input graph
nodes and that belong to nCFA states combined in a common \dCFAstate,
are mapped to the same parameter by the algorithm. For technical simplicity,
we use nodes as parameters rather than introducing a special class of symbols
for parameters. An \emph{item} is then an nCFA state together with a parameter mapping:

\begin{defn}[Item]\label{def:dCFA:pitem}
  An \emph{item} $\pstate(q,\sigma)$ consists of a state
  $q = \qrule{\literal A \to \alpha \CDOT \beta}$ of our (fixed) nCFA and an injective
  partial \emph{parameter mapping} $\sigma \colon X \pto X$ with
  $X(\alpha)\subseteq \dom(\sigma) \subseteq X(\alpha) \cup X(\literal A)$.
  $\items$ denotes the set of all items.
\end{defn}

Note that the parameter mapping maps only those nodes of the item
which occur in literals preceding the dot (or in $\literal A$).

If an nCFA processes a graph and reaches a state~$q$, it can also reach
those states reachable from $q$ by closure transitions. Of course,
nodes must be renamed appropriately, as we are dealing with items instead
of pure states. An item that is reachable from another
item by a closure transition is called \emph{closure item} of the latter. 
The formal definition reads as follows:

\begin{defn}[Closure of Items]\label{def:dCFA:closure}
  We call an item $\pstate(q, \tau)$ a \emph{closure item} of $\pstate(p, \sigma)$, written $\pstate(p, \sigma) \clps \pstate(q, \tau)$, 
  if $p = \qrule{\literal A \to \alpha \CDOT
  \literal B \beta}$ and $q = \qrule{\literal C \to \CDOT \delta}$,
  $\literal C^\tau = \literal B^\sigma$, and $\dom(\tau) \subseteq
  X(\literal C)$.

  The \emph{closure} of a set~$I$ of items is the smallest set $J$
  that contains all members of $I$ and, for each item in $J$, all its
  closure items.
\end{defn}

The closure of a given set of
items can be computed in the usual way by adding all closure items to the set
and repeating this procedure as long as new items are added to the set:

\begin{fact}\label{fact:closure}
  Let $J$ be the closure of a set $I$ of items. Then, for every item 
  $\pstate(q', \sigma') \in J$, there is an 
  item $\pstate(q, \sigma) \in I$ such that 
  $\pstate(q, \sigma) \clps^* \pstate(q', \sigma')$.
\end{fact}

In the definition of dCFAs below, states will be (finite) subsets $Q$ of $2^\items$. Such a set is said to be \emph{closed} if it is its own closure, and its set of \emph{parameters} is
\[
  \params(Q) = \bigcup_{\pstate(q, \sigma) \in Q} \sigma(X).
\]

For a renaming~$\mu$ with $\dom(\mu) \supseteq \params(Q)$, we let
$$Q^\mu = \set{ \pstate(q,\mu \circ \sigma) \mid \pstate(q,\sigma) \in Q}$$
be the set of items obtained by renaming parameters according to $\mu$.
Sets~$Q, Q'\subseteq 2^\items$ are equivalent, written $Q \similar Q'$, if
$Q^\mu = Q'$ for some renaming $\mu$.

We are now ready to give the formal definition of dCFAs.
Each \dCFAstate\ is a closed set of items. In particular, the initial
state $\startstate$ is the closure of the initial state of the nCFA.
Transitions in the dCFA are labeled with pairs that consist of a
literal and a node mapping. The literal is the one that triggers the
state transition; its nodes are parameters of the source state
of the transition and new parameters whose ``values'' will be the corresponding
nodes of the literal processed during parsing. The node mapping of the transition will later be
used to set the ``values'' of the target state parameters.

\begin{defn}
  A \emph{deterministic characteristic finite automaton} (dCFA) 
  $\dCFA = (\mathcal Q, \startstate, \Delta)$ consists of a finite set 
  $\mathcal Q\subseteq 2^\items$ of so-called \emph{\dCFAstate{}s}, an initial state
  $\startstate \in \mathcal Q$, and a 
  transition relation 
  $\Delta$
  with the following properties:
  \begin{enumerate}

  \item For all $Q, Q' \in \mathcal Q$, $Q \similar Q'$ implies
    $Q = Q'$.

  \item $\startstate$ is the closure of $\set{ \pstate(q_0, \iota)}$,
    where $q_0=(\pseudoLit \to \CDOT \startLit)$ is the initial state
    of the nCFA.
      
  \item $\Delta$ is a finite set of transition rules $Q \atrans{(\literal e, \mu)} Q'$, such that $Q,Q'\in\mathcal Q$, $\literal e\in\Lit\Lab$, and $\mu\colon\params(Q') \to \params(Q)\cup X(\literal e)$.
  \end{enumerate}
\end{defn}

If it is obvious that we speak of the \dCFAstate{}s of a dCFA, and thus there is no risk of confusing them with the concrete states to be defined soon, we may simply call them \emph{states}.
\begin{algorithm}[tb]
  \caption{Converting the nCFA $\nCFA$ into a deterministic CFA.}
  \label{alg:deterministic}
  \Input{Nondeterministic CFA $\nCFA$} 
  \Output{Equivalent deterministic CFA $\dCFA$}
  let $q_0=(\pseudoLit \to \CDOT \startLit)$ be the initial state of 
    $\nCFA$ \;
  compute $\startstate$ as the closure of $\set{ \pstate(q_0, \iota)}$ and let
    $\dCFA$ be the automaton with initial state $\startstate$ and no 
    further states yet\;\label{an:start-state}
  $W \leftarrow \set{\startstate}$\;
  \While{$W \neq \emptyset$}{
    select and remove any state $S$ from $W$\;
    \ForEach{$\literal l \in \leave(S)$}{
      obtain literal $\literal e$ from $\literal l$ by replacing
      each occurrence of `$\unknown$' in $\literal l$ by a new node not
      used anywhere else\;\label{an:new-parameter}
      $\Tgt \leftarrow \emptyset$\;
      \ForEach{$\pstate(q,\sigma) \in S$ with 
               $\literal l \in \leave(q,\sigma)$}{
          let $q = \qrule{\literal A \to \alpha \CDOT \literal f \beta}$\;
          let $\nu \colon X \pto X$ be injective
          with $\sigma \sqsubseteq \nu$, 
          $\literal f^\nu = \literal e$,
          and $\dom(\nu) = \dom(\sigma) \cup X(\literal f)$\; \label{an:nu}
          add $\pstate(\literal A \to \alpha \literal f \CDOT \beta, \nu)$ 
          to $\Tgt$\label{an:add-to-N}
      }
      compute the closure $\Tgt'$ of $\Tgt$\;\label{an:closure-N}
      \eIf{$\dCFA$ has a state $Q \similar \Tgt'$\label{an:equiv}}{
        add a transition $S \atrans{(\literal e, \mu)} Q$ to $\dCFA$ 
        where $\Tgt' = Q^\mu$\label{an:reuse}
      }{
        add $\Tgt'$ as a new state to $\dCFA$ and $W$\;
        add a transition $S \atrans{(\literal e, \id)} \Tgt'$ to $\dCFA$
        \label{an:new-state}
    }
  }
  }
\end{algorithm}

\algref{alg:deterministic} converts an nCFA into a corresponding dCFA.
To determine the set of all transitions leaving a \dCFAstate\ $S$,
it considers each element $\literal l$ of the set 
$$
\leave(S) = \bigcup_{\pstate(q,\sigma) \in S} \leave(q,\sigma)
$$
where $\leave(\literal A \to \alpha \CDOT,\sigma) = \emptyset$ and
$\leave(\literal A \to \alpha \CDOT \literal e \beta,\sigma) =
\set{\literal e^\sigma}$, i.e., $\leave(S)$ contains mapped images of
those literals that are labels of goto transitions leaving the
corresponding nCFA states. These literals are mapped by the parameter
mapping, i.e., they are in general pseudo-literals whose ``nodes'' are either parameters of~$S$,
or~`$\unknown$' if they are not (yet) mapped in~$S$.

Parameter names can be chosen arbitrarily as long as the parameter mappings are
injective. That way, \algref{alg:deterministic} frequently
creates new sets of items which should become states of the 
dCFA~$\dCFA$, but
are equivalent to sets that have already been added as states to $\dCFA$
and should thus not be added again. \algref{alg:deterministic}
avoids equivalent states (\alineref{an:equiv}) and ``reuses'' existing
ones instead (\alineref{an:reuse}). The node mapping being part of
transition labels is the parameter renaming that must be applied to reuse
an already existing state.


\begin{example}[The dCFA for the Tree-Generating Grammar\label{x:tree-dCFA}]\hspace*{0pt plus 1em}%
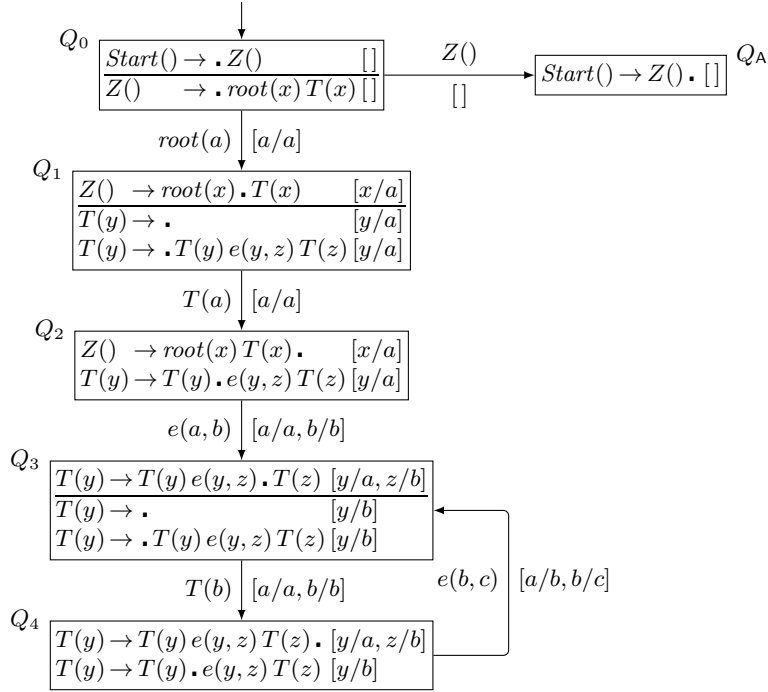
\begin{figure}[tb]
  \centering
\begin{tikzpicture}[font=\small, 
                    auto, 
                    ->,
                    node distance=8mm, label distance=-0.5mm]
  \tikzset{
    state/.style={draw, rectangle,inner sep=2pt},
  }

  \node (init) {};

  \node [state, below=5mm of init] (M0) {
    $\begin{array}{@{}l@{\,\to\,}l@{\,}l@{}}
      \pseudoLi & \CDOT \startLi & [\,]\\ \hline
      \startLi & \CDOT \startedgeL(x) \, \TnontLi x & [\,]\\
    \end{array}$
  };

  \node [state, right=2cm of M0] (M5) {
    $\begin{array}{@{}l@{\,\to\,}l@{\,}l@{}}
      \pseudoLi & \startLi \CDOT & [\,]\\
    \end{array}$
  };

  \node [state, below=of M0] (M1) {
    $\begin{array}{@{}l@{\,\to\,}l@{\,}l@{}}
      \startLi & \startedgeL(x) \CDOT \TnontLi x & [x/\parama]\\ \hline
      \TnontLi y & \CDOT & [y/\parama]\\
      \TnontLi y & \CDOT \TnontLi y \, \edgeLi yz \, \TnontLi z & [y/\parama]
    \end{array}$
  };

  \node [state, below=of M1] (M2) {
    $\begin{array}{@{}l@{\,\to\,}l@{\,}l@{}}
      \startLi & \startedgeL(x) \, \TnontLi x \CDOT & [x/\parama]\\
      \TnontLi y & \TnontLi y \CDOT \edgeLi yz \, \TnontLi z & [y/\parama]
    \end{array}$
  };

  \node [state, below=of M2] (M3) {
    $\begin{array}{@{}l@{\,\to\,}l@{\,}l@{}}
      \TnontLi y & \TnontLi y \, \edgeLi yz \CDOT \TnontLi z & [y/\parama, z/\paramb] \\ \hline
      \TnontLi y & \CDOT & [y/\paramb]\\
      \TnontLi y & \CDOT \TnontLi y \, \edgeLi yz \, \TnontLi z & [y/\paramb]
    \end{array}$
  };

  \node [state, below=of M3] (M4) {
    $\begin{array}{@{}l@{\,\to\,}l@{\,}l@{}}
      \TnontLi y & \TnontLi y \, \edgeLi yz \, \TnontLi z \CDOT & [y/\parama, z/\paramb] \\
      \TnontLi y & \TnontLi y \CDOT \edgeLi yz \, \TnontLi z & [y/\paramb]
    \end{array}$
  };

  \node [left=0mm of M0.north west, anchor=east] {$Q_0$};
  \node [left=0mm of M1.north west, anchor=east] {$Q_1$};
  \node [left=0mm of M2.north west, anchor=east] {$Q_2$};
  \node [left=0mm of M3.north west, anchor=east] {$Q_3$};
  \node [left=0mm of M4.north west, anchor=east] {$Q_4$};
  \node [right=0mm of M5.north east, anchor=west] {$\acceptstate$};

  \draw (init) -- (M0);
  \draw (M0) -- node[left] {$\startedgeL(\parama)$} 
                node[right] {$[\parama/\parama]$} (M1);
  \draw (M0) -- node[above] {$\startLi$} 
                node[below] {$[\,]$} (M5);
  \draw (M1) -- node[left] {$\TnontLi \parama$} 
                node[right] {$[\parama/\parama]$} (M2);
  \draw (M2) -- node[left] {$\edgeLi \parama\paramb$} 
                node[right] {$[\parama/\parama, \paramb/\paramb]$} (M3);
  \draw[rounded corners=2pt]
                (M3.south) -- 
                             node[left] {$\TnontLi \paramb$} 
                             node[right] {$[\parama/\parama, \paramb/\paramb]$} 
                             (M4.north);
  \draw[rounded corners=2pt]
                (M4.east) -| ++(10mm,10mm) 
                             node[left] {$\edgeLi \paramb\paramc$} 
                             node[right] {$[\parama/\paramb, \paramb/\paramc]$} |- (M3.east);
                                                         
\end{tikzpicture}
  \caption{Deterministic CFA created by \algref{alg:deterministic} from
    the nCFA in \figref{fig:nCFA}}
  \label{fig:tree-dCFA}
\end{figure}
\figref{fig:tree-dCFA} shows the dCFA obtained by
\algref{alg:deterministic} from the nondeterministic CFA
(\figref{fig:nCFA}) for the tree-generating grammar in \exref{x:tree:HR}.  It
consists of the states $Q_0,\dots,Q_4,\acceptstate$, which are sets of
items. Each item $\pstate(q,\sigma)$
is written in a single line that shows its nCFA state~$q$ on the left
and its parameter mapping~$\sigma$ on the right (in brackets). Each
pair $u/v$ denotes that node $u$ of the item is mapped to the
parameter $v$, i.e., $\sigma(u) = v$; $\sigma$~is undefined for all
other nodes. Transitions are labeled by pairs of literals and node
mappings. The latter are represented analogously.

Note that the three transitions leading from $Q_0$ via $Q_1$ and $Q_2$ to
$Q_3$ have identities as node mappings because the target states of these
transitions were new states when \algref{alg:deterministic} created
the transition (\alineref{an:new-state}). When it created the
transition leaving $Q_4$, however, it constructed the set
\begin{align*}
  \Tgt' = \{ \;      
   & \pstate({\TnontLi y \to \TnontLi y \, \edgeLi yz \CDOT \TnontLi z} , {[y/\paramb, z/\paramc]}), \\
   & \pstate({\TnontLi y \to \CDOT} , {[y/\paramc]}),\\
   & \pstate({\TnontLi y \to \CDOT \TnontLi y \, \edgeLi yz \, \TnontLi z} , {[y/\paramc]}) \; \}
\end{align*}
of items, for the new parameter $\paramc$ created in \alineref{an:new-parameter}.
But this set is equal to $Q_3^\mu$ where $\mu = [\parama/\paramb, \paramb/\paramc]$ is the 
corresponding node mapping, and \algref{alg:deterministic} has reused 
state $Q_3$ when adding the transition in \alineref{an:reuse}.
%
%
\qed
\end{example}


\begin{assumption}\label{a:dCFA}
Since \algref{alg:deterministic} constructs the dCFA from an nCFA,
and as we have assumed a fixed HR grammar~$\Gamma$ and a fixed nCFA, we
also assume a fixed dCFA~$\dCFA$ in the following.
\end{assumption}

The dCFA approves graphs in a similar way as the nCFA does. However, we
have to deal with \inputbinding{}s~$\tau$ that map parameters to nodes of the
input graph $g$. Formally, an \emph{\inputbinding} for a \dCFAstate\ $Q$ is a total
injective function $\tau\colon\params(Q)\to X(g)$, where $g$ is the input
graph. Here, $g$ may not explicitly be given, but is implicitly present as
the graph on which the dCFA is run. The item set $Q^\tau$ is then a concrete state,
which we simply call a \emph{state}. In other words, a state is a finite set of
items in which the parameters are mapped to (pairwise distinct) nodes of the
input graph.

We define
dCFA configurations, dCFA moves, and approval by a dCFA
analogously to their nCFA counterparts
(Definitions~\ref{def:auto-configuration}, \ref{def:auto-conf},
and~\ref{def:auto-approve}). The primary difference is that dCFA
moves do not have to take closure moves into account:

\begin{defn}\label{def:dCFA-conf}
  A \emph{dCFA configuration} $\cfaconf(\phi,Q^\tau)$ consists of a
  graph $\phi \in \Gr\Lab$ and a state~$Q^\tau$. 
  
  A dCFA transition 
  $\psrtr =  (Q \atrans{(\literal e, \mu)} Q')$ turns  
  $\cfaconf(\phi,Q^\tau)$ into 
  $\cfaconf(\phi \literal e^\nu, Q'^{\nu \circ \mu})$ with 
  $\tau \sqsubseteq \nu$,
  $\dom(\nu) = \dom(\tau) \cup X(\literal e)$,
  and $X(\literal e^\nu) \cap X(\phi) \subseteq X(\literal e^\tau)$.
  We write such a \emph{dCFA move} as 
  $
    \cfaconf(\phi,Q^\tau) 
    \lcfastep{\psrtr } 
    \cfaconf(\phi \literal e^\nu, {Q'}^{\nu \circ \mu})
  $,
  where the subscript $\psrtr $ may be omitted.

  The dCFA \emph{approves} a graph $\phi \in \Gr \Lab$
  if there is a dCFA configuration 
  $C = \cfaconf(\phi,Q^\tau)$ such that 
  $\cfaconf(\emptyseq, \startstate^\iota) \cfastep^* C$.
\end{defn}


\begin{figure}[bt] \renewcommand{\arraystretch}{1.4}
  \[ \begin{array}{rcr}
     \cfaconf(\emptyseq, Q_0^{\,})
     & \cfastep & 
     \cfaconf(\startedgeLi 1, Q_1^{\parama/1}\quad\;~) 
     \\ & \cfastep & 
     \cfaconf(\startedgeLi 1 \, \TnontLi 1,Q_2^{\parama/1}\quad\;~) 
     \\ & \cfastep & 
     \cfaconf(\startedgeLi 1 \, \TnontLi 1 \, \edgeLi 12, Q_3^{\parama/1,\paramb/2}) 
     \\& \cfastep & 
     \cfaconf(\startedgeLi 1 \, \TnontLi 1 \, \edgeLi 12 \, \TnontLi 2, Q_4^{\parama/1,\paramb/2}) 
     \\ & \cfastep & 
     \cfaconf(\startedgeLi 1 \, \TnontLi 1 \, \edgeLi 12 \, \TnontLi 2 \, \edgeLi 24, Q_3^{\parama/1,\paramb/2}) 
     \\ & \cfastep & 
     \cfaconf(\startedgeLi 1 \, \TnontLi 1 \, \edgeLi 12 \, \TnontLi 2 \, \edgeLi 24 \, \TnontLi 4,Q_4^{\parama/2,\paramb/4}) 
  \end{array}
  \]
  \caption{Moves of the dCFA in \figref{fig:tree-dCFA}. \Inputbinding{}s
    $\tau$ of the \concretestate{}s $Q_i^\tau$ with
    $\tau(a_1) = y_1, \dots, \tau(a_k) = y_k$ are represented by exponents
    $a_1/y_1, \dots, a_k/y_k$ 
    .}
  \label{fig:tree-approval}
\end{figure}
\begin{example}[Moves of the dCFA for the Tree-Generating
  Grammar\label{x:tree-approval}]
  \figref{fig:tree-approval} shows moves of the deterministic CFA
  in \figref{fig:tree-dCFA}.
  They approve the same graph as the moves (shown in
  \figref{f:nCFA-moves}) of the nondeterministic CFA of
  \figref{fig:nCFA}. Note that the states $Q_i^\tau$ in 
  \figref{fig:tree-approval} are in fact concrete states.%
  \qed
\end{example}


To prove that the dCFA approves the same graphs as the nCFA, we
need the following lemma. It shows that constructing closure items
is tightly related to performing closure moves. 

\begin{lemma}\label{lemma:closure}
  $\pstate(p, \sigma) \clps^* \pstate(q, \tau)$ implies 
  $\autoconf(p, \mu \circ \sigma, \phi) 
   \expandstep^* 
   \autoconf(q, \mu \circ \tau, \phi)$
  for all items $\pstate(p, \sigma)$ and $\pstate(q, \tau)$, every injective partial function $\mu \colon X \pto X$ with $\dom(\mu \circ \sigma) = \dom(\sigma)$, and every graph $\phi \in \Gr\Lab$ such that $\autoconf(p, \mu \circ \sigma, \phi)$ is a valid nCFA configuration.
\end{lemma}

\begin{proof}
  Consider any items $\pstate(p, \sigma)$ and $ \pstate(q, \tau)$ such that 
  $\pstate(p, \sigma) \clps^n \pstate(q, \tau)$, and $\mu$ as well as $\phi$ 
  as in the lemma. We prove $\dom(\mu \circ \tau) = \dom(\tau)$ and  
  $\autoconf(p, \mu \circ \sigma, \phi) 
  \expandstep^n 
  \autoconf(q, \mu \circ \tau, \phi)$ 
  by induction on~$n$.

  For $n = 0$, $\pstate(p, \sigma) = \pstate(q, \tau)$, and therefore,
  $\autoconf(p, \mu \circ \sigma, \phi) 
  = 
  \autoconf(q, \mu \circ \tau, \phi)$ 
  as well as $\dom(\mu \circ \tau) = \dom(\tau)$, as claimed.

  For $n > 0$, we have 
  $\pstate(p, \sigma) \clps^{n-1} \pstate(p', \sigma') \clps \pstate(q, \tau)$,
  and by the induction hypothesis,
  $\autoconf(p, \mu \circ \sigma, \phi) 
  \expandstep^{n-1} 
  \autoconf(p', \mu \circ \sigma', \phi)$ 
  and $\dom(\mu \circ \sigma') = \dom(\sigma')$.
  Then $p'$ and $q$ are of the form 
  $p' = \qrule{\literal A \to \alpha \CDOT \literal B \beta}$,
  $q = \qrule{\literal C \to \CDOT \delta}$, 
  $\literal C^{\tau} = \literal B^{\sigma'}$, 
  $\dom(\tau) \subseteq X(\literal C)$.
  
  We first show that 
  $\dom(\mu \circ \tau) = \dom(\tau)$.
  The inclusion $\dom(\mu \circ \tau) \subseteq \dom(\tau)$ follows 
  immediately from the definition of the composition of partial functions. 
  In order to show the opposite inclusion, consider any node 
  $x \in \dom(\tau) \subseteq X(\literal C)$. 
  There must be a node $y \in X(\literal B)$ such that
  $\tau(x) = \sigma'(y)$ because $\literal C^{\tau} = \literal B^{\sigma'}$, 
  and therefore $y \in \dom(\sigma') = \dom(\mu \circ \sigma')$. In other words,
  $\tau(x) = \sigma'(y) \in \dom(\mu)$, i.e., $x \in \dom(\mu \circ \tau)$,
  which proves $\dom(\tau) \subseteq \dom(\mu \circ \tau)$.

  As a consequence, the equalities
  $\literal C^{\mu \circ \tau} = (\literal C^{\tau})^\mu = 
   (\literal B^{\sigma'})^\mu = \literal B^{\mu \circ \sigma'}$
  and $\dom(\mu \circ \tau) = \dom(\tau) \subseteq X(\literal C)$ hold. As
  $\autoconf(p', \mu \circ \sigma', \phi)$ is a valid nCFA configuration, and by 
  \defref{def:auto-conf},  
  $\autoconf(p', \mu \circ \sigma', \phi) 
   \expandstep 
   \autoconf(q, \mu \circ \tau, \phi).$
  \qed
\end{proof}

The following \lemmaref{lemma:A-to-C-nCFA} shows that each graph approved by the nCFA
is also approved by the dCFA, and \lemmaref{lemma:C-to-A-nCFA} shows
the opposite direction. But these lemmata are even more specific: by picking
the right item of each state from an approving sequence of \dCFAstate{}s,
an approving sequence of nCFA states can be found
``within'' the sequence of \dCFAstate{}s.
In other words, the relation between the two automata is similar to that
between an ordinary nondeterministic finite automaton and its powerset
automaton.

\begin{lemma}\label{lemma:A-to-C-nCFA}
  For each sequence 
  $$
    \autoconf(q_0, \iota, \emptyseq) \autostep^* \autoconf(q, \rho, \phi)
  $$ 
  of nCFA moves, there is a \concretestate\ $Q^\tau$ such that
  $\pstate(q,\rho) \in Q^\tau$ and
  $$\cfaconf(\emptyseq, \startstate^\iota) \cfastep^* \cfaconf(\phi, Q^\tau).$$
\end{lemma}

\begin{proof}
  We prove the lemma by induction on the number~$n$ of moves in 
  $\autoconf(q_0, \iota, \emptyseq) \autostep^n \autoconf(q, \rho, \phi)$. 
  For $n = 0$, it follows immediately from the definition of initial nCFA  
  configurations and~$\startstate$. 
  
  For $n > 0$ and the last move being a closure move, the considered sequence of moves is of the form
  $$
  \autoconf(q_0, \iota, \emptyseq)
  \autostep^{n-1} 
  \autoconf(\literal A \to \alpha \CDOT \literal B \beta, \kappa, \phi)
  \expandstep
  \autoconf(\literal C \to \CDOT \delta, \rho, \phi)  
  $$
  with $\literal B^\kappa = \literal C^\rho$ and
  $\dom(\rho) \subseteq X(\literal C)$. By the induction hypothesis, there
  is a \concretestate\ $Q^\tau$ such that
  $\cfaconf(\emptyseq, \startstate^\iota) \cfastep^* \cfaconf(\phi, Q^\tau)$
  and  $\pstate(\literal A \to \alpha \CDOT \literal B \beta,\kappa) \in Q^\tau$.
  Therefore, there is an injective $\eta \colon X \pto X$
  with $\kappa = \tau \circ \eta$ and 
  $\pstate(\literal A \to \alpha \CDOT \literal B \beta,\eta) \in Q$. Since
  each \dCFAstate\ is closed 
  (lines~\ref{an:start-state} and~\ref{an:closure-N}), 
  we also have $\pstate(\literal C \to \CDOT \delta, \xi) \in Q$ with 
  $\literal B^\eta = \literal C^\xi$ and $\dom(\xi) \subseteq X(\literal C)$.
  Therefore, $\literal C^\rho = \literal B^\kappa = (\literal B^\eta)^\tau =
  (\literal C^\xi)^\tau$. And because of injectivity,
  $\rho = \tau \circ \xi$, and therefore 
  $\pstate(\literal C \to \CDOT \delta, \rho) \in Q^\tau$.

  For $n > 0$ and the last move being a goto move, the considered sequence of moves is of the form
  $$
  \autoconf(q_0, \iota, \emptyseq) 
  \autostep^{n-1} 
  \autoconf(\literal A \to \alpha \CDOT \literal f \beta, \kappa, \phi)
  \consumestep
  \autoconf(\literal A \to \alpha \literal f \CDOT \beta, \rho, 
            \phi \literal f^\rho)
  $$
  with 
  \begin{equation}
    \label{eq:A-to-C-nCFA-a}
    \kappa \sqsubseteq \rho, \text{ } 
    \dom(\rho) = \dom(\kappa) \cup X(\literal f), \text{ and }
    X(\literal f^\rho) \cap X(\phi) \subseteq X(\literal f^\kappa).
  \end{equation} 
  By the induction hypothesis, there
  is a \dCFAstate\ $S$ and an \inputbinding\  
  $\chi$ such that
  $\cfaconf(\emptyseq, \startstate^\iota) \cfastep^* \cfaconf(\phi, S^\chi)$
  and  $\pstate(\literal A \to \alpha \CDOT \literal f \beta,\kappa) \in S^\chi$.
  Therefore, there is an injective $\sigma \colon X \pto X$
  with $\kappa = \chi \circ \sigma$, 
  $\pstate(\literal A \to \alpha \CDOT \literal f \beta,\sigma) \in S$, and
  $\literal f^\sigma \in \leave(S)$. (Note that the identifiers used here match 
  those in \algref{alg:deterministic}.) \algref{alg:deterministic}, therefore, 
  obtained a literal $\literal e$ from $\literal f^\sigma$ by replacing
  each occurrence of `$\unknown$' by a new node not
  used anywhere else (\alineref{an:nu}). 
  It has also obtained an injective partial function $\nu$ with 
  \begin{equation}
    \label{eq:A-to-C-nCFA-b}
    \sigma \sqsubseteq \nu, \text{ } 
    \literal f^\nu = \literal e, \text{ and }
    \dom(\nu) = \dom(\sigma) \cup X(\literal f).
  \end{equation} 
  \algref{alg:deterministic} added a transition 
  $S \atrans{(\literal e, \mu)} Q$ to $\dCFA$ 
  (\alineref{an:reuse} or~\ref{an:new-state})
  by constructing a set $\Tgt'$ of items such that 
  $\pstate(\literal A \to \alpha \literal f \CDOT \beta, \nu) \in \Tgt'$ 
  (\alineref{an:add-to-N} and~\ref{an:closure-N})
  and $Q^\mu = \Tgt'$.
  By the construction of~$\literal e$, by~\eqref{eq:A-to-C-nCFA-a} as well 
  as~\eqref{eq:A-to-C-nCFA-b}, and because $\mu$ is injective, there
  is an injective $\xi \colon X \pto X$ with 
  $\chi \sqsubseteq \xi$ and
  $\rho = \xi \circ \nu$, and therefore 
  $\pstate(\literal A \to \alpha \literal f \CDOT \beta, \rho) \in {\Tgt'}^\xi = 
  Q^{\xi \circ \mu}$.
  Since $\literal f^\rho = \literal e^\xi$, we can conclude 
  $\cfaconf(\phi, S^\chi) \cfastep 
   \cfaconf(\phi \literal f^\rho, Q^{\xi \circ \mu})$. Thus, the lemma holds
   with $\tau=\xi \circ \mu$.
  \qed
\end{proof}

\begin{lemma}\label{lemma:C-to-A-nCFA}
  For each sequence
  $$
  \cfaconf(\emptyseq,\startstate^\iota) \cfastep^* \cfaconf(\phi, Q^\tau)
  $$ 
  and each item
  $\pstate(q, \theta) \in Q^\tau$,
  there exists a sequence
  $$ 
    \autoconf(q_0, \iota, \emptyseq) \autostep^* \autoconf(q, \theta, \phi)
  $$
  of nCFA moves.
\end{lemma}

\begin{proof}
  Let  
  $\cfaconf(\emptyseq,\startstate^\iota) \cfastep^n \cfaconf(\phi, Q^\tau)$
  be any sequence of dCFA moves and $\pstate(q, \theta)$ any item 
  with $\pstate(q, \theta) \in Q^\tau$. We prove 
  $
  \autoconf(q_0, \iota, \emptyseq) \autostep^* \autoconf(q, \theta, \phi)
  $ by induction on~$n$. 
  
  For $n = 0$, the proposition follows from $Q = \startstate$, 
  $\phi = \emptyseq$,  
  $\tau = \iota$, and therefore $Q^\tau = \set{\pstate(q_0, \iota)}$, 
  i.e., $q = q_0$ and $\theta= \iota$.
  
  For $n > 0$, there is a sequence of moves
  $$
    \cfaconf(\emptyseq,\startstate^\iota) \cfastep^{n-1} \cfaconf(\phi, S^\chi) 
    \cfastep \cfaconf(\phi \literal e^\xi, Q^{\xi \circ \mu}) 
  $$
  with $\tau=\xi \circ \mu$, $\chi \sqsubseteq \xi$, 
  $\dom(\xi) = \dom(\chi) \cup X(\literal e)$ 
  and the last move using transition $S \atrans{(\literal e, \mu)} Q$. 
  \algref{alg:deterministic} added this transition to $\dCFA$ after
  computing a set $Q^\mu$. 
  (Again, the identifiers used here match those in \algref{alg:deterministic}.)
  As $\pstate(q, \theta) \in Q^{\xi \circ \mu}$, there is a $\pi$ with 
  $\theta = \xi \circ \pi$ and $\pstate(q, \pi) \in Q^\mu = \Tgt'$. Since 
  $\Tgt'$ was computed as the closure of $\Tgt$ (\alineref{an:closure-N}),
  and by \factref{fact:closure}, there is a item 
  $\pstate(\literal A \to \alpha \literal f \CDOT \beta, \nu) \in \Tgt$ 
  that was added to $\Tgt$ in \alineref{an:add-to-N}, and 
  \begin{equation}
    \pstate(\literal A \to \alpha \literal f \CDOT \beta, \nu) 
    \clps^* 
    \pstate(q, \pi).\label{eq:closure-to-q}
  \end{equation}
  In fact, $\pstate(\literal A \to \alpha \literal f \CDOT \beta, \nu)$ 
  was added to $\Tgt$ after choosing an item 
  $\pstate(\literal A \to \alpha \CDOT \literal f \beta, \sigma) \in S$, which 
  was turned into 
  $\pstate(\literal A \to \alpha \literal f \CDOT \beta, \nu)$.
  Literal $\literal e$ was obtained from $\literal f^\sigma$ by replacing
  each occurrence of `$\unknown$' by a new node not
  used anywhere else,
  and the injective partial function $\nu \colon X \pto X$
  was chosen such that $\sigma \sqsubseteq \nu$, 
  $\literal f^\nu = \literal e$,
  and $\dom(\nu) = \dom(\sigma) \cup X(\literal f)$. By the
  induction hypothesis, 
  $\autoconf(q_0, \iota, \emptyseq) \autostep^* \autoconf(q, \kappa, \phi)$ 
  for each item $\pstate(q, \kappa) \in S^\chi$, and in particular for
  $\pstate(\literal A \to \alpha \CDOT \literal f \beta, \kappa) \in S^\chi$ 
  with $\kappa = \chi \circ \sigma$. Let us define $\rho = \xi \circ \nu$, 
  and therefore 
  $\pstate(\literal A \to \alpha \literal f \CDOT \beta, \rho) \in \Tgt^\xi 
  \subseteq \Tgt'^\xi = Q^{\xi \circ \mu}$ and 
  $\literal f^\rho = \literal e^\xi$. By this 
  construction, $\kappa \sqsubseteq \rho$, 
  $\dom(\rho) = \dom(\kappa) \cup X(\literal f)$, and 
  $X(\literal f^\rho) \cap X(\phi) \subseteq X(\literal f^\kappa)$, and 
  therefore 
  $$
    \autoconf(\literal A \to \alpha \CDOT \literal f \beta, \kappa, \phi)
    \consumestep
    \autoconf(\literal A \to \alpha \literal f \CDOT \beta, \rho, 
              \phi \literal f^\rho)
    =
    \autoconf(\literal A \to \alpha \literal f \CDOT \beta, \xi \circ \nu, 
              \phi \literal e^\xi).
  $$
  By the construction of~$\xi$ and~$\nu$, we have 
  $\dom(\xi \circ \nu) = \dom(\nu)$. And, by~\eqref{eq:closure-to-q} and 
  \lemmaref{lemma:closure}, 
  $$
  \autoconf(\literal A \to \alpha \literal f \CDOT \beta, \xi \circ \nu, 
              \phi \literal e^\xi)
  \expandstep^*
  \autoconf(q, \xi \circ \pi, \phi \literal e^\xi) =
  \autoconf(q, \theta, \phi \literal e^\xi),
  $$
  which completes the proof of the lemma.
  \qed
\end{proof}

An immediate consequence of these lemmata is the following:

\begin{theorem}\label{thm:dCFA-prefix}
  A graph is approved by the dCFA if and only if it is approved by the nCFA if and only if it is a viable prefix.
\end{theorem}

\thmref{thm:dCFA-prefix} implies that the naïve parser can
reach the accepting configuration with some appropriate remaining input if and only if
the dCFA approves its current stack. 


\def\SeqText{\mathit{Seq}} 
\def\StmtText{\mathit{Stmt}}
\def\beginText{\mathit{begin}} 
\def\endText{\mathit{end}}
\def\predText{\mathit{pred}} 
\def\actionText{\mathit{act}}
\def\S(){\startSym()} \def\Seq(#1,#2){\SeqText(#1, #2)}
\def\Stmt(#1,#2){\StmtText(#1, #2)} 
\def\beg(#1){\beginText(#1)}
\def\stop(#1){\endText(#1)} 
\def\action(#1,#2){\actionText(#1, #2)}
\def\pred(#1,#2,#3){\predText(#1, #2, #3)} 
Before we describe assisted shift-reduce parsers using the deterministic CFA 
(\sectref{s:ASR}), let us observe that 
\algref{alg:deterministic} may fail to terminate for some
HR grammars. An HR grammar for the visual language of \emph{structured flowcharts} is used to demonstrate this. 

\begin{figure}[tb]
  \begin{minipage}[t]{0.55\textwidth}
  \centering
  \begin{tikzpicture}[->, font=\footnotesize, x=8mm, y=7mm]
    \tikzset{
      base/.style={draw, align=center},
      proc/.style={base, rectangle},
      test/.style={base, diamond, aspect=2},
      term/.style={proc, ellipse}
    }
    \node [term] (begin)  at (2,5)  {begin};
    \node [proc] (p1) at (2,3.8) {read $n$};
    \node [test] (t1) at (2,2) {$n \ne 0$?};
    \node at (0.5,2.2) {yes};
    \node at (3.4,2.2) {no};
    \node [term] (p2) at (4,1) {end};
    \node [test] (t2) at (0,0.5) {$n$ even?};
    \node at (-1.5,0.7) {yes};
    \node at (1.5,0.7) {no};
    \node [proc] (p3) at (-1.8,-0.5) {$n \leftarrow n/2$};
    \node [proc] (p4) at (1.8,-0.5) {$n \leftarrow 3n+1$};
    \draw (begin) -- (p1);
    \draw (p1) -- (t1);
    \draw (t1.west) -| (t2.north);
    \draw (t1.east) -| (p2.north);
    \draw (t2.west) -| (p3.north);
    \draw (t2.east) -| (p4.north);
    \draw (p4.south) -- ++(0,-0.5) -- ++(-4.8,0) |- (2,3.1);
    \draw (p3.south) -- ++(0,-0.5);
  \end{tikzpicture}
  \caption{A structured flowchart.}
  \label{fig:fc}
  \end{minipage}
  \hfill
  \begin{minipage}[t]{0.4\textwidth}
    \centering
    \begin{tikzpicture}[font=\footnotesize, auto, 
                        node distance=3mm, label distance=-0.5mm]
      \tikzset{
        edge/.style={draw, align=center, rectangle, fill=gray!30},
        node/.style={circle,draw,inner sep=0.0pt,minimum size=2mm}
      }
      \node [edge] (begin)  {$\beginText$};
      \node [node, below=of begin, label=left:$a$] (a) {};
      \node [edge, below=of a] (act1)   {$\actionText$};
      \node [node, below=of act1, label={[label distance=-1mm]175:$b$}] (b) {};
      \node [edge, below=of b] (pred1)  {$\predText$};
      \node [node, left=of pred1, label=left:$c$] (c) {};
      \node [node, right=of pred1, label=right:$d$] (d) {};
      \node [edge, below=of d] (end)    {$\endText$};
      \node [edge, below=of c] (pred2)  {$\predText$};
      \node [node, left=of pred2, label={[label distance=-1mm]180:$e$}] (e) {};
      \node [edge, below=of e] (act2)   {$\actionText$};
      \node [node, right=of pred2, label={[label distance=-1mm]0:$f$}] (f) {};
      \node [edge, below=of f] (act3)   {$\actionText$};
      \draw (begin) -- (a) -- (act1) -- (b) -- (pred1) -- (d) -- (end);
      \draw (pred1) -- (c) -- (pred2) -- (e) -- (act2);
      \draw (pred2) -- (f) -- (act3);
      \draw[rounded corners=2pt] (act2.south) -- ++(0,-3mm) -- ++(-7mm,0) |- (b);
      \draw[rounded corners=2pt] (act3.south) -- ++(0,-3mm) -- ++(16mm,0) |- (b);
    \end{tikzpicture}
    \caption{Graph representation of the structured flowchart in 
      \figref{fig:fc}.}
    \label{fig:fc-graph}
    \end{minipage}  
\end{figure}
\begin{example}[Structured Flowcharts\label{x:flowcharts}]
Structured flowcharts consist of rectangles containing actions,
diamonds that indicate conditions, and ovals indicating begin and end
of the program. Arrows indicate control flow; see \figref{fig:fc} for an
example.
\def\su#1{[#1]}%
\def\suu(#1,#2){#1/#2}%
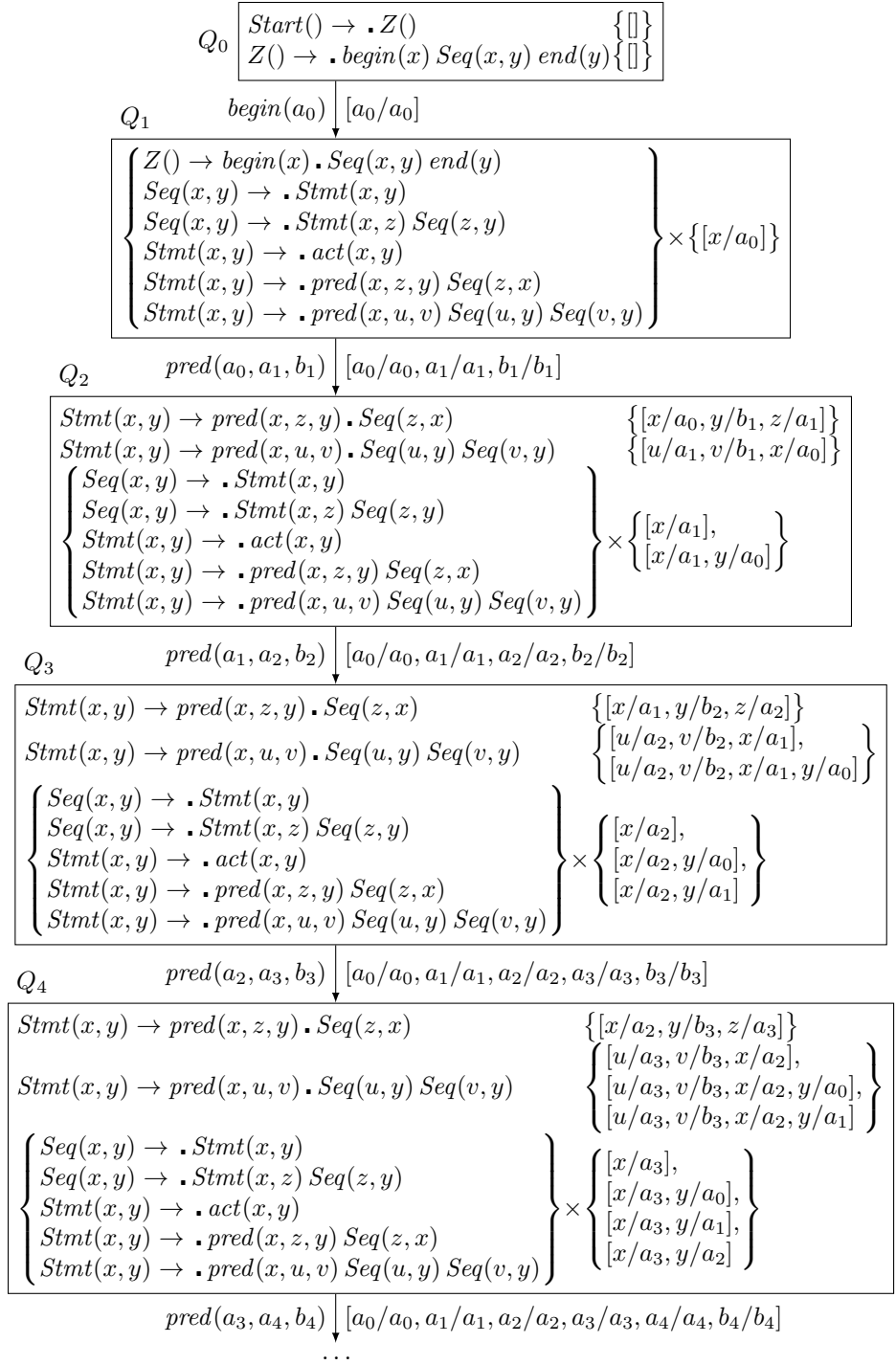
\begin{figure}[tbhp]
\def\substs#1{\left\{\begin{array}{@{}l@{}}  #1 \end{array}\right\}}%
\def\psaa{\S() \to \CDOT \beg(x) \, \Seq(x,y) \, \stop(y)}%
\def\psab{\S() \to \beg(x) \CDOT \Seq(x,y) \, \stop(y)}%
\def\psba{\Seq(x,y) \to \CDOT \Stmt(x,y)}%
\def\psca{\Seq(x,y) \to \CDOT \Stmt(x,z) \, \Seq(z,y)}%
\def\psda{\Stmt(x,y) \to \CDOT \action(x,y)}%
\def\psea{\Stmt(x,y) \to \CDOT \pred(x,u,v) \, \Seq(u,y) \, \Seq(v,y)}%
\def\pseb{\Stmt(x,y) \to \pred(x,u,v) \CDOT \Seq(u,y) \, \Seq(v,y)}%
\def\psfa{\Stmt(x,y) \to \CDOT \pred(x,z,y) \, \Seq(z,x)}%
\def\psfb{\Stmt(x,y) \to \pred(x,z,y) \CDOT \Seq(z,x)}%
\def\PSTATE(#1,#2){#1 && \substs{#2}\\}%
\def\PSTATES(#1,#2){\substs{#1}&\times&\substs{#2}\\}%
\centering
\begin{tikzpicture}[auto, ->]
  \node [draw, label=left:$\startstate$] (s0) {
    $\begin{array}{@{}l@{}c@{}l@{}}
      \PSTATE(\pseudoSym()\to\CDOT\S(),\su{})
      \PSTATE(\psaa,\su{})
    \end{array}$
  };


  \node [draw, below=8mm of s0] (s1) {
    $\begin{array}{@{}l@{}c@{}l@{}}
      \PSTATES(\psab\\\psba\\\psca\\\psda\\\psfa\\\psea,
               \su{\suu(x,a_0)})
    \end{array}$
  };

  \node [draw, below=8mm of s1] (s2) {
    $\begin{array}{@{}l@{}c@{}l@{}}
      \PSTATE(\psfb,
              \su{\suu(x,a_0), \suu(y,b_1), \suu(z,a_1)})
      \PSTATE(\pseb,
              \su{\suu(u,a_1), \suu(v,b_1), \suu(x,a_0)})
      \PSTATES(\psba\\\psca\\\psda\\\psfa\\\psea,
               \su{\suu(x,a_1)}, \\
               \su{\suu(x,a_1), \suu(y,a_0)})
    \end{array}$
  };

  \node [draw, below=8mm of s2] (s3) {
    $\begin{array}{@{}l@{}c@{}l@{}}
      \PSTATE(\psfb,
              \su{\suu(x,a_1), \suu(y,b_2), \suu(z,a_2)})
      \PSTATE(\pseb,
              \su{\suu(u,a_2), \suu(v,b_2), \suu(x,a_1)},\\ 
              \su{\suu(u,a_2), \suu(v,b_2), \suu(x,a_1), \suu(y,a_0)})
      \PSTATES(\psba\\\psca\\\psda\\\psfa\\\psea,
               \su{\suu(x,a_2)}, \\
               \su{\suu(x,a_2), \suu(y,a_0)}, \\
               \su{\suu(x,a_2), \suu(y,a_1)})
    \end{array}$
  };

  \node [draw, below=8mm of s3] (s4) {
    $\begin{array}{@{}l@{}c@{}l@{}}
      \PSTATE(\psfb,
              \su{\suu(x,a_2), \suu(y,b_3), \suu(z,a_3)})
      \PSTATE(\pseb,
              \su{\suu(u,a_3), \suu(v,b_3), \suu(x,a_2)},\\ 
              \su{\suu(u,a_3), \suu(v,b_3), \suu(x,a_2), \suu(y,a_0)}, \\
              \su{\suu(u,a_3), \suu(v,b_3), \suu(x,a_2), \suu(y,a_1)})
      \PSTATES(\psba\\\psca\\\psda\\\psfa\\\psea,
               \su{\suu(x,a_3)}, \\
               \su{\suu(x,a_3), \suu(y,a_0)}, \\
               \su{\suu(x,a_3), \suu(y,a_1)},\\ 
               \su{\suu(x,a_3), \suu(y,a_2)})
    \end{array}$
  };
  \node [above=0mm of s1.north west, anchor=south west] {$Q_1$}; 
  \node [above=0mm of s2.north west, anchor=south west] {$Q_2$}; 
  \node [above=0mm of s3.north west, anchor=south west] {$Q_3$}; 
  \node [above=0mm of s4.north west, anchor=south west] {$Q_4$}; 

  \path[draw]
    ($(s0.south)+(-2,0)$) -- ($(s1.north)+(-2,0)$) 
              node[pos=0.5,left] {$\beg(a_0)$} 
              node[pos=0.5,right] {$\su{\suu(a_0,a_0)}$};
  \path[draw]
    ($(s1.south)+(-2,0)$) -- ($(s2.north)+(-2,0)$) 
              node[pos=0.5,left] {$\pred(a_0,a_1,b_1)$}  
              node[pos=0.5,right] {$\su{\suu(a_0,a_0) 
                               , \suu(a_1,a_1), \suu(b_1,b_1)}$}; 
  \path[draw]
    ($(s2.south)+(-2,0)$) -- ($(s3.north)+(-2,0)$) 
              node[pos=0.5,left] {$\pred(a_1,a_2,b_2)$}  
              node[pos=0.5,right] {$\su{\suu(a_0,a_0) 
                               , \suu(a_1,a_1) 
                               , \suu(a_2,a_2), \suu(b_2,b_2)}$}; 
  \path[draw] 
    ($(s3.south)+(-2,0)$) -- ($(s4.north)+(-2,0)$)
              node[pos=0.5,left] {$\pred(a_2,a_3,b_3)$}  
              node[pos=0.5,right] {$\su{\suu(a_0,a_0) 
             , \suu(a_1,a_1) 
             , \suu(a_2,a_2) 
             , \suu(a_3,a_3), \suu(b_3,b_3)}$}; 
  \path[draw]
    ($(s4.south)+(-2,0)$) -- ($(s4.south)+(-2,-0.85)$)
              node[pos=0.5,left] {$\pred(a_3,a_4,b_4)$}  
              node[pos=0.5,right] {$\su{\suu(a_0,a_0) 
             , \suu(a_1,a_1) 
             , \suu(a_2,a_2) 
             , \suu(a_3,a_3) 
             , \suu(a_4,a_4), \suu(b_4,b_4)}$}
             node[pos=1.3,left=-4mm] {\ldots};

\end{tikzpicture}
\caption{Excerpt of the infinite dCFA of the flowchart grammar.}
\label{fig:fc-dCFA}
\end{figure}
Such flowcharts can be represented by graphs using terminal
symbols $\beginText, \endText, \actionText$, and $\predText$ where
binary $\actionText$ edges represent actions (rectangles) and ternary
$\predText$ edges conditions (diamonds). Nodes correspond to arrows
where edges are attached to the same node if the corresponding
components (rectangle, diamond, or oval) are connected by an arrow. The
example flowchart in \figref{fig:fc} can be represented by the graph
$$
  \beg(a) \, \action(a,b) \, \pred(b,c,d) \, \pred(c,e,f) \,
  \action(e,b) \, \action(f,b) \, \stop(d)\, ,
$$
which can be drawn as the diagram shown in \figref{fig:fc-graph}.
For instance, literal $\action(a,b)$ represents the rectangle ``read
$n$'', $\action(f,b)$ the rectangle ``$n \leftarrow 3n+1$'', and
$\pred(c,e,f)$ the diamond ``$n$ even?''. An HR grammar for the graph
representation of structured flowcharts has four nonterminal symbols $\pseudoSym$, $\startSym$,
$\SeqText$ (for ``sequence''), $\StmtText$ (for
``statement'') and the following rules:
$$
\begin{array}{r@{\;}r@{\;}l@{\qquad}l}
  \pseudoSym()       & \to & \S() \\
  \S()       & \to & \beg(x) \, \Seq(x,y) \, \stop(y) \\
\Seq(x,y)    & \to & \Stmt(x,z) \, \Seq(z,y) & \text{(sequence)}\\
\Seq(x,y)    & \to & \Stmt(x,y) & \text{(end of sequence)}\\
\Stmt(x,y) & \to & \action(x,y) & \text{(single action)}\\
\Stmt(x,y) & \to & \pred(x,z,y) \, \Seq(z,x) & \text{(while loop)}\\
\Stmt(x,y) & \to & \pred(x,u,v) \, \Seq(u,y) \, \Seq(v,y) & \text{(selection)}\\
\end{array}
$$

The dCFA of this grammar is infinite, i.e.,
\algref{alg:deterministic} does not terminate. To see this, consider
the excerpt of the dCFA in \figref{fig:fc-dCFA}. To save
space, we use a compact notation. If a \dCFAstate\ contains items
$\pstate(q,\sigma_1), \ldots, \pstate(q, \sigma_n)$ that share the nCFA
state~$q$, we write $q \set{\sigma_1, \ldots, \sigma_n}$ where the
parameter mappings $\sigma_1, \ldots, \sigma_n$ are denoted as
introduced earlier. And, if a \dCFAstate\ contains the full Cartesian product of
a set~$Q$ of nCFA states and a set~$P$ of parameter mappings, we write
$Q \times P$.
This notation even allows to represent~$Q_4$ with its 24~items.

\figref{fig:fc-dCFA} shows the states $Q_0, \ldots, Q_4$ of the dCFA
and only the transitions between them. One can see that $Q_2$,
$Q_3$, and $Q_4$ are identical when ignoring the parameter mappings.
Moreover, when renaming parameters, $Q_2$ is properly contained in $Q_3$, which in turn
is properly contained in $Q_4$. These three states are the
first states of an infinite sequence $Q_2, Q_3, Q_4, \ldots$, which
makes the entire dCFA infinite.

As a consequence, \algref{alg:deterministic} may fail to terminate. However, a modified algorithm not described here can
recognize and handle this situation. For this, it represents the infinite dCFA
in a finite way by equipping states with variables that may contain
sets of arbitrarily many parameters and using these variables in
transitions. This algorithm has been implemented in the \emph{Grappa}
tool. In this paper, we have described the
simpler algorithm in \algref{alg:deterministic} instead of the more
general one, because the latter is rather technical. In fact, the
HR grammar for structured flowcharts is the only HR grammar
with an infinite dCFA known to us---and it is not even \PSR
parsable because the finitely represented infinite dCFA contains a state with 
conflicts (see \tabref{t:testresults}). We describe the concept of conflicts in \sectref{s:PSR}.
\end{example}




\section{CFA-Assisted Shift-Reduce Parsing}\label{s:ASR}

Let us now discuss how the naïve shift-reduce parser discussed in 
\sectref{s:naive} can read off all permissible moves
from the current \dCFAstate\ in order to reach the accepting configuration with 
some rest graph, as we shall henceforth call the remaining input.
Recall that the naïve parser just maintains a stack of literals.
The extended parser, instead, 
maintains a stack as an alternating sequence of \concretestate{}s and literals
and makes sure that its stack (when ignoring the \concretestate{}s
on the stack) is always approved by the dCFA. The top stack element is
always the \emph{current state}, which is the uniquely determined \concretestate\ reached when approving the stack. The stack prior to a move is
called \emph{current stack}, and the next one is the \emph{successor} stack,
thus defining a \emph{successor
state}. The successor stack together with the successor state
will then be the current stack and the current state,
respectively, at the next move.
 
When performing a shift move,
the parser selects a literal from the remaining input that matches the
label of a transition leaving the current state (on top of the stack).
This literal is then pushed onto the stack, together with the successor
state reachable by this transition. The successor stack thus
consists of the current stack, followed by the shifted literal
and the successor state.

A reduce move removes the (bound) right-hand side
of a rule from the stack, together with the corresponding \concretestate{}s,
yielding some intermediate stack with a \concretestate\ on top. The parser
then selects a transition which leaves the corresponding \dCFAstate\ and
carries a label matching
the reduced nonterminal literal under the same \inputbinding\ as before. Next, the
literal is pushed onto the stack, together with the successor state
reachable by this transition. The successor stack thus consists
of the intermediate stack, followed by the reduced nonterminal literal and
the successor state.




The parser \emph{accepts} the input graph processed so far when 
the \concretestate\ on top of the stack is
$\acceptstate=\set{\pstate(\pseudoLit \to \startLit \CDOT, \iota)}$,
which is called the \emph{accepting state}.
%
Note that the entire input graph is accepted that way if there are 
no unprocessed input literals left when reaching this \concretestate, i.e., the
rest graph is empty.

Let us now define the extension of the naïve shift-reduce parser more 
precisely. We call this parser
\emph{dCFA-assisted shift-reduce parser} or simply
\emph{assisted shift-reduce parser}, abbreviated as \emph{ASR parser}.

\begin{defn}[ASR Parser]\label{def:CFA-parser}
  Let $\mappedStates$ denote the set of all (concrete) states. An \emph{(ASR parser) configuration} $\buc(\psrstack S, g)$ consists of a 
  \emph{parse stack} 
  $\psrstack S 
   \subseteq 
   \mappedStates \cdot (\Lit\Lab \cdot \mappedStates)^*$ 
  and a graph $g \in \Gr \TLab$.

  Let $\psrtop{\psrstack S}$ denote the rightmost element of a parse
  stack, which is always a \concretestate.
  The graph obtained by
  removing all \concretestate{}s from $\psrstack S$ is denoted by
  $\psrgraph{\psrstack S}$. A configuration $\buc(\psrstack S, g)$ is
  \emph{accepting} if $\psrtop{\psrstack S}$ is the accepting state
  $\acceptstate$ of the dCFA.

  An \emph{ASR move} turns $\buc(\psrstack S, g)$ into $\buc(\psrstack S', g')$ 
  and is either an ASR shift move or an ASR reduce move, defined as follows.

  Let $Q^\tau=\psrtop{\psrstack S}$ for a \concretestate\ $Q\in\mathcal Q$ and an \inputbinding~$\tau\colon\params(Q)\to X$.
  \begin{itemize}
  \item Suppose that there is a literal $\literal e \in \Lit\TLab$ and a
  \concretestate\ $\bar Q $ such that
  $\cfaconf(\psrgraph{\psrstack S},Q^\tau) \lcfastep{\psrtr }  \cfaconf(\psrgraph{\psrstack S} \literal e, \bar Q)$ and
  $X(\literal e) \cap X(g) \subseteq X(\psrgraph{\psrstack S})$. Then there is an \emph{ASR shift move}
  $\buc(\psrstack S, g) \lpstep{\psrtr} \buc(\psrstack S \literal e \bar Q, g \literal e)$.
  
  \item Suppose that $Q$ contains an item $\psrit = \pstate(\literal A \to \rho \CDOT, \sigma)$ and
  one can remove $2 \cdot |\rho|$ elements from the top of $\psrstack S$ 
  to obtain a parse stack $\psrstack S''$ with $R = \psrtop{\psrstack S''}$ such that
  there exists a \concretestate\ $\bar Q$ with
  $\cfaconf(\psrgraph{\psrstack S''},R) \cfastep \cfaconf(\psrgraph{\psrstack S''} \literal A^{\tau \circ \sigma}, \bar Q)$.
  Then there is an \emph{ASR reduce move}  $\buc(\psrstack S, g) \lpstep{\psrit} \buc(\psrstack S'' \literal A^{\tau \circ \sigma} \bar Q,g)$.
  \end{itemize}
  We may write $\buc(\psrstack S, g) \pstep \buc(\psrstack S', g')$ if 
  $\buc(\psrstack S, g) \lpstep{\psrtr} \buc(\psrstack S', g')$ for a transition $\psrtr$ or
  $\buc(\psrstack S, g) \lpstep{\psrit} \buc(\psrstack S', g')$ for an item $\psrit$.
 
  A configuration $\buc(\psrstack S, g)$ can be \emph{reached} if 
  $
  \buc(\startstate^\iota, \emptyseq)
  \pstep^*
  \buc(\psrstack S, g).
  $
  An ASR parser \emph{accepts} a graph $g \in \Gr\TLab$ if it can reach an 
  accepting configuration $\buc(\psrstack S, g')$ for a graph 
  $g' \perm g$.
\end{defn}

Note that shift and reduce moves of the ASR parser always push (concrete) \concretestate{}s onto the
stack that are reachable from their immediate predecessor \concretestate{}s on the
stack. This is expressed in the following fact:

\begin{fact}\label{fact:parse-stack}
  $
  \buc(\startstate^\iota, \emptyseq)
  \pstep^*
  \buc(\psrstack S, g)
  $
  implies 
  $
  \cfaconf(\emptyseq,\startstate^\iota) 
  \cfastep^* 
  \cfaconf(\psrgraph{\psrstack S'},\psrtop{\psrstack S'})
  $
  for every ASR parser configuration $\buc(\psrstack S, g)$ and
  every parse stack $\psrstack S'$ being a prefix of~$\psrstack S$.
\end{fact}


\begin{figure}[t]
\small
\[ \renewcommand{\arraystretch}{1.1}
   \begin{array}{lr@{\,}c@{\,}ll}
           & \text{stack}
                & \CDOT & \text{read input} &  \text{match }
     \\
  \hline 
          & Q_0 
             & \CDOT & \emptyseq & \\
  \pstep & Q_0 \startEdge 1 Q_1^1
            \underline{~}
             & \CDOT & \startEdge 1 &   \\
  \pstep[3] & Q_0 \startEdge 1 Q_1^1 \nontT 1 Q_2^1 
             & \CDOT & \startEdge 1  & y/1 \\
  \pstep & Q_0 \startEdge 1 Q_1^1\nontT 1 Q_2^1 \edge 12 Q_3^{12}  
            \underline{~}
             & \CDOT & \startEdge 1 \edge 12  &  \\
  \pstep[3] & Q_0 \startEdge 1 Q_1^1\nontT 1 Q_2^1 \edge 12
            Q_3^{12} \nontT 2 Q_4^{12}   
             & \CDOT & \startEdge 1 \edge 12  & y/2 \\
  \pstep & Q_0 \startEdge 1 Q_1^1\nontT 1 Q_2^1 \edge 12
            Q_3^{12} \nontT 2 Q_4^{12} \edge 24 Q_3^{24} \underline{~}
             & \CDOT & \startEdge 1 \edge 12 \edge 24 & x/2, y/4 \\
  \pstep[3] & Q_0 \startEdge 1 Q_1^1\nontT 1 Q_2^1 \edge 12
            Q_3^{12} \underline{\nontT 2 Q_4^{12} \edge 24
                                        Q_3^{24} \nontT 4 Q_4^{24} }
             & \CDOT & \startEdge 1 \edge 12 \edge 24 & y/4 \\
  \pstep[2] & Q_0 \startEdge 1 Q_1^1
            \underline{\nontT 1 Q_2^1 \edge 12 Q_3^{12} \nontT 2 Q_4^{12}}   
             & \CDOT & \startEdge 1 \edge 12 \edge 24 & x/2, y/4 \\
  \pstep[2] & Q_0 \startEdge 1 Q_1^1\nontT 1 Q_2^1    
             & \CDOT & \startEdge 1 \edge 12 \edge 24 & x/1, y/2 \\
  \pstep & Q_0 \startEdge 1 Q_1^1\nontT 1 Q_2^1 \edge 13 Q_3^{13}
            \underline{~}
             & \CDOT & \startEdge 1 \edge 12 \edge 24 \edge 13
                                                            &  \\
  \pstep[3] & Q_0 \startEdge 1 Q_1^1
             \underline{\nontT 1 Q_2^1 \edge 13 Q_3^{13}\nontT 3 Q_4^{13}}
             & \CDOT & \startEdge 1 \edge 12 \edge 24 \edge 13 
                                                            & y/3 \\
  \pstep[2] & Q_0 \underline{\startEdge 1 Q_1^1 \nontT 1 Q_2^1}
             & \CDOT & \startEdge 1 \edge 12 \edge 24 \edge 13
                                                            & x/1, y/3 \\
  \pstep[1] & Q_0 \, \startSym \, \acceptstate
             & \CDOT & \startEdge 1 \edge 12 \edge 24 \edge 13
                                                            & 
\end{array} \]
\caption{Moves of the ASR parser recognizing the tree in
  \exref{x:tree:HR}. 
  Places on the stack
  where reductions occur are underlined. 
  Rules used in reduce moves are indicated as subscripts in the
  leftmost column, and their corresponding matches appear in the
  rightmost column.}
  \label{f:psrmoves}
\end{figure}

\begin{example}[An ASR Parse of a Tree]\label{x:tree-psr-parser}
  \figref{f:psrmoves} shows the moves of the ASR parser when
  recognizing the tree in \exref{x:tree:HR}.%
  \footnote{In this example and in \figref{f:psrmoves}, we use
    following abbreviated notation: literals $\ell(x_1, \dots,x_k)$ are
    denoted as $\ell^{x_1 \dots x_k}$, and (concrete) states of the dCFA that
    were written as $Q_i^{[a_1/y_1,\dots,a_k/y_k]}$ in
    \figref{fig:tree-approval} are abbreviated as
    $Q_i^{y_1 \dots y_k}$. } %
  Moves~1--6 in \figref{f:psrmoves} correspond to the moves
  of the dCFA shown in \figref{fig:tree-approval} of
  \exref{x:tree-approval} in the way stated in \factref{fact:parse-stack}:
  \begin{itemize}
  \item The initial configuration of the ASR parser agrees with the
    initial state of the dCFA.
  \item The literal and state pushed in move~$i$ agrees with the literal
    processed, and the state reached, by move~$i$ of the dCFA.  In
    three steps, terminal literals are pushed by shift moves; the
    other moves push nonterminal literals with label $\nontA$ in the course of
    reducing rule~2 or rule~3. The former has a right-hand side of length three; six elements are thus popped off the stack, whereas the latter has no literals on its right-hand side so
    that nothing has to be popped off the stack.
  \item After move~$i$, the symbols on the stack of the parser
    (ignoring the states) agree with the viable prefix approved in
    move~$i$ of the dCFA.  \qed
  \end{itemize}
\end{example}

Note also that a reduce move of the naïve shift-reduce parser must check a rather
complex condition in order to select a reduce move
(\defref{def:sr-steps}); it must examine whether the stack contains the
right hand side of the rule (under an appropriate match), and it must
additionally check condition $X(\alpha) \cap X(\rho^\mu) \subseteq X(\literal A^\mu)$
of \defref{def:sr-steps} to make sure that the corresponding derivation step is valid.
The ASR parser, instead, just inspects the top
\concretestate\ on the stack and checks whether this \concretestate\ contains an item with
the dot at the end of the rule; it can thus read off from the dCFA
whether it can select a reduce move. The following lemma states this
formally. It
will be used for proving the correctness of the ASR parser later.

\begin{lemma}\label{lemma:reduce-correct}
  For every rule $\literal A \to \rho$ with $\literal A \neq \pseudoLit$ and 
  every sequence 
  $\cfaconf(\emptyseq, \startstate^\iota)
  \cfastep^* 
  \cfaconf(\phi, Q^\tau)$,
  $Q$~contains an item $\pstate(\literal A \to \rho \CDOT, \sigma)$ 
  if and only if 
  there is a graph $\alpha \in \Gr\Lab$ such that 
  $\phi = \alpha \rho^{\tau \circ \sigma}$, 
  $\alpha \literal A^{\tau \circ \sigma}$ is also approved by the dCFA, and
  $X(\alpha) \cap X(\rho^{\tau \circ \sigma}) \subseteq 
   X(\literal A^{\tau \circ \sigma})$.
\end{lemma}

\begin{proof}
  For the \emph{only-if} direction, consider a sequence 
  $\cfaconf(\emptyseq, \startstate^\iota)
  \cfastep^* 
  \cfaconf(\phi, Q^\tau)$ and an item 
  $\pstate(\literal A \to \rho \CDOT, \sigma) \in Q$. 
  We have $\pstate(\literal A \to \rho \CDOT, \mu) \in Q^\tau$ with
  $\mu = \tau \circ \sigma$, and by \lemmaref{lemma:C-to-A-nCFA}, 
  there is a sequence 
  \begin{equation}\label{eq:reduce-seq}
    \autoconf(q_0, \iota, \emptyseq)
    \autostep^*
    \autoconf(\literal A \to \rho \CDOT, \mu, \phi).
  \end{equation}
  The dot in $\literal A \to \rho \CDOT$ must have been moved 
  there by goto moves, starting at $\literal A \to \CDOT \rho$, an nCFA state
  that was reached by a closure move.  Therefore, 
  \eqref{eq:reduce-seq} reads
  $$
    \autoconf(q_0, \iota, \emptyseq)
    \autostep^*
    \autoconf(\literal B \to \gamma \CDOT \literal C \delta, \nu, \alpha)
    \expandstep
    \autoconf(\literal A \to \CDOT \rho, \mu', \alpha)
    \consumestep^*
    \autoconf(\literal A \to \rho \CDOT, \mu, \alpha \rho^\mu)
  $$
  with $\phi = \alpha \rho^\mu = \alpha \rho^{\tau \circ \sigma}$,  
  $\literal C^\nu = \literal A^{\mu'}$, $\mu' \sqsubseteq \mu$, 
  $X(\literal A) \subseteq X(\rho) = \dom(\mu)$.
  Thus $\literal A^\mu$ is a literal, and there is an injective 
  $\nu' \colon X \pto X$ with 
  $\nu \sqsubseteq \nu'$, $\dom(\nu') = \dom(\nu) \cup X(\literal C)$,
  and $\literal C^{\nu'} = \literal A^\mu$. As a consequence, there is also a 
  sequence
  $$
    \autoconf(q_0, \iota, \emptyseq)
    \autostep^*
    \autoconf(\literal B \to \gamma \CDOT \literal C \delta, \nu, \alpha)
    \consumestep
    \autoconf(\literal B \to \gamma \literal C \CDOT \delta, \nu', 
              \alpha \literal A^\mu)
  $$
  showing that $\alpha \literal A^\mu$ is approved by the nCFA and consequently, using
  \lemmaref{lemma:A-to-C-nCFA}, by the dCFA.
  Moreover, 
  $X(\alpha) \cap X(\rho^\mu) \subseteq X(\rho^{\mu'}) = X(\literal A^{\mu'})
   \subseteq X(\literal A^\mu)$ using \lemmaref{lemma:n-gotos}.

  For the \emph{if} direction, let $\literal A \to \rho$ be a rule with 
  $\literal A \neq \pseudoLit$,  and consider a sequence
  $\cfaconf(\emptyseq, \startstate^\iota)
  \cfastep^* 
  \cfaconf(\alpha \rho^\mu, Q^\tau)$ with an injective 
  $\mu \colon X \pto X$, and 
  $
    \cfaconf(\emptyseq, \startstate^\iota)
    \cfastep^* 
    \cfaconf(\alpha \literal A^\mu, {\widehat Q}^\xi)
  $
  for some \dCFAstate\ ${\widehat Q}$ and injective
  $\xi \colon X \pto X$. By \lemmaref{lemma:C-to-A-nCFA},
  $$
    \autoconf(q_0, \iota, \emptyseq)
    \autostep^*
    \autoconf(\literal B \to \gamma \CDOT \literal C \delta, \nu', 
              \alpha)
    \consumestep
    \autoconf(\literal B \to \gamma \literal C \CDOT \delta, \nu, 
              \alpha \literal A^\mu)
  $$
  with $\literal A^\mu = \literal C^{\nu}$, $\nu' \sqsubseteq \nu$, and
  $\dom(\nu) = \dom(\nu') \cup X(\literal C)$. Therefore, we also have
  $$
    \autoconf(q_0, \iota, \emptyseq)
    \autostep^*
    \autoconf(\literal B \to \gamma \CDOT \literal C \delta, \nu', \alpha)
    \expandstep
    \autoconf(\literal A \to \CDOT \rho, \mu', \alpha)
    \consumestep^*
    \autoconf(\literal A \to \rho \CDOT, \mu, \alpha \rho^\mu)
  $$  
  with $\literal C^{\nu'} = \literal A^{\mu'}$ and, by 
  \lemmaref{lemma:A-to-C-nCFA},
  $\cfaconf(\emptyseq, \startstate^\iota)
  \cfastep^* 
  \cfaconf(\alpha \rho^\mu, {Q'}^{\tau'})$ 
  with a \dCFAstate\ $Q'$, injective $\tau' \colon X \pto X$, and 
  $\pstate(\literal A \to \rho \CDOT, \mu) \in {Q'}^{\tau'}$. In fact,
  $Q = Q'$ and $\tau = \tau'$ since the dCFA is deterministic. 
  Hence, $Q$ contains an item $\pstate(\literal A \to \rho \CDOT, \sigma)$
  with $\mu = \tau \circ \sigma$.
  \qed
\end{proof}

We are now ready to prove that the ASR parser is in fact an improved
version of the naïve shift-reduce
parser (\defref{def:sr-steps}) that always makes sure that its stack is a viable prefix:

\begin{lemma}\label{lemma:ASR-naive}
  For every ASR parser configuration $\buc(\psrstack S, g)$ with 
  $\psrgraph{\psrstack S} \neq \pseudoLit$ 
  and every $n \in \Nat$, 
  $
  \buc(\startstate^\iota, \emptyseq)
  \pstep^n
  \buc(\psrstack S, g)
  $
  if and only if
  $\cfaconf(\emptyseq,\startstate^\iota) \cfastep^* \cfaconf(\phi,R)$ and
  $
  \buc(\emptyseq, \emptyseq) \cstep^n \buc(\phi, g)
  $
  where $R = \psrtop{\psrstack S}$ and $\phi = \psrgraph{\psrstack S}$.
\end{lemma}

\begin{proof}
  We prove the lemma by induction on $n$.
  For $n=0$, it immediately follows
  from the fact that $\psrstack S = \psrtop{\psrstack S} = \startstate^\iota$ 
  and $\psrgraph{\psrstack S} = g = \emptyseq$.

  For the inductive step, let $n\ge 0$. We show that the statement holds
  for $n+1$ under the assumption that it holds for all shorter configuration sequences
  of length up to $n$.
  We show the \emph{only-if} and the \emph{if} direction separately:
  

  \begin{enumerate}[(1)]
  \item To show the \emph{only-if} direction, we assume any sequence
  $$
    \buc(\startstate^\iota, \emptyseq)
    \pstep^n
    \buc(\psrstack S', g')
    \pstep
    \buc(\psrstack S, g).
  $$
  Let $R = \psrtop{\psrstack S'}$ and $\phi = \psrgraph{\psrstack S'}$. 
  The last move is either a shift move or a reduce move. 
  \begin{enumerate}[({1}a)]
  \item If it is a shift 
    move, there exist a literal $\literal e \in \Lit\TLab$ and a
    \concretestate\ $T$ with
    \begin{align}
      \cfaconf(\phi,R) &
        \cfastep \cfaconf(\phi \literal e, T)
        \label{eq:p535-1}\\
      X(\literal e) \cap X(g) &\subseteq X(\phi)
        \label{eq:p535-2}\\
      \psrstack S &= \psrstack S' \literal e T
        \label{eq:p535-3}\\
      g &= g' \literal e.
        \label{eq:p535-4}
    \end{align}
    Now,
    $
      \cfaconf(\emptyseq,\startstate^\iota) 
      \cfastep^* 
      \cfaconf(\phi,R)
      \cfastep
      \cfaconf(\phi \literal e, T)
    $
    follows from \eqref{eq:p535-1} and the induction hypothesis, and  
    $
    \buc(\emptyseq, \emptyseq) 
    \cstep^n 
    \buc(\phi, g')
    \shift
    \buc(\phi \literal e, g)
    $
   from the induction hypothesis, \eqref{eq:p535-2}, 
    \eqref{eq:p535-4}, and \defref{def:sr-steps}. This concludes case~(1a)
    because $\psrtop{\psrstack S'} = T$ and 
    $\psrgraph{\psrstack S'} = \phi \literal e$.

  \item If the last move is a reduce move, there is a rule
    $\literal A \to \rho$, and one can obtain a parse stack $\psrstack S''$
    by removing  $2 \cdot |\rho|$ elements from the end of $\psrstack S'$.
    Let $\psi = \psrgraph{\psrstack S''}$ and $Q = \psrtop{\psrstack S''}$.
    By \defref{def:CFA-parser}, there is a \concretestate\ $T$ and a \dCFAstate\ $Q_i$ containing an item $\pstate(\literal A \to \rho \CDOT, \sigma)$ such that
    \begin{align}
      \cfaconf(\psi,Q) & 
        \cfastep \cfaconf(\psi \literal A^{\tau \circ \sigma}, T) 
        \label{eq:p535-5}\\
      \psrstack S &= 
        \psrstack S'' \literal A^{\tau \circ \sigma} T 
        \label{eq:p535-6}\\
      g &= g'. \label{eq:p535-6a}
    \end{align}
    By \lemmaref{lemma:reduce-correct}, there is a graph 
    $\alpha \in \Gr\Lab$ and a \concretestate\ $T'$ such that 
    \begin{align}
      \phi & = \alpha \rho^{\tau \circ \sigma} 
        \label{eq:p535-7}\\
      \cfaconf(\emptyseq,\startstate^\iota) 
        & \cfastep^* \cfaconf(\alpha \literal A^{\tau \circ \sigma},T')  
        \label{eq:p535-8}\\
      X(\alpha) \cap X(\rho^{\tau \circ \sigma}) 
        &\subseteq X(\literal A^{\tau \circ \sigma}). 
        \label{eq:p535-9}
    \end{align}
    Now,
    $\alpha = \psi$ follows from the construction of $\psrstack S''$ and
    $$
    \cfaconf(\emptyseq,\startstate^\iota) 
    \cfastep^* 
    \cfaconf(\alpha \literal A^{\tau \circ \sigma}, T')
    = \cfaconf(\psi \literal A^{\tau \circ \sigma}, T)
    $$
    from~\eqref{eq:p535-5}, \eqref{eq:p535-8}, and the fact that the dCFA 
    is deterministic. Finally,
    $$
    \buc(\emptyseq, \emptyseq) 
    \cstep^n
    \buc(\phi, g')
    =
    \buc(\psi \rho^{\tau \circ \sigma},g)
    \lstep{\literal A^{\tau \circ \sigma} \der\rho^{\tau \circ \sigma}}
    \buc(\psi A^{\tau \circ \sigma}, g)
    $$
    using the induction hypothesis, \eqref{eq:p535-6a}, \eqref{eq:p535-7}, 
    \eqref{eq:p535-9} and \defref{def:sr-steps}. This concludes case~(1b)
    because $\psrtop{\psrstack S} = T$ and 
    $\psrgraph{\psrstack S} = \psi A^{\tau \circ \sigma}$.
  \end{enumerate}

  \item To show the \emph{if} direction, we now assume any sequence 
  \begin{equation}
    \buc(\emptyseq, \emptyseq) 
    \cstep^n
    \buc(\phi', g')
    \cstep
    \buc(\phi, g).
    \label{eq:p535-10}
  \end{equation}
  of moves and 
  \begin{equation}
    \cfaconf(\emptyseq,\startstate^\iota) \cfastep^* \cfaconf(\phi,R)
    \label{eq:p535-11}
  \end{equation}
  for a \concretestate\ $R$.
  The last move in~\eqref{eq:p535-10} is either a shift or a reduce move.
  \begin{enumerate}[({2}a)]
  \item If it is a shift move, there exists a literal 
    $\literal e \in \Lit\TLab$ such that
    \begin{align}
      \phi &= \phi' \literal e
        \label{eq:p535-12}\\
      g &= g' \literal e
        \label{eq:p535-13}\\
      X(\literal e) \cap X(g') &\subseteq X(\phi')
        \label{eq:p535-14}
    \end{align}
    Because of \eqref{eq:p535-12}, we can write \eqref{eq:p535-11} as
    \begin{equation}
      \cfaconf(\emptyseq,\startstate^\iota) 
      \cfastep^* 
      \cfaconf(\phi',Q)
      \cfastep
      \cfaconf(\phi' \literal e,R)
      \label{eq:p535-15}
    \end{equation}
    for some \concretestate\ $Q$. Therefore, the induction hypothesis applies and yields
    $
    \buc(\startstate^\iota, \emptyseq)
    \pstep^n
    \buc(\psrstack S', g')
    $
    with $\psrtop{\psrstack S'} = Q$ and $\psrgraph{\psrstack S'} = \phi'$.
    Finally, because of~\eqref{eq:p535-13}, \eqref{eq:p535-14} 
    and~\eqref{eq:p535-15}, there is a shift move
    $$
    \buc(\psrstack S', g') 
    \pstep 
    \buc(\psrstack S' \literal e R, g' \literal e)
    =
    \buc(\psrstack S, g)
    $$
    with $\psrstack S = \psrstack S' \literal e R$ and, therefore,
    $\psrtop{\psrstack S} = R$ and $\psrgraph{\psrstack S} = \phi' \literal e = \phi$ because of~\eqref{eq:p535-12}, which concludes case~(2a). 

  \item If the last move is a reduce move, there is a rule 
    $\literal A \to \rho$, a match $\mu \colon X \to X$, and a graph
    $\alpha \in \Gr\Lab$ such that
    \begin{align}
      \phi' &= \alpha \rho^\mu
        \label{eq:p535-16}\\
      \phi &= \alpha \literal A^\mu
        \label{eq:p535-17}\\
      g &= g'
        \label{eq:p535-18}\\
      X(\alpha) \cap X(\rho^\mu) &\subseteq X(\literal A^\mu)
      \label{eq:p535-18a}
    \end{align}
    and \eqref{eq:p535-11} can be written as 
    \begin{equation}
      \cfaconf(\emptyseq,\startstate^\iota) 
      \cfastep^* 
      \cfaconf(\alpha \literal A^\mu,R).
      \label{eq:p535-11a}
    \end{equation}
    The graph
    $\phi = \alpha \literal A^\mu$ is a viable prefix because 
    of~\eqref{eq:p535-11}, \thmref{thm:nCFA-prefix}, and 
    \thmref{thm:dCFA-prefix}. 
    Therefore, $\phi' = \alpha \rho^\mu$ is also a viable
    prefix because of 
    $\phi = \alpha \literal A^\mu \drm \alpha \rho^\mu = \phi'$.
    Since the grammar is reduced, there must 
    be \concretestate{}s $Q, Q'$ such that 
    \begin{equation}
      \cfaconf(\emptyseq,\startstate^\iota) 
      \cfastep^*
      \cfaconf(\alpha,Q')
      \cfastep^*
      \cfaconf(\alpha \rho^\mu,Q)
      =
      \cfaconf(\phi',Q).
      \label{eq:p535-19}
    \end{equation}
    Because of~\eqref{eq:p535-10}, there is also a sequence
    $
    \buc(\emptyseq, \emptyseq) 
    \cstep^k
    \buc(\alpha, g'')
    $
    for some prefix $g''$ of $g = g'$ and $k \le n$. Therefore, the induction
    hypothesis applies, and we can conclude
    $$
    \buc(\startstate^\iota, \emptyseq)
    \pstep^k
    \buc(\psrstack S'', g'')
    $$ 
    for a parse stack $\psrstack S''$ with $\psrtop{\psrstack S''} = Q'$ and 
    $\psrgraph{\psrstack S''} = \alpha$.
    Using the same argument, we can also conclude 
    $$
    \buc(\startstate^\iota, \emptyseq)
    \pstep^n
    \buc(\psrstack S', g')
    $$
    for a parse stack $\psrstack S'$ with $\psrtop{\psrstack S'} = Q$ and
    $\psrgraph{\psrstack S'} = \alpha \rho^\mu = \phi'$.

    Let us assume that $\psrstack S''$ is not a prefix of $\psrstack S'$. There must be a parse stack
    $\psrstack{\hat S}$, literal $\literal l$ and \concretestate{}s 
    $P', P''$, $P' \neq P''$, such that
    $\psrstack{\hat S} \literal l P'$ is a prefix of $\psrstack S'$ and
    $\psrstack{\hat S} \literal l P''$ a prefix of $\psrstack S''$. 
    Let $\psi = \psrgraph{\psrstack{\hat S} \literal l P'} = \psrgraph{\psrstack{\hat S} \literal l P''}$. We can conclude   
    $
    \cfaconf(\emptyseq,\startstate^\iota) \cfastep^* \cfaconf(\psi,P')$
    and
    $\cfaconf(\emptyseq,\startstate^\iota) \cfastep^* \cfaconf(\psi,P'')$
    using \factref{fact:parse-stack}, and $P' = P''$ using the fact that the 
    dCFA is deterministic, contradicting our assumption. $\psrstack S''$ is 
    thus a prefix of $\psrstack S'$, and $\psrstack S''$ can be obtained
    from $\psrstack S'$ by removing $2 \cdot |\rho|$ elements from its end.
    
    Because of~\eqref{eq:p535-18a}, \eqref{eq:p535-11a}, \eqref{eq:p535-19}, 
    and \lemmaref{lemma:reduce-correct}, there is a \dCFAstate~$Q_i$,
    an \inputbinding~$\tau$ and an item 
    $\pstate(\literal A \to \rho \CDOT, \sigma) \in Q_i$ such that
    \begin{align}
      Q &= Q_i^\tau
        \label{eq:p535-20}\\
      \mu &= \tau \circ \sigma 
        \label{eq:p535-21}.
    \end{align}

    Moreover, we know that
    $$
    \cfaconf(\alpha,Q') \cfastep \cfaconf(\alpha \literal A^\mu,R)
    $$
    by~\eqref{eq:p535-11a} and~\eqref{eq:p535-19}, using the fact that
    the dCFA is deterministic.
    Therefore, using \defref{def:CFA-parser},
    $$
    \buc(\startstate^\iota, \emptyseq)
    \pstep^n
    \buc(\psrstack S', g') 
    \pstep 
    \buc(\psrstack S'' \literal A^\mu R, g').
    $$
    This concludes case~(2b) and with it the proof of the lemma because of~\eqref{eq:p535-18}
    and~\eqref{eq:p535-21}, choosing
    $\psrstack S = \psrstack S'' \literal A^{\tau \circ \sigma} R$.
    \qed
  \end{enumerate}  
\end{enumerate}  
\end{proof}

We are now ready to prove the correctness of the ASR parser.

\begin{theorem}\label{t:ASR}%
  Let $g \in \Gr\TLab$.
  The ASR parser can reach an accepting configuration $\buc(\psrstack S, g)$ 
  if and only if $\startLit \drm^* g$.
  Moreover, for every reachable configuration
  $\buc(\psrstack S, g)$, there is a graph $g' \in \Gr\TLab$ and
  an accepting configuration $\buc(\psrstack S', gg')$ such that 
  $
  \buc(\psrstack S, g)
  \pstep^*
  \buc(\psrstack S', gg').
  $
\end{theorem}

\begin{proof}
  Consider any graph $g \in \Gr\TLab$.
  
  For the first part of the theorem,  by \thmref{the:sr:correct} it
  holds that $\startLit \drm^* g$ if and only if 
  $\buc(\emptyseq,\emptyseq) \cstep^* \buc(\startLit,g)$. 
  By \lemmaref{lemma:ASR-naive}, the latter is the case if an only
  if 
  $
  \buc(\startstate^\iota, \emptyseq)
  \pstep^*
  \buc(\startstate^\iota\startLit \acceptstate, g)
  $,
  because the dCFA approves the viable prefix $\startLit$ via
  $\cfaconf(\emptyseq,\startstate) \cfastep^* \cfaconf(\startLit,\acceptstate)$.

  To prove the second part of the theorem, consider any configuration 
  $\buc(\psrstack S, g)$ with 
  $
  \buc(\startstate^\iota, \emptyseq)
  \pstep^*
  \buc(\psrstack S, g)
  $.
  By \lemmaref{lemma:ASR-naive}, \thmref{thm:nCFA-prefix}, and
  \thmref{thm:dCFA-prefix}, $\psrgraph{\psrstack S}$ is a viable prefix.
  Moreover, 
  $
  \buc(\emptyseq,\emptyseq) \cstep^* \buc(\psrgraph{\psrstack S},g)
  $.
  By \lemmaref{lemma:prefix-to-parse}, there is a graph $g' \in \Gr\TLab$
  such that 
  $
  \buc(\psrgraph{\psrstack S},g) \cstep^* \buc(\startLit, gg')
  $.
  Thus, the same argument as above yields 
  $
  \buc(\psrstack S, g)
  \pstep^*
  \buc(\startstate^\iota\startLit \acceptstate, gg')
  $, and its final configuration is accepting.
  \qed
\end{proof}

It is worthwhile pointing out that the ASR parser is still nondeterministic, despite
the ``assistance'' by the dCFA. In fact, there are two sources
of nondeterminism. First,
the \concretestate\ on top of the stack may contain several items that fulfill the conditions
of shift or reduce moves and thus enable several possible moves. There may be items
leading to shifts of different literals, items that result in reductions according
to different rules, and items of which one triggers a shift move whereas the other
triggers a reduce move. For example, in state $Q_1$ of the dCFA in
\figref{fig:fc-dCFA} (under some \inputbinding), the parser may choose among three shift moves.

The second source of nondeterminism lies in the choice of the edge to be read
by a shift move, as there may be several literals $\literal e$ in the input graph
that fulfill the conditions.

Naturally, the ``right'' choice must be made in order to ensure that the parser
accepts a given input graph. Note that this does not contradict
\thmref{t:ASR} which states that, regardless of the choice made,
there exists a possible rest graph with which the parser can reach an accepting
configuration. Clearly, that
rest graph can differ from the actual rest graph in the input. Looking at the
ASR parser, this observation should not come as a surprise, because the parser does not
inspect the rest graph in any way (except for selecting a literal to be shifted
whenever a shift move is made). The extension of the ASR parser by an appropriate
inspection of the rest graph to \emph{predict} the necessary move will be
discussed next. It leads to the main notion proposed
in this paper, the predictive shift-reduce parser.

\section{Predictive Shift-Reduce Parsing\label{s:PSR}}

Intuitively, a move of the parser is appropriate if it keeps it on its
way towards accepting the input graph $g$, provided that $g$ is valid. (Naturally,
if $g$ is not valid, every possible move is appropriate as $g$ will eventually be
rejected anyway.) To identify such a
move, the parser needs criteria that it can check by inspecting the rest graph.
These criteria should preferably only require a fixed number of patterns to be
checked, in order to ensure that an appropriate move can be selected in constant
time. While the desired patterns will obviously have to depend on $\Gamma$,
they should be computable from the grammar by the parser generator.
Such criteria do exist only if the dCFA is conflict-free and if $\Gamma$ 
has the free edge choice property in a sense to be made 
precise in this section. Thus, in contrast
to the pure ASR parser, which works for every HR grammar, the resulting
\emph{predictive shift-reduce parser} exists only for a subset of all HR grammars,
i.e., the parser generator may fail to construct a parser, reporting the existence of
a conflict or failing to have the free edge choice property instead.

For the following considerations, suppose that the ASR parser is in
the process of parsing a valid input graph $g$ and has reached a
configuration $\buc(\psrstack S,g')$, but has not yet processed the
rest graph $g''$ of $g$ where $g \perm g' g''$.%
\footnote{Note that we can represent the rest graph by any permutation
  of $g''$ since none of its literals have been processed by the
  parser yet.}
The top of $\psrstack S$ is
$\psrtop{\psrstack S} = Q^\tau$ with a \dCFAstate~$Q$ and an \inputbinding~$\tau$.

The parser must now choose between shift and reduce moves until the input graph has been accepted or no further move is possible. Shift moves are
caused by transitions leaving $Q$,
and reduce moves by items within $Q$
with a dot at the end of their right-hand side. Let us call such an item a \emph{reduce item}.
Each transition and each reduce item is called a \emph{trigger}
that causes the corresponding move. Note that acceptance is
also caused by a reduce item, which is the only item in the accepting
state~$\acceptstate$.%

We now describe a decision procedure which inspects the rest
graph~$g''$ to select the trigger that causes an appropriate move, i.e.,
a move which turns the parser into a new configuration from which it
can still reach an accepting configuration by reading the remaining 
rest graph. Let us call a sequence of
moves that ends in an accepting configuration a \emph{successful sequence},
even if it does not process the entire rest graph.
\thmref{t:ASR} states that such a sequence always exists
when the parser has reached $\buc(\psrstack S,g')$. %
The decision
procedure must thus select a trigger that causes the first move of a successful
sequence that processes the entire rest graph.

The idea for selecting the right trigger is as follows:
Suppose 
that the rest graph~$g''$ is not yet empty. The procedure
now checks for each trigger whether $g''$ contains a literal $\literal e$ 
that may
be processed next by some successful sequence caused by this
trigger. There must be a trigger with this property since $g$ is
valid. If this trigger is the only possible one that causes a successful 
sequence that eventually reads $\literal e$, this trigger must be the
one causing the right move; the parser thus selects this trigger. 
In the following, we will show that this idea makes an effective decision 
procedure if the dCFA is conflict-free.

Let us consider more closely when a literal is processed next by a
successful sequence caused by a trigger. If the trigger is a
transition, this literal is just the one that is processed by the
corresponding shift move. If the trigger, however, is a reduce item, it
must be the one processed by the first shift move in the move
sequence following the reduce move. This shift move may of course not be the first move of the
sequence, as it can be preceded by further reduce moves.

Suppose now that the parser has processed the input graph entirely, i.e.
the rest graph~$g''$ is empty. The procedure then checks for each reduce
item whether there is a successful sequence that consists of
reduce moves only. The parser then selects any 
reduce item that causes such a successful sequence.

We will now discuss the decision procedure more precisely. To this
end, we consider all successful sequences caused by a trigger.
Recall that we assume that the parser has reached configuration
$\buc(\psrstack S,g')$ with $\psrtop{\psrstack S} = Q^\tau$.

Suppose the trigger is a transition $\mathit{tr} = (Q \atrans{(\literal e,
\mu)} Q')$ of the dCFA. \defref{def:CFA-parser} implies that the shift move
induced by~$\mathit{tr}$ is 
$
\buc(\psrstack S,g')
\lpstep{\mathit{tr}}
\buc(\psrstack S \literal{\bar e} Q'',g' \literal{\bar e})
$
for an appropriate literal $\literal{\bar e} \in \Lit\TLab$ and \concretestate\ $Q''$. And by \thmref{t:ASR}, there is a graph 
$v \in \Gr\TLab$ such that 
$
\buc(\psrstack S \literal{\bar e} Q'',g' \literal{\bar e})
\pstep^*
\buc(\acceptstack,g' \literal{\bar e} v)
$
with $\psrtop{\acceptstack} = \acceptstate$. This means that the parser
accepts $g' \literal{\bar e} v$ or, in other words, $\literal{\bar e} v$ is the
graph processed by this successful sequence. Let us denote the
set of all graphs processed by any successful sequence
caused by~$\mathit{tr}$ as $\Rest(Q^\tau, g', \mathit{tr})$.

Suppose now that the trigger is a reduce item 
$\mathit{it} = \pstate(\literal A \to \rho \CDOT,\sigma) \in Q$. 
\defref{def:CFA-parser} implies that the reduce move
induced by~$\mathit{it}$ is 
$
\buc(\psrstack S,g')
\lpstep{\mathit{it}}
\buc(\psrstack S' \literal A^{\tau \circ \sigma} Q',g')
$
with an appropriate parse stack $\psrstack S'$ and \concretestate\ $Q'$. And by \thmref{t:ASR}, there is a graph 
$v \in \Gr\TLab$ such that 
$
\buc(\psrstack S' \literal A^{\tau \circ \sigma} Q',g')
\pstep^*
\buc(\acceptstack,g' v)
$
with $\psrtop{\acceptstack} = \acceptstate$. This means that $v$ is the graph
processed by this successful sequence. Let us denote the set of
all graphs processed by any successful sequence caused
by~$\mathit{it}$ as $\Rest(Q^\tau, g', \mathit{it})$.

Before utilizing the sets $\Rest(Q^\tau, g', t)$ for a trigger~$t$, let
us introduce some terminology. For a graph $h = \literal e_1
\cdots \literal e_n$ with $n>0$ literals, let $\Fi(h) = \literal e_1$
be the first literal of $h$. In the special case $n=0$, we let $\Fi(\emptyseq) = \eoi$ where the
special symbol $\eoi$ indicates that there are no literals at all. For
a set $S \subseteq \Gr\Lab$ of graphs, let $\Fi(S) = \set{\Fi(h) \mid h
\in S}$.

For a trigger~$t$, now consider the set
$$
\Fi(\Rest(Q^\tau, g', t)).
$$
This set contains all literals that can be processed next by
successful sequences caused by $t$, and it contains~\eoi{} if there is a
successful sequence caused by $t$ without any shift move. The decision
procedure outlined above thus has to select the trigger~$t$ such that
$\Fi(\Rest(Q^\tau, g', t))$ contains a literal of the rest graph, 
or~\eoi{} if $g'' = \emptyseq$. However, this does not make a practical
decision procedure as these sets are usually infinite. We 
turn them into finite sets by mapping their members to 
\emph{pseudo-literals} as described next.

Note first that every node in any literal of any of these sets
falls into one of three categories: It is either (1) a node assigned
to a parameter of~$Q$ by
$\tau$, (2) a node not occurring in~$X(g')$, or
(3) a node in $X(g')$ not assigned to a parameter of $Q$ by $\tau$.
We now define a function that maps nodes of
category (1) to their corresponding parameter, nodes of category (2) to
`$\unknown$', and all others to `$\Known$'.
$$
f_Q^{\tau,g'}(x) = \left\{
  \begin{array}{ll}
    y & \text{if there exists } y \in \params(Q) \text{ such that } \tau(y) = x \\
    \unknown & \text{if } x \notin X(g')\\
    \Known & \text{otherwise} 
  \end{array}
  \right.
$$

We extend function $f_Q^{\tau,g'}$ to literals and sets of
literals in the obvious way. Literals are thus turned into
\emph{pseudo-literals}, which are similar to literals, but may be
attached to `$\unknown$' and `$\Known$' instead of
nodes.\footnote{Note that these pseudo-literals are a generalized
version of those introduced in \sectref{s:nCFA}.} 

Function $f_Q^{\tau,g'}$ applied to $\Fi(\Rest(Q^\tau, g', t))$ turns
this set into a finite set, because the number of terminal
labels and the number of parameters in $Q$ are finite. However, this set cannot be
computed from the HR grammar alone as it depends on $g'$. Recall that the
\inputbinding~$\tau$ is uniquely determined by~$g'$ because the dCFA
approves $g'$ by
$
\cfaconf(\emptyseq,\startstate^\iota) 
\cfastep^* 
\cfaconf(g',Q^\tau)
$
and the dCFA is deterministic.
To simplify things, let us define the finite set
\begin{equation}
  \Fo(Q,t) := \bigcup_{g' \in \Gr\Lab} f_Q^{\tau,g'}(\Fi(\Rest(Q^\tau, g', t)))
  \label{eq:follow}
\end{equation}
by building the union over all terminal graphs~$g'$. Clearly, only
the graphs $g'$ approved by the dCFA as mentioned above contribute to
this set. This set just depends on the \dCFAstate\ $Q$ and one of its
triggers $t$, and is thus static information independent of the
input graph. While $\Fo(Q,t)$ cannot directly be computed using
\eqref{eq:follow}, one can compute it by analyzing the
dCFA in a way very similar to the computation of the \emph{follower}
symbols for string grammars (\sectref{s:PSRintro}). 

\begin{example}\label{ex:follow}
  Consider the dCFA for the tree-generating grammar in 
  \figref{fig:tree-dCFA} and in particular its state $Q_4$, which has triggers $\psrtr$ and $\psrit$:
  $\psrtr$ is the transition from $Q_4$ to $Q_3$, and $\psrit$ is 
  the reduce item $\pstate({T(y) \to T(y)\,e(y,z)\,T(z)\CDOT},{[y/\parama,z/\paramb]})$.
  $Q_4$ has the parameters $\parama$ and $\paramb$. Function $f_{Q_4}^{\tau,g'}$, when 
  applied to nodes, thus maps into the set $\set{\unknown,\Known,\parama,\paramb}$. In fact
  \begin{align*}
    \Fo(Q_4,\psrtr) &= \set{ e(\paramb,\unknown) }\\
    \Fo(Q_4,\psrit) &= \set{ e(\parama,\unknown), e(\Known,\unknown), \eoi }
  \end{align*}
  It is clear that any successful sequence caused by transition $\psrtr$
  must begin with a shift move and that the read literal must match 
  edge $e(\paramb,\paramc)$, which is ascribed to the transition. However, the ``new'' 
  parameter $\paramc$ is mapped to $\unknown$.
  
  Reduce item $\psrit$ can cause a successful sequence without any shift move, 
  indicated by \eoi. To see 
  this, consider, e.g., a parse stack $\psrstack S$ with 
  $\psrtop{\psrstack S} = Q_4^\tau$ and 
  $\psrgraph{\psrstack S} = \startedgeL(1) \, T(1) \, e(1,2) \, T(2)$. The 
  reduce move will yield a stack $\startedgeL(1) \, T(1)$, which can be further
  reduced to $\startSym()$.

  Moreover, $e(\parama,\unknown)$ and $e(\Known,\unknown)$ indicate that the literal 
  read next must be an $e$-literal attached to node $\tau(\parama)$ or to any node 
  that has been processed already, but that is not kept track of by a parameter 
  in $Q_4$, indicated by $\Known$, and  a node that has not yet been processed, 
  indicated by $\unknown$.
  \qed
\end{example}

Now let $\literal e$ be a literal  of the rest graph $g''$. 
The definition of $\Fo(Q,t)$ implies that 
\begin{equation}
  f_Q^{\tau,g'}(\literal e) \in \Fo(Q,t)
  \label{eq:select-problem-x}
\end{equation} 
is a necessary and easily verifiable condition 
for $\literal e$ to be a literal that can be
processed next by a successful sequence caused by trigger $t$.
Similarly, $\eoi \in \Fo(Q,t)$ can be used to check whether $t$ can cause a 
successful sequence without any shift move. 
But this information is not yet sufficient to make an effective decision 
procedure. To see this, 
recall that function $f_Q^{\tau,g'}$ may map many literals to the same
pseudo-literal. Moreover, $\Fo(Q,t)$ contains the pseudo-literals of
any literal that may be read next in some successful sequence,
not necessarily only those that process the rest graph $g''$ entirely. As a
consequence, \eqref{eq:select-problem-x} does not yet make sure 
that $t$ is the right trigger.
However, if we notice somehow---and additionally to
\eqref{eq:select-problem-x}---that $\literal e$ can never be processed
by any successful sequence not caused by $t$, we know for certain that $t$ 
is the only candidate for the right trigger. This observation leads the way to an effective procedure for selecting the right trigger.

Let us determine which literals can be processed by a successful
sequence caused by a trigger $t$. We are not only interested in the 
literals that are processed first, but also in those literals that are 
processed \emph{eventually}. Instead of a function $\Fi$, we will use a 
function $\Any$ which is defined as follows:
For a graph $h = \literal e_1 \cdots \literal e_n$ with $n>0$ literals,
let $\Any(h) = \set{\literal e_1, \ldots, \literal e_n}$ the set of all
of its literals. For the empty graph, let $\Any(\emptyseq) =
\set{\eoi}$. For a set $S \subseteq \Gr\Lab$ of graphs, let $\Any(S) =
\bigcup_{h \in S} \Any(h)$. We then define the finite set
\begin{equation}
  \FoA(Q,t) := \bigcup_{g' \in \Gr\Lab} f_Q^{\tau,g'}(\Any(\Rest(Q^\tau, g', t))).
  \label{eq:follow-all}
\end{equation}
Note the close resemblance to~\eqref{eq:follow}; the only difference is
the use of $\Any$ instead of $\Fi$, i.e., $\FoA(Q,t)$ contains the
$f_Q^{\tau,g'}$-images of all literals that occur eventually in some
graph processed by a successful sequence caused by $t$, and
it contains \eoi{} if there is a successful sequence caused by $t$ 
that does not contain any shift move.

Again, this definition cannot be used for computing $\FoA(Q,t)$
directly, but one can compute it by analyzing the dCFA in a similar way as
for $\Fo(Q,t)$.

\begin{example}\label{ex:followAll}
  We continue \exref{ex:follow} and consider again the dCFA for the tree-generating grammar shown in 
  \figref{fig:tree-dCFA} and in particular its state $Q_4$ with its two triggers $\psrtr$ and $\psrit$. In addition to
  \begin{align*}
    \Fo(Q_4,\psrtr) &= \set{ e(\paramb,\unknown) }\\
    \Fo(Q_4,\psrit) &= \set{ e(\parama,\unknown), e(\Known,\unknown), \eoi }
  \end{align*}
  we have
  \begin{align*}
    \FoA(Q_4,\psrtr) &= \set{ e(\paramb,\unknown), e(\Known,\unknown), e(\unknown,\unknown) }\\
    \FoA(Q_4,\psrit) &= \set{ e(\parama,\unknown), e(\Known,\unknown), e(\unknown,\unknown), \eoi }
  \end{align*}
  We can see that any literal that matches the only pseudo-literal 
  $e(\paramb,\unknown)$ in $\Fo(Q_4,\psrtr)$ can never be processed in any successful 
  sequence caused by $\psrit$, even if the rest graph contains literals 
  matching the pseudo-literals $e(\parama,\unknown)$ or $e(\Known,\unknown)$, which 
  are members of $\Fo(Q_4,\psrit)$. 
  This can be concluded from the fact that  $e(\paramb,\unknown)$ does not occur 
  in $\FoA(Q_4,\psrit)$. As a consequence, $\psrit$ cannot be the right trigger if
  we find a literal that matches $e(\paramb,\unknown)$.

  However, we can see that a literal that matches
  $e(\Known,\unknown) \in \Fo(Q_4,\psrit)$ may indeed be processed later when 
  transition $\psrtr$ is chosen. The existence of a literal matching any 
  pseudo-literal in $\Fo(Q_4,\psrit)$ does thus not help to eliminate $\psrtr$ from 
  the candidates of right triggers. 

  As a consequence, a procedure can reliably predict the next move in state
  $Q_4$ by first checking whether there is a rest graph literal $\literal e$ with
  $f_{Q_4}^{\tau,g'}(\literal e) = e(\paramb,\unknown)$. If there is such a literal, 
  $\psrtr$ is guaranteed to be the right trigger because 
  $e(\paramb,\unknown) \notin \FoA(Q_4,\psrit)$. If such a literal, however, does not 
  exist, $\psrtr$ cannot be the right trigger. One then checks whether the rest graph contains any literal 
  $\literal e'$ that matches a pseudo-literal of $\Fo(Q_4,\psrit)$, i.e., with
  $f_{Q_4}^{\tau,g'}(\literal e') \in \Fo(Q_4,\psrit)$. If there is such a literal,
  one chooses the reduce move caused by $\psrit$. If there is no such
  $\literal e'$, it is guaranteed that there is no successful sequence 
  caused by $\psrit$ that processes the rest graph entirely, and the parser
  can terminate with a failure. 
  \qed
\end{example}

This example motivates that one must compare the $\Fo$ and $\FoA$ sets of 
the different triggers and that one must determine which trigger should be 
considered first when looking for rest graph literals that match any 
pseudo-literals in the $\Fo$ set of this trigger:

\begin{defn}\label{def:prec}
  A trigger $t$ \emph{precedes} a trigger $t'$, written $t \prec t'$, 
  if $t$ and $t'$ are triggers of the same \dCFAstate~$Q$, 
  $t \neq t'$, and 
  $
    \FoA(Q,t) \cap \Fo(Q,t') \neq \emptyset.
  $
\end{defn}

Note that $\prec$ is not an ordering because it is in general not
transitive. But $t \prec t'$ indicates that one must check $t$ prior to
$t'$. However, $t \prec t'$ does not help to find an order if there is
a $\prec$-chain $t \prec t' \prec \cdots \prec t$. This motivates the
definition of \emph{conflicting} triggers. We will see in the
following that an effective decision procedure for identifying the
right trigger requires conflict-freeness:

\begin{defn}\label{def:conflict}
  Let $Q$ be a \dCFAstate\ and $T_Q$ the set of its triggers. A subset
  $T \subseteq T_Q$ is \emph{in conflict} 
  if there is a sequence $t_1 \prec t_2 \prec \cdots \prec t_k \prec t_1$
  with $T = \set{t_1, t_2, \ldots, t_k}$.
  $Q$ is \emph{conflict-free} if no subset of its triggers is in 
  conflict.
\end{defn}

If a \dCFAstate\ is conflict-free, one can sort the triggers so that their order 
respects~$\prec$, which will be necessary for the effective decision procedure:

\begin{lemma}\label{lemma:trigger-seq}
  For every conflict-free \dCFAstate{} $Q$, there is an ordered sequence
  $t_1, \ldots, t_n$ of its triggers such that 
  $
     \Fo(Q,t_i) \cap \FoA(Q,t_j) = \emptyset
  $
  for every pair of indices $i, j$ with $i < j$.
\end{lemma}

\begin{proof}
  Let $T_Q$ be the set of triggers of $Q$. $T_Q$ can be considered
  as a directed graph with triggers acting as nodes and having an
  edge from $t$ to $t'$ iff $t \prec t'$. A cycle in $T_Q$ would indicate
  a conflict of the members of the cycle. Therefore, one
  can sort the transitions topologically into an ordered sequence $t_1,
  \ldots, t_k$ such that $T_Q = \set{t_1, \ldots, t_k}$ and $t_i \prec
  t_j$ implies $i < j$ for every pair of indices $i,j$. As a
  consequence, $j < i$ implies $t_i \not\prec t_j$, which is equivalent
  to $\Fo(Q,t_j) \cap \FoA(Q,t_i) = \emptyset$.
  \qed
\end{proof}

Before we will use this ordered sequence of triggers to devise an 
efficient procedure that reliably identifies an appropriate
trigger for the next move together with the yet unread literal to
be shifted if the selected trigger is a transition, we show that the
number of moves that an ASR parser needs to check the validity of a
graph depends indeed linearly on the size of the graph. We will need
this result later in the correctness proof for PSR parsers and for
discussing their performance.

\begin{lemma}\label{lemma:linear-bounds}
  If no $Q \in \mathcal{Q}$ has conflicts, there is a constant
  $c \in \Nat$ so that $n < c \cdot (|g_n|-|g_0|+1)$ for every sequence
  $
    \buc(\startstate^\iota, \emptyseq)
    \pstep^*
    \buc(\psrstack S_0, g_0)
    \pstep
    \buc(\psrstack S_1, g_1)
    \pstep
    \cdots
    \pstep
    \buc(\psrstack S_n, g_n)
  $
  of ASR moves where $\psrstack S_0$ is a prefix of $\psrstack S_i$ for 
  $i = 1, \ldots, n$.
\end{lemma}

\begin{proof}
    Let us assume that none of the \dCFAstate{}s has a conflict.
    We prove the lemma by induction on $|g_n| - |g_0|$. 
    
    For $g_n = g_0$, let us assume the contrary, i.e., there is, for any 
    $m \in \Nat$, a sequence
    $
    \buc(\startstate^\iota, \emptyseq)
    \pstep^*
    \buc(\psrstack S_0, g_0)
    \pstep
    \buc(\psrstack S_1, g_0)
    \pstep
    \cdots
    \pstep
    \buc(\psrstack S_n, g_0)
    $
    where $\psrstack S_0$ is a prefix of $\psrstack S_i$ for 
    $i = 1, \ldots, n$, and $n \ge m$. Let 
    $\psrtop{\psrstack S_i} = \bar Q_i^{\tau_i}$ with \dCFAstate{}s 
    $\bar Q_i \in \mathcal{Q}$ and \inputbinding{}s~$\tau_i$ 
    for $i = 1, \ldots, n$. All moves in 
    $
    \buc(\psrstack S_0, g_0)
    \pstep
    \buc(\psrstack S_1, g_0)
    \pstep
    \cdots
    \pstep
    \buc(\psrstack S_n, g_0)
    $
    are reduce moves, and by the definition of reduce moves, we have 
    $\tau_i(X) \subseteq \tau_0(X)$ for $i = 1, \ldots, n$. Because 
    $\mathcal{Q}$ and $\tau_0(X)$ are finite and because $n \ge m$,
    there must be some $j,k$ with $0 \le j < k \le n$
    and $\bar Q_j^{\tau_j} = \bar Q_k^{\tau_k}$, i.e., the subsequence of moves
    $
    \buc(\psrstack S_j, g_0)
    \pstep
    \buc(\psrstack S_{j+1}, g_0)
    \pstep
    \cdots
    \pstep
    \buc(\psrstack S_k, g_0)
    $
    forms a cycle. Because the grammar is reduced, there must be a chance to 
    leave the cycle, i.e., there is an index $\hat\iota$ with $j \le \hat\iota \le k$ such that
    $\bar Q_{\hat\iota}$ must have at least two outgoing transitions: $t$ belongs to the 
    cycle and goes into the next state of the cycle whereas 
    $t'$ leads out of the cycle. 
    Let us choose any $\ell \in \Fo(\bar Q_{\hat\iota},t')$, i.e., $\ell$ is 
    either a pseudo-literal or $\ell = \eoi$. In either case, 
    $\ell \in \Fo(\bar Q_{\hat\iota},t)$, too, because
    $\bar Q_{\hat\iota}^{\tau_{\hat\iota}}$ can be reached again when using 
    transition $t$. And because 
    $\Fo(\bar Q_{\hat\iota},t) \subseteq \FoA(\bar Q_{\hat\iota},t)$ and 
    $\Fo(\bar Q_{\hat\iota},t') \subseteq \FoA(\bar Q_{\hat\iota},t')$,
    we can conclude $t \prec t' \prec t$, i.e., $\bar Q_{\hat\iota}$ has 
    a conflict, in contradiction to the assumption that none of the \dCFAstate{}s
    has conflicts.
      
    For $|g_n|-|g_0| > 0$, we consider any sequence
    \begin{equation}
    \buc(\startstate^\iota, \emptyseq)
    \pstep^*
    \buc(\psrstack S_0, g_0)
    \pstep
    \buc(\psrstack S_1, g_1)
    \pstep
    \cdots
    \pstep
    \buc(\psrstack S_n, g_n)
    \label{eq:bound-a}
    \end{equation}
    of ASR moves where $\psrstack S_0$ is a prefix of $\psrstack S_i$ for 
    $i = 1, \ldots, n$.
    Let $k$ be the index where $|g_k| + 1 = |g_{k+1}| =|g_n|$,
    i.e., there is a literal $\literal e$ so that 
    $g_k \literal e = g_{k+1} = g_{k+1} = \cdots = g_n$.
    Sequence \eqref{eq:bound-a} can thus be written in two ways:
    \begin{align}
      \buc(\startstate^\iota, \emptyseq)
      &\pstep^*
      \buc(\psrstack S_0, g_0)
      \pstep
      \buc(\psrstack S_1, g_1)
      \pstep
      \cdots
      \pstep
      \buc(\psrstack S_k, g_k)
      \label{eq:bound-b}
      \\
      \buc(\startstate^\iota, \emptyseq)
      &\pstep^*
      \buc(\psrstack S_{k+1}, g_k \literal e)
      \pstep
      \buc(\psrstack S_{k+2}, g_k \literal e)
      \pstep
      \cdots
      \pstep
      \buc(\psrstack S_n, g_k \literal e)
      \label{eq:bound-c}
    \end{align}
    By the induction hypothesis, we have 
    $k < c \cdot (|g_k|-|g_0|+1) = c(|g_n|-|g_0])$ for \eqref{eq:bound-b}. 
    And by 
    the same arguments as for the case $g_n = g_0$, we have $n-k-1 < c$ for \eqref{eq:bound-c}. Therefore, $n \le c \cdot (|g_n|-|g_0]+1)$.
    \qed
\end{proof}

The following corollary is an immediate consequence of the previous
lemma by choosing $\psrstack S_0 = \startstate^\iota$, $\psrstack S_n =
\psrstack S$, $g_0 = \emptyseq$, and $g_n = g$:

\begin{corollary}\label{c:linear-bounds}
  If no $Q \in \mathcal{Q}$ has conflicts, there is a constant
  $c \in \Nat$ so that $n < c \cdot |g| + c$ for every sequence  
  $
    \buc(\startstate^\iota, \emptyseq)
    \pstep^n
    \buc(\psrstack S, g)
  $
  of ASR moves.
\end{corollary}

This linear bound on the number of parser moves is in fact a
consequence of the required absence of conflicts. Note that
there is no such bound on the length of a derivation $\startLit \drm^*
g$ for an arbitrary HR grammar.

We are now going to devise the
procedure~\funref{alg:select-edge-action} below that reliably
identifies an appropriate trigger for the next move together with the yet
unread literal to be shifted if the selected trigger is a transition.
This procedure will control the parser and what parser move is executed
next. We will show that this procedure is effective as well as
efficient for certain HR grammars. These are grammars whose \dCFAstate{}s
are all conflict-free and that have an additional property, the
\emph{free edge choice property}, which we are going to define below.
\funref{alg:select-edge-action} is effective in the sense that it will
always select an appropriate move that leads to a (successful) sequence of
parser moves ending in an accepting configuration if the input graph is
valid. It is also efficient as constant time suffices for computing
the results.

The latter might come as a surprise because
\funref{alg:select-edge-action} must perform a kind of graph matching
task. The procedure is in fact called after each move of the ASR parser (which 
we will then extend to the PSR parser) when it is in a
configuration $\buc(\psrstack S, g')$ with $\psrtop{\psrstack S} =
Q^\tau$ with a \dCFAstate\ $Q \in \mathcal Q$ and an input binding~$\tau$.
\funref{alg:select-edge-action} must then select the appropriate trigger
$t$ for $Q$ by checking whether the unread part $g''$ of the input
graph contains a literal whose $f^{\tau,g'}_Q$-image is a member of
$\Fo(Q,t)$ as it has been discussed above. 
\funref{alg:select-edge-action} in fact uses a simpler criterion and
looks for literals of $g''$ that are attached to certain nodes
determined by some pseudo-literal $\ell \in \Fo(Q,t)$. We will show
later that this simpler criterion still makes the procedure effective
if the grammar has the free edge choice property. The look-up criterion
uses the binary relation \emph{fits} defined as follows:
 
\begin{defn}\label{def:fits}
Let $\ell = a(y_1, \ldots, y_k)$ be a pseudo-literal and 
$\literal e \in \Lit\TLab$ a literal. Furthermore, let $Q$ be a 
\dCFAstate\ and $\tau$ an \inputbinding. We say that $\literal e$ \emph{fits} $\ell$ 
in $Q^\tau$, written $\literal e \fits{\tau}{Q} \ell$, if and only if
$\literal e = a(x_1, \ldots, x_k)$ and $x_i = \tau(y_i)$ for each 
$i = 1, \ldots, n$ with $y_i \in \params(Q)$.
\end{defn}

Note that, given a pseudo-literal $\ell$, there is no requirement on the 
attached nodes of $\literal e$ that correspond to any $y_i$ which are not 
parameters of $Q$. Note also that, by the definition of $\fits \tau Q$ and 
$f_Q^{\tau,g'}$,
$\literal e \fits \tau Q f_Q^{\tau,g'}(\literal e)$ holds for every 
literal~$\literal e$, graph $g'$, \dCFAstate~$Q$, and
\inputbinding~$\tau$.

\funref{alg:select-edge-action} below shows the pseudo-code of the decision 
procedure using the \emph{fits} relation. It returns the appropriate trigger and,
also, the literal to be read next, or $\eoi$ if $g'' = \emptyseq$; it
returns `\textit{failure}' if it is guaranteed that there is no
successful sequence processing $g''$ entirely. Note that the procedure
requires an ordered sequence of all triggers as described in
\lemmaref{lemma:trigger-seq}, i.e., it does not work if there are
conflicting triggers.

\begin{procedure}[ht]
  \caption{SelectTrigger($Q,\tau,g''$))}
  \label{alg:select-edge-action}
  \Input{State $Q \in \mathcal Q$, \inputbinding~$\tau$, rest graph $g''$} 
  \Output{a pair $(t, \eoi)$ or $(t,\literal e)$ where $t$ is a trigger and  $\literal e$ a literal of $g''$, \\
  or `\textit{failure}' }
  let $t_1, \ldots, t_n$ be a sequence of triggers of $Q$ as in 
     \lemmaref{lemma:trigger-seq}\label{an:seq}\;
  \For{$i\leftarrow 1$ \KwTo $n$\label{an:for}}{
    \uIf{$g'' \neq \emptyseq$}{
      \ForEach{pseudo-literal $\ell \in \Fo(Q,t_i)$}{
        look for a literal $\literal e$ of $g''$ such that 
        $\literal e \fits{\tau}{Q} \ell$\;\label{an:lookup}
        \lIf{$\literal e$ exists}{
          \Return $(t_i, \literal e)$
        }
      }
    }
    \lElseIf{$\eoi \in \Fo(Q,t_i)$}{
      \Return $(t_i, \eoi)$\label{an:eoi}
    }
  }
  \Return `\textit{failure}'
\end{procedure}

We will now show the effectiveness of \funref{alg:select-edge-action}
in the sense that it always returns the appropriate trigger together with
the literal to be read next. It is clear from \alineref{an:seq} that
the procedure cannot work if any state of the dCFA has conflicts. But
the grammar must have yet another property, which we are going to
define next. The rationale behind this property is the following:
There may be several unread literals that may be read by a transition
selected as the appropriate trigger. Some of those literals may lead to a
successful parse whereas others do not. The \emph{free edge choice}
property simply requires that it does not matter which of the fitting
literals is selected. The property additionally requires that the
simpler criterion based on the \emph{fits} relation (\defref{def:fits})
may in fact be used.

\begin{defn}[Free edge choice property]\label{def:free-edge-choice}
  Let $g', g'' \in \Gr\TLab$ be any graphs, $g'' \neq \emptyseq$, 
  $\buc(\psrstack S,g')$ a configuration reached by the ASR parser, 
  $\psrtop{\psrstack S} = Q^\tau$ where $Q$ is a \dCFAstate\ and 
  $\tau$ an \inputbinding. The return value $(t, \literal e)$ of 
  \funref[Q, \tau, g'']{alg:select-edge-action} is called \emph{useful} if 
  \begin{enumerate}[(i)]
    \item \label{it:fec-a}%
          $f_Q^{\tau,g'}(\literal e) \in \Fo(Q,t)$,
          and
    \item \label{it:fec-b}%
          there is a graph $h \in \Gr\TLab$ such that $\literal e h \perm g''$ 
          and
       \begin{equation}
          \buc(\psrstack S, g')
          \lpstep{t} 
          \buc(\psrstack S', g' \literal e)
          \pstep^*
          \buc(\acceptstack, g' \literal e h)
       \end{equation} 
       with $\psrtop{\acceptstack} = \acceptstate$ if 
       \begin{equation}
          \buc(\psrstack S, g')
          \lpstep{t} 
          \buc(\psrstack S', g' \literal e')
          \pstep^*
          \buc(\acceptstack, g' \literal e' h')
       \end{equation}
       for any literal $\literal e'$ and graph $h' \in \Gr\TLab$ so
       that $\literal e' h' \perm g''$.
  \end{enumerate}
  The grammar is said to have the \emph{free edge choice property} 
  if  the return value of \funref[Q, \tau, g'']{alg:select-edge-action} 
  is always useful.
  \qed
\end{defn}

This property is in fact similar to the property defined
in~\cite{Drewes-Hoffmann-Minas:15} under the same name for PTD parsing
whereas we now need this property for PSR parsing. There are
sufficient criteria that can be used by static grammar analysis that
guarantee that a grammar has the free edge choice property and that
have been realized in the \emph{Grappa} tool (see below). A discussion
of how this property can effectively be tested is out of scope of the
present paper, however.

We will show below that \funref{alg:select-edge-action} returns
reliable results only for grammars with the free edge choice property.
We thus extend \assref{a:dCFA}:

\begin{assumption}\label{a:conflict-free-and-fec}
  For the remainder of the paper, we further assume that every
  state $Q$ of the dCFA is conflict-free and that $\Gamma$ has the free edge choice property.
\end{assumption}

The following lemma states that \funref{alg:select-edge-action}, under
this assumption, can reliably identify the unique right trigger
together with the unread literal to be read next:

\begin{lemma}\label{lemma:chooses-right-action}
  Let $\buc(\psrstack S,g')$ be any configuration reached by the ASR parser and 
  $\psrtop{\psrstack S} = Q^\tau$ where $Q$ is a \dCFAstate\ and 
  $\tau$ an \inputbinding.

  For every graph $g'' \in \Gr\TLab$ such that 
  $
  \buc(\psrstack S, g')
  \pstep^*
  \buc(\acceptstack, g'g'')
  $
  with $\psrtop{\acceptstack} = \acceptstate$, \funref[Q,\tau,g'']{alg:select-edge-action} returns a pair $(t, \literal e)$ with the following properties:
  \begin{itemize}
    \item If $t = \pstate(\pseudoLit \to \startLit \CDOT, \iota)$ is the reduce 
    item causing acceptance, then $\psrstack S = \acceptstack$, $\literal e = \eoi$,
    and $g'' = \emptyseq$.
    \item  If $t \neq \pstate(\pseudoLit \to \startLit \CDOT, \iota)$
    is any other reduce item, there is a stack~$\psrstack S'$ with
    \begin{equation}
      \buc(\psrstack S, g')
      \lpstep{t} 
      \buc(\psrstack S', g')
      \pstep^*
      \buc(\acceptstack, g'g'')
      \label{eq:lemma-cra-r}
    \end{equation}
    and $g'' = \emptyseq$ iff $\literal e = \eoi$.
    \item If $t$ is a transition, then $\literal e$ is a literal of $g''$ 
    and 
    \begin{equation}
    \buc(\psrstack S, g')
    \lpstep{t} 
    \buc(\psrstack S \literal e Q', g' \literal e)
    \pstep^*
    \buc(\acceptstack, g' \literal e h)
    \label{eq:lemma-cra-s}
    \end{equation}
    for some \concretestate\ $Q'$ 
    and a graph $h \in \Gr\TLab$ with
    $g'' \perm \literal e h$.
    \qed
  \end{itemize}
\end{lemma}

\begin{proof}
  Let $\buc(\psrstack S,g')$, $Q$, $\tau$, and $g''$ be as in the lemma. We 
  distinguish three cases:
  \begin{enumerate}[(1)]
    \item $g'' = \emptyseq$ and $Q = \acceptstate$. $\acceptstate$ 
    consists of just the reduce item 
    $\psrit = \pstate(\pseudoLit \to \startLit \CDOT, \iota)$ and 
    $\eoi \in \Fo(Q_A,\psrit)$. Thus \funref{alg:select-edge-action} returns 
    $(\psrit, \eoi)$,
    and the parser terminates by accepting~$g'$.

    \item $g'' = \emptyseq$ and $Q \neq \acceptstate$. There must
    be a nonempty sequence $s$ of reduce moves leading to $\buc(\acceptstack, g')$.
    Because of $g'' = \emptyseq$, the procedure returns  
    $(t, \literal e) = (t_i, \eoi)$ in \alineref{an:eoi}, and 
    $\eoi \in \Fo(Q,t_i)$. In fact, $t_i$ is the only trigger in $Q$ with 
    $\eoi \in \Fo(Q,t_i)$. Otherwise, if  $\eoi \in \Fo(Q,t_j)$ for $j < i$, 
    \funref{alg:select-edge-action} would have returned $(t_j, \eoi)$. And if 
    $\eoi \in \Fo(Q,t_j)$ for $j > i$, we would have $\eoi \in \FoA(Q,t_j)$, 
    in contradiction to the construction of sequence $t_1, \ldots, t_n$. 
    Therefore, $s$ must be caused by $t$ as in~\eqref{eq:lemma-cra-r}.
    
    \item $g'' \neq \emptyseq$. There is a successful sequence since 
    \begin{equation}
    \buc(\psrstack S, g')
    \pstep^*
    \buc(\acceptstack, g'g'')
    \label{eq:accept-seq-l},
    \end{equation}
    which contains at least one shift move. Let $t_j$ be the trigger causing this particular sequence and $\literal{\bar e}$ the
    literal processed by its first shift move.
    Then
    \begin{equation}
      f_Q^{\tau,g'}(\literal{\bar e}) \in \Fo(Q,t_j).\label{eq:accept-seq-jm}
    \end{equation}
    \funref{alg:select-edge-action} cannot return any pair
    $(t_i, \literal e')$ where $i < j$. To see this, let us assume the 
    contrary, i.e., $g''$ contains a literal $\literal e'$ with
    $\literal e' \fits \tau Q \ell$ for some $\ell \in \Fo(Q,t_i)$. Because of 
    the free edge choice property, in 
    particular \defref{def:free-edge-choice}(\ref{it:fec-a}), we can conclude 
    that $f_Q^{\tau,g}(\literal e') \in \Fo(Q,t_i)$ and, therefore, 
    $f_Q^{\tau,g}(\literal e') \notin \FoA(Q,t_j)$ by the 
    construction of $t_1, \ldots, t_n$. But this means that 
    $\literal e'$ cannot be processed by any shift move in 
    \eqref{eq:accept-seq-l}, in contradiction to \eqref{eq:accept-seq-l} 
    being a successful sequence. 

    Because of \eqref{eq:accept-seq-jm} and 
    $\literal{\bar e} \fits \tau Q f_Q^{\tau,g'}(\literal{\bar e})$, 
    \funref{alg:select-edge-action} can find $\literal{\bar e}$ in 
    \alineref{an:lookup} as soon as $t_j$ is selected in the outer for-loop.
    But $g''$ may also contain a different literal $\literal e$ that fits
    a pseudo-literal $\ell \in \Fo(Q,t_j)$, and \funref{alg:select-edge-action}
    returns $(t_j, \literal e)$. Trigger $t_j$ may be either a reduce item or a 
    transition. If it is a reduce item, \eqref{eq:accept-seq-l} has in fact
    the form~\eqref{eq:lemma-cra-r}, which proves the lemma. But if $t_i$ is a 
    transition, there is a successful sequence 
    \begin{equation}
          \buc(\psrstack S, g')
          \lpstep{t_j} 
          \buc(\psrstack S', g' \literal e)
          \pstep^*
          \buc(\acceptstack, g' \literal e h)
          \label{eq:accept-seq-n}
    \end{equation}
    where $h \in \Gr\TLab$ is a graph such that $\literal e h \perm g''$
    because of \eqref{eq:accept-seq-l} and the free edge choice property, in 
    particular \defref{def:free-edge-choice}(\ref{it:fec-b}). This proves the 
    lemma because \eqref{eq:accept-seq-n} has the form~\eqref{eq:lemma-cra-s}.
    \qed
    \end{enumerate}  
\end{proof}

\funref{alg:select-edge-action} can now be used to predict the next move in every 
configuration reachable by the ASR parser. This leads to the 
\emph{predictive shift-reduce (PSR)} parser, the
main notion proposed in this paper, which is in fact the ASR parser equipped
with \funref{alg:select-edge-action} for selecting the next
move:

\begin{defn}[PSR Parser]\label{def:PSR-parser}
  A \emph{(PSR parser) configuration} $\psrbuc(\psrstack S, g, {\bar g})$ is an ASR 
  parser configuration $\buc(\psrstack S, g)$ together with a rest graph 
  ${\bar g}\in \Gr\TLab$. $\psrbuc(\psrstack S, g, {\bar g})$ is \emph{accepting}
  if ${\bar g} = \emptyseq$ and $\buc(\psrstack S, g)$ is an accepting ASR parser 
  configuration.

  A \emph{PSR move} turns $\psrbuc(\psrstack S, g, {\bar g})$ into $\psrbuc(\psrstack S', g',{\bar g}')$,
  written $\psrbuc(\psrstack S, g, {\bar g}) \psrstep \psrbuc(\psrstack S', g',{\bar g}')$,
  if 
  \begin{enumerate}[(i)]
    \item \funref[Q,\tau,\bar g]{alg:select-edge-action} returns 
          $(t, \literal e)$, where 
          $\psrtop{\psrstack S} = Q^\tau$ for a state 
          $Q \in \mathcal{Q}$ and input binding~$\tau$,
    \item $g' = g \literal e$ if $t$ is a transition and $g' = g$ otherwise,  
    \item $g{\bar g} \perm g'{\bar g}'$, and\label{i:PSR-parser-c}
    \item $\buc(\psrstack S, g) \lpstep{t} \buc(\psrstack S', g')$ is an
          ASR move. 
  \end{enumerate}
   
  A PSR parser can \emph{reach} a configuration $\psrbuc(\psrstack S', g',{\bar g}')$ 
  from $\psrbuc(\psrstack S, g,{\bar g})$ if 
  $
  \psrbuc(\psrstack S, g,{\bar g})
  \psrstep^*
  \psrbuc(\psrstack S', g',{\bar g}').
  $
  A PSR parser \emph{accepts} a graph $g \in \Gr\TLab$ if it can reach an 
  accepting configuration $\psrbuc(\psrstack S, g', \emptyseq)$ 
  from $\psrbuc(\startstate^\iota, \emptyseq, g)$.
\end{defn}

We are now ready for the main result of this paper. It shows that a PSR
parser, for an HR grammar satisfying \assref{a:conflict-free-and-fec},
accepts a graph if and only if it is valid. Moreover, a PSR parser
cannot run into dead ends, i.e., whenever a PSR parser fails, there
cannot be a different sequence of parser moves that is successful:

\begin{theorem}\label{t:psr-correct}
  Let 
  $
     \psrbuc(\startstate^\iota, \emptyseq, g) 
     \psrstep^*
     \psrbuc(\psrstack S, g',{\bar g}')
  $
  be any sequence of PSR moves. The PSR parser can reach an accepting 
  configuration from $\psrbuc(\psrstack S, g',{\bar g}')$ if and only if 
  $g$ is valid.
\end{theorem}

 \begin{proof}
  Let us assume any HR grammar satisfying \assref{a:conflict-free-and-fec}, 
  graph $g \in \Gr\TLab$ and sequence 
  $
     \psrbuc(\startstate^\iota, \emptyseq, g) 
     \psrstep^n
     \psrbuc(\psrstack S, g',{\bar g}')
  $
  of PSR moves. 
  
  To show the \emph{only-if} direction, let us assume that the parser can reach 
  an accepting configuration 
  $\psrbuc(\psrstack S', g'', \emptyseq)$ from
  $\psrbuc(\psrstack S, g',{\bar g}')$. Then $\startLit \drm^* g''$ follows from
  \thmref{t:ASR} and the fact that each sequence of PSR moves is also a sequence
  of ASR moves. Moreover, $g \perm g''$ follows from 
  \defref{def:PSR-parser}(\ref{i:PSR-parser-c}), i.e., $g \in \L(\Gamma)$.
  
  To show the \emph{if} direction, let us assume that $g$ is valid. We 
  first show, as an auxiliary result, that the ASR parser can reach an accepting ASR parser configuration
  $\buc(\psrstack S',g'')$ from $\buc(\psrstack S, g')$ and $g'' \perm g$
  by induction on $n$.
  For $n=0$, the ASR parser can reach an accepting configuration
  $\buc(\psrstack S',g'')$ by \thmref{t:ASR}.
  For $n>0$, there is a sequence 
  \begin{equation}
    \psrbuc(\startstate^\iota, \emptyseq, g) 
    \psrstep^{n-1}
    \psrbuc(\psrstack S'', h,\bar h)
    \psrstep
    \psrbuc(\psrstack S, g',{\bar g}')
    \label{eq:psr-correct-a}
  \end{equation}
  of PSR moves and, using the induction hypothesis, an accepting ASR
  parser configuration $\buc(\psrstack S'',h')$ that the ASR parser can reach 
  from $\buc(\psrstack S'', h)$. The last PSR move in 
  \eqref{eq:psr-correct-a} uses \funref{alg:select-edge-action}. Therefore, 
  \lemmaref{lemma:chooses-right-action} applies, and the ASR parser can
  reach an accepting configuration $\buc(\psrstack{\bar S}'',h'')$ where 
  $h'' \perm g$. This proves our auxiliary result, which we use now 
  to show the \emph{if} direction. By using 
  \lemmaref{lemma:chooses-right-action} again, the auxiliary result implies 
  that every sequence of PSR moves starting at 
  $\psrbuc(\startstate^\iota, \emptyseq, g)$
  either ends with an accepting configuration, or can be extended by yet 
  another PSR move. This process can be repeated, but it must end eventually 
  because there cannot be an infinite sequence of ASR moves and, thus, of 
  PSR moves because of \corelref{c:linear-bounds}. This proves the 
  \emph{if} direction since the final configuration of this sequence must 
  be an accepting one.
  \qed
\end{proof}

Note, however, that the parser is still nondeterministic, despite the
fact that it chooses the trigger causing the next parser move for
every configuration deterministically. The reason is that 
\funref{alg:select-edge-action} does not uniquely determine
the literal to be processed by the shift move to be made. 
For instance, the ASR parser moves shown in \figref{f:psrmoves}, which
are also valid PSR parser moves, choose edge $e^{12}$ in the third
move, but could have chosen $e^{13}$ instead, keeping $e^{12}$ for
later. There are thus two different sequences of parser moves that
both prove the validity of the given input graph, i.e., the PSR parser
is nondeterministic. However, this nondeterminism is harmless as it
does not make a difference when it comes to acceptance because of the
free edge choice of the grammar.

The \emph{Grappa} tool implemented by Mark Minas generates \PSR
parsers based on the construction of the dCFA and the analysis of
the criteria for \assref{a:conflict-free-and-fec} outlined above. 
\tabref{t:testresults} summarizes
results for some HR grammars.  The columns under ``Grammar''
indicate the size of the grammar in terms of the maximal arity of
nonterminals (A), number of nonterminals (N), number of terminals (T)
and number of rules (R). The columns under ``dCFA'' indicate the size
of the generated dCFA in terms of the number of states (S), the overall
number of items (I) and the number of transitions ($\Delta$). The number of
conflicting sets in the dCFA is shown in the column ``Conflicts''. The last 
column indicates whether the grammar satisfies the free edge choice property.
Note that the PSR parser can successfully be generated for the grammars
without any conflicts and satisfying the free edge choice property. 
For the others, the parser generator fails with
a message pointing out the reason of the failure. 
\begin{table}[htb]
\caption{PSR-Parsability of some HR grammars.}
\label{t:testresults}
\begin{center}
\small
\begin{tabular}{|l|*{3}{c@{\hspace*{1.5mm}}}c|*{2}{c@{\hspace*{2mm}}}c|c|c|}
\hline
& \multicolumn{4}{c|}{Grammar} & \multicolumn{3}{c|}{dCFA} & Conflicts & Free edge \\
\raisebox{3ex}[-3ex]{Example} & A & N & T & R & S & I & $\Delta$ &  & choice prop \\[1pt] \hline
Persuade (\exref{x:boygirl:HR}) & 4 & 1 & 3 & 5 & 9 & 36 & 20 & -- & yes \\
Trees (\exref{x:tree:HR}) & 1 & 2 & 1 & 3 & 4 & 10 & 4 & -- & yes \\
$a^nb^nc^n$ \cite{Drewes-Hoffmann-Minas:15} & 4 & 3 & 3 & 5 & 14 & 22 & 14 & --  & yes \\
\parbox{3cm}{Nassi-Shneiderman diagrams~\cite{minas97}} & 4 & 3 & 3 & 6 & 12 & 78 & 59 & --  & yes \\
Palindromes \cite{Drewes-Hoffmann-Minas:15} & 2 & 2 & 2 & 7 & 12 & 32 & 19 & --  & yes \\
Arithmetic expressions & 2 & 4 & 5 & 7 & 12 & 34 & 22 & --  & yes \\
Flowcharts (\exref{x:flowcharts}) & 2 & 3 & 4 & 6 & 14 & 75 & 50 &  1  & yes \\
Series-parallel graphs & 2 & 2 & 1 & 4 & 7 & 63 & 32 &  6  & no \\\hline
\end{tabular}
\end{center}
\end{table}

We refer the reader to \cite[Sect.~6]{hoffmann-minas:17b} for runtime
measurements of PSR parsers that confirm that they run in linear time,
for all practical purposes. Let us finally outline why this is the case
for an HR grammar satisfying \assref{a:conflict-free-and-fec}: The
number of moves required by a PSR parser grows linearly with the number
of literals of the input graph, i.e., its size, and independent of the
graph being valid or not (\corelref{c:linear-bounds}). For each move,
the parser must first call \funref{alg:select-edge-action} and then
execute the selected action. The latter part depends on the grammar
only and is in particular independent of the size of the input graph,
i.e., it always takes constant time.

Let us now discuss the efficiency of \funref{alg:select-edge-action},
i.e., consider how long an invocation of this procedure takes. The
number of triggers limits the number of how often the loop starting in
\alineref{an:for} is executed. This number is bounded by a constant
that depends on the grammar only. The runtime of
\funref{alg:select-edge-action} is thus determined by the time required
to execute \alineref{an:lookup}. It must find a literal~$\literal e$ in
the rest graph such that it ``fits'' some member of $\Fo(Q,t_i)$. Note
that the size of the latter set is also bounded by a constant that
depends on the grammar only.

We now consider any pseudo-literal in $\Fo(Q,t_i)$, and argue that
looking up a literal of the rest graph that fits this pseudo-literal
always takes constant time if the input graph has been pre-processed in
an appropriate way. Note that a certain number of attached nodes of
such a literal is already fixed by the pseudo-literal via the $\fits
\tau Q$ relation. If no node is fixed, selecting such a literal is
easy: One picks an arbitrary unread literal with the appropriate label.
This must be the right choice because of the free edge choice
property. If exactly one node is fixed, one picks any unread literal
that is attached to this node. This can be done in constant time if
each node maintains a list of all unread literals attached to this
node. Such association lists can be set up prior to parsing in time
$O(n+m)$ if the input graph has $n$~literals and $m$~nodes. And if
two or more nodes are fixed, selecting any unread literal attached to
these nodes can also be done in (worst case) constant time by
maintaining association lists of yet unread literals for each of these
node combinations and storing these association lists in hash tables
that allow efficient lookup by these node combinations, for instance
using \emph{Cuckoo Hashing} \cite{Pagh-Rodler:2001}, which supports
(worst case) constant time access during parsing, and an expected linear
time set-up once prior to parsing.

As a consequence, a PSR parser
always has a (worst case) linear parsing time. It additionally needs a
set-up of some search data structures in either worst case linear or
expected linear time, though. Expected linear time processing is only
needed if \funref{alg:select-edge-action} must look up unread literals
with two or more nodes fixed. But we have argued
in~\cite[Sect.~5]{hoffmann-minas:17b} that one can improve
preprocessing even for many of those grammars: Even though
\funref{alg:select-edge-action} would have to look up unread literals
with two or more nodes fixed, static analysis of the grammar may reveal
that it is in fact sufficient to consider just one of these fixed
nodes. The PSR parser for grammars with this property thus need not
manage association lists in hash tables, which would require expected
linear time for preprocessing, but can attach them to nodes, which
requires linear time in the worst case for preprocessing.
In fact, we had to particularly handcraft an HR grammar (the
``blowball'' grammar in~\cite{hoffmann-minas:17b}) that indeed requires
hash tables for efficient parsing and thus preprocessing in
expected linear time. All PSR parsers for grammars without conflicts in
\tabref{t:testresults} need linear time in the worst case though.



\section{Conclusions}\label{s:concl}
We have devised a predictive shift-reduce (\PSR) parsing algorithm for
HR grammars, along the lines of \SLR(1) string parsing, thus continuing
the work begun in~\cite{Drewes-Hoffmann-Minas:17} by formalizing
the construction of PSR parsers and proving its correctness.
%
%
\PSR parsing is somewhat complementary to predictive top-down (\PTD) 
parsing that lifts \SLL(1) parsing for string grammars to HR 
grammars~\cite{Drewes-Hoffmann-Minas:15}. Checking \PSR-parsability 
is complicated enough, but easier than for \PTD, as we do not need 
to consider HR rules that merge nodes of their
left-hand sides, which is necessary for \PTD parsing.
\PSR parsers also work more efficiently than \PTD parsers: while PTD
parsers require quadratic time in the worst case, PSR parsers run in
linear time for all practical purposes; see the discussion in \sectref{s:PSR},
and in particular \corelref{c:linear-bounds}. The reader is encouraged to download the
\emph{Grappa} generator of \PTD and \PSR parsers and to conduct own
experiments.\addtocounter{footnote}{-1}%
\footnote{\label{grappa} The \emph{Grappa} tool is available at
  \href{https://www.unibw.de/inf2/grappa/}{www.unibw.de/inf2/grappa};
  the examples mentioned in \tabref{t:testresults} can be found there
  as well.} %

\subsection*{Related Work} 
Much related work on graph parsing has been done for graph grammars
based on context-free node replacement \cite{Engelfriet-Rozenberg:97}.
In these grammars, a node $v$ is replaced by a graph $R$, where
embedding instructions specify what happens to the edges incident in
$v$; in general, such an edge can just be deleted, or turned around, or
replicated and directed towards different nodes of $R$. Node
replacement has greater generative power, but is difficult to handle
for general embedding instructions.
So papers on parsing for node replacement graph grammars restrict these instructions.
The earliest ones (to our knowledge), by T.~Pavlides, T.W.~Pratt, and P. Della Vigna and C. Ghezzi,
\cite{Pavlidis:72,Pratt:71,DellaVigna-Ghezzi:78}, appeared well before
visual user interfaces supported input and processing of diagrams by
computers.
R.~Franck \cite{Franck:1978} extended precedence string parsing 
to graphs, in order to implement a ``\emph{two-dimensional programming
  language}'' based on Algol-68. W.~Kaul corrected and extended this
idea of parsing \cite{Kaul:86}. His parser is linear, and can cope
with ambiguous grammars, but fails to parse some languages that are
both \PSR- and \PTD-parsable, like the trees of \exref{x:tree:HR}.

A parsing algorithm following the idea of the well-known Cocke-Younger-Kasami algorithm
was proposed and investigated by C.~Lautemann~\cite{Lautemann:90} who gave a sufficient condition
under which this algorithm is polynomial. However, even if the condition is met, the degree
of the polynomial depends on the grammar. The algorithm was recently refined by D.~Chiang et
al.~\cite{chiang-et-al:2013}, making it more practical but without changing its general
characteristics. An alternative algorithm developed by W.~Vogler
in~\cite{Vogler:91} and generalized by F.~Drewes in~\cite{Drewes:93c} guarantees a cubic running
time at the expense of employing a severe connectedness requirement. Due to this requirement
it seems fair to say that this algorithm is mainly of theoretical interest. A promising approach
for certain types of applications, especially for graph languages appearing in computational
linguistics, has recently been proposed by S.~Gilroy, A.~Lopez, and
S.~Maneth~\cite{Gilroy-Lopez-Maneth:17}. This parsing algorithm applies to Courcelle's ``regular''
graph grammars~\cite{Courcelle:91b} and runs in linear time.

Over the years, M.~Flasiński and his group have developed top-down
and bottom-up parsing techniques for pattern recognition
\cite{Flasinski:89,Flasinski:98,Flasinski-Flasinski:14}. The graph
classes they consider are very restricted: rooted directed acyclic
graphs with ordered nodes. Their parsers are also linear, but this is
achieved by forbidding all concepts that make graph parsing essentially
different from string parsing.
According to our knowledge, another early attempt at
\textit{LR}-like graph parsers by H.J.~Ludwigs \cite{Ludwigs:80} has
never been completed.

G.~Costagliola's positional grammars \cite{costagliola97a} are used to
specify visual languages, but they can also describe certain HR
languages. Although they are parsed in an LR-like fashion, many
decisions are deferred until the parser is actually executed, in order
to avoid complex analyses of the grammar when the parsers are
generated. In contrast, the \PSR parser generator implemented
in the \emph{Grappa} tool performs an elaborate static analysis of the
grammar. It includes the detection of conflicts that prevent the parser
from running into situations where, despite the use of a dCFA,
a nondeterministic choice must be
made (i.e., backtracking must be employed). It also checks and makes use
of other properties, such as the so-called
free-edge-choice property, and the existence of uniquely determined
start nodes. As mentioned before, the precise 
discussion of these analysis techniques will be presented in a follow-up
paper.

The CYK-style parsers for unrestricted HR grammars (plus
edge-embedding rules) implemented in DiaGen \cite{minas97} work for
practical input with hundreds of nodes and edges, although their
worst-case complexity is exponential. A closer comparison with \PTD and
\PSR parsers shows its limits with larger input
\cite[Sect.~6]{hoffmann-minas:17b}.

\subsection*{Future Work} 
So far, PSR parsing has only been tested on small HR grammars; most of
them are not relevant for practical modeling of graph and diagram
languages (see \tabref{t:testresults}).
Indeed, HR grammars cannot generate graph languages like Petri nets
or UML diagrams -- not even the language of all graphs over some
alphabet! For instance, HR languages do always have bounded
treewidth.
Therefore the authors have proposed a modest extension of this
formalism \cite{Drewes-Hoffmann:15,Drewes-Hoffmann-Minas:12}:
\emph{contextual HR grammars} allow to connect the right-hand side of
a rule to a node that does not occur in the left-hand side
nonterminal, but exists elsewhere in the context. In this way,
some of the structural restrictions
of HR languages can be overcome.
Fortunately, PSR parsing for HR grammars can be extended to
contextual HR grammars in a straightforward way: when shifting an
edge attached to a context node, the parser must just be ready to
match this node to a rule that has already been read.
(This has already been implemented for the PTD
  parsing of graphs~\cite{Drewes-Hoffmann-Minas:15}.)

However, the grammar in \exref{x:boygirl:HR} indicates that HR
grammars do have practical applications: for \emph{abstract meaning
  representations} of natural language, HR grammars are being used in
the natural language community
\cite{chiang-et-al:2013}. Unfortunately the grammars occurring in this
domain are not only huge, with thousands of rules, but also ambiguous,
so that they do not lend themselves to PSR parsing.
We are currently working on a \emph{generalized} PSR parsing algorithm
that pursues several PSR parses in parallel if states do have
conflicts or if the free edge choice property is violated~\cite{hoffmann-minas:19}.



%

We will also study the relationship between \PTD and \PSR parsing.
Bottom-up string parsing is known to be more powerful than top-down
string parsing in the sense that all top-down parsable languages are
also bottom-up parsable, but not vice versa. \PTD and \PSR parsing are
extensions of these string parsing approaches. Therefore, it appears to
be an obvious assumption that \PSR is more powerful than \PTD. So far,
we have no conclusive answers to this question. One of the challenges
is to find an HR (or a contextual HR) language that has a \PSR parser,
but no \PTD parser. The corresponding example for \LL(k) and \LR(k)
string languages exploits that strings are always parsed from left to
right---but this is not the case for \PTD and \PSR parsers.


\section*{References}
\bibliography{refs}

\end{document}